\documentclass[11pt]{article}
\usepackage{arxiv}
\newcommand{\N}{\mathcal{N}}
\newcommand{\G}{G_{\theta}}
\newcommand{\ImagePopulation}{33,140}
\newcommand{\TrainIdentities}{29,816}
\newcommand{\ValidationIdentities}{2,000}
\newcommand{\AugmentedViews}{900,000}

\newcommand{\ImageValidationRMSE}{0.0817}

\newcommand{\PretrainedRegionalRMSE}{0.0470}
\newcommand{\RandomRegionalRMSE}{0.1345}
\newcommand{\DCTRegionalRMSE}{0.0496}
\newcommand{\PretrainedOpenFWIRMSE}{0.0524}
\newcommand{\RandomOpenFWIRMSE}{0.3775}
\newcommand{\DCTOpenFWIRMSE}{0.2130}
\newcommand{\TypicalOpenFWIRMSE}{0.0536}
\newcommand{\PretrainedOpenFWIGradient}{0.513}

\newcommand{\OpenFWIRandomRatio}{7.58}
\newcommand{\OpenFWIDCTRatio}{4.13}

\newcommand{\RandomRegionalSeedSD}{0.0053}
\newcommand{\RandomOpenFWISeedSD}{0.0044}
\newcommand{\RandomRegionalCorrelation}{0.8648}
\newcommand{\RandomOpenFWICorrelation}{0.8555}
\newcommand{\RandomRegionalGradient}{0.4778}
\newcommand{\RandomOpenFWIGradient}{0.0920}
\newcommand{\RandomOpenFWISeedMin}{0.3727}
\newcommand{\RandomOpenFWISeedMax}{0.3841}

\newcommand{\HMCCoverageVP}{0.882}
\newcommand{\HMCCoverageVS}{0.913}
\newcommand{\HMCCoverageRho}{0.884}
\newcommand{\HMCSDVP}{0.259}
\newcommand{\HMCSDVS}{0.159}
\newcommand{\HMCSDRho}{0.126}
\newcommand{\VIHMCSDRatioVP}{2.02}
\newcommand{\VIHMCSDRatioVS}{1.97}
\newcommand{\VIHMCSDRatioRho}{2.29}
\newcommand{\VILatentMeanSD}{0.803}
\newcommand{\HMCLatentMeanSD}{0.882}
\newcommand{\HMCLatentMeanSDRatio}{1.27}
\newcommand{\HMCLatentMedianSDRatio}{1.04}
\newcommand{\HMCLatentMaximumCorrelation}{0.780}
\newcommand{\HMCFullGaussianMaximumSDGap}{1.4\%}
\newcommand{\HMCDiagonalGaussianMaximumSDGap}{10.4\%}
\newcommand{\HMCSpatialPITMeanVP}{0.522}
\newcommand{\HMCSpatialPITMeanVS}{0.492}
\newcommand{\HMCSpatialPITMeanRho}{0.423}
\newcommand{\HMCSpatialCoverageNinetyFiveVP}{0.939}
\newcommand{\HMCSpatialCoverageNinetyFiveVS}{0.957}
\newcommand{\HMCSpatialCoverageNinetyFiveRho}{0.951}
\newcommand{\HMCSpatialCoverageMaxGapVP}{0.019}
\newcommand{\HMCSpatialCoverageMaxGapVS}{0.041}
\newcommand{\HMCSpatialCoverageMaxGapRho}{0.079}

\newcommand{\HMCRhatMax}{1.007}

\newcommand{\SensitivityRatioVP}{1.13}
\newcommand{\SensitivityRatioVS}{1.37}
\newcommand{\SensitivityRatioRho}{1.18}
\newcommand{\SensitivityCorrelationVP}{-0.156}
\newcommand{\SensitivityCorrelationVS}{-0.494}
\newcommand{\SensitivityCorrelationRho}{-0.394}

\newcommand{\HMCStabMeanTwoFiveSix}{0.00571}
\newcommand{\HMCStabSDTwoFiveSix}{0.00530}
\newcommand{\HMCStabLowTwoFiveSix}{0.01070}
\newcommand{\HMCStabHighTwoFiveSix}{0.01816}
\newcommand{\HMCStabInclusionTwoFiveSix}{0.01697}
\newcommand{\HMCStabRatioTwoFiveSix}{0.05248}
\newcommand{\HMCStabMeanOneZeroTwoFour}{0.00215}
\newcommand{\HMCStabSDOneZeroTwoFour}{0.00216}
\newcommand{\HMCStabLowOneZeroTwoFour}{0.00457}
\newcommand{\HMCStabHighOneZeroTwoFour}{0.00768}
\newcommand{\HMCStabInclusionOneZeroTwoFour}{0.00623}
\newcommand{\HMCStabRatioOneZeroTwoFour}{0.00724}
\newcommand{\HMCStabMeanTwoZeroFourEight}{0.00159}
\newcommand{\HMCStabSDTwoZeroFourEight}{0.00138}
\newcommand{\HMCStabLowTwoZeroFourEight}{0.00321}
\newcommand{\HMCStabHighTwoZeroFourEight}{0.00463}
\newcommand{\HMCStabInclusionTwoZeroFourEight}{0.00488}
\newcommand{\HMCStabRatioTwoZeroFourEight}{0.00814}

\newcommand{\CheckpointHash}{37d9494a4b9ccffb6f3c7ca31aefc207576dcbc6a03a28ec8ae814da6b2daf13}

\newcommand{\FieldTravelTrain}{192}
\newcommand{\FieldTravelDevelopment}{64}
\newcommand{\FieldTravelHoldout}{128}
\newcommand{\FieldGravityTrain}{192}
\newcommand{\FieldGravityDevelopment}{64}
\newcommand{\FieldGravityHoldout}{64}
\newcommand{\FieldSurfaceTrain}{143}
\newcommand{\FieldSurfaceDevelopment}{28}
\newcommand{\FieldSurfaceHoldout}{40}
\newcommand{\FieldTravelPriorRMSE}{1.255}
\newcommand{\FieldTravelPosteriorRMSE}{0.726}
\newcommand{\FieldTravelReduction}{42.1\%}
\newcommand{\FieldTravelCRPSReduction}{28.3\%}
\newcommand{\FieldGravityPriorRMSE}{7.635}
\newcommand{\FieldGravityPosteriorRMSE}{4.278}
\newcommand{\FieldGravityReduction}{44.0\%}
\newcommand{\FieldGravityCRPSReduction}{42.5\%}

\newcommand{\FieldTravelPredictiveInclusion}{1.000}
\newcommand{\FieldGravityPredictiveInclusion}{0.922}
\newcommand{\FieldSurfacePredictiveInclusion}{0.900}
\newcommand{\FieldRepeatMinimumCorrelation}{0.9996}
\newcommand{\FieldCVMHCorrelationVP}{0.608}
\newcommand{\FieldCVMHCorrelationVS}{0.619}
\newcommand{\FieldCVMHCorrelationRho}{0.690}

\newcommand{\TwoStepSourceCurves}{4,076}
\newcommand{\TwoStepProfileSites}{25}
\newcommand{\TwoStepPeriods}{17}
\newcommand{\TwoStepTrainSites}{15}
\newcommand{\TwoStepDevelopmentSites}{5}
\newcommand{\TwoStepHoldoutSites}{5}
\newcommand{\TwoStepUsedSourceCurves}{31}
\newcommand{\TwoStepLocalMeanRMSE}{0.0144}
\newcommand{\TwoStepLocalMaxRMSE}{0.0317}
\newcommand{\TwoStepPriorRMSE}{0.208}
\newcommand{\TwoStepPosteriorRMSE}{0.151}
\newcommand{\TwoStepRMSEDecrease}{27.7\%}

\newcommand{\TwoStepPredictiveInclusion}{0.988}
\newcommand{\TwoStepDiscrepancy}{0.0000}
\newcommand{\TwoStepModularScale}{1.496}

\newcommand{\SharedVSRMSE}{0.217}

\newcommand{\SharedVSContraction}{71.5\%}
\newcommand{\RawSharedVSCoverage}{0.551}
\newcommand{\RawSharedVSWidth}{0.434}
\newcommand{\StrongSharedVSRMSE}{0.187}

\newcommand{\StrongSharedVSContraction}{76.6\%}
\newcommand{\StrongRawSharedVSCoverage}{0.575}
\newcommand{\StrongRawSharedVSWidth}{0.366}
\newcommand{\IndependentFullVSRMSE}{0.226}

\newcommand{\IndependentFullVSContraction}{$-2.1$\%}
\newcommand{\IndependentFullVSCoverage}{0.994}
\newcommand{\IndependentFullVSWidth}{1.585}

\newcommand{\CouplingGuardSharedWeight}{0.875}
\newcommand{\CouplingGuardVSRMSE}{0.188}

\newcommand{\CouplingGuardVSContraction}{48.8\%}
\newcommand{\CouplingGuardVSCoverage}{0.927}
\newcommand{\CouplingGuardVSWidth}{0.705}
\newcommand{\CouplingGuardStressCoverage}{0.938--0.963}
\newcommand{\StrongCouplingGuardSharedWeight}{0.875}
\newcommand{\StrongCouplingGuardVSRMSE}{0.162}

\newcommand{\StrongCouplingGuardVSContraction}{52.5\%}
\newcommand{\StrongCouplingGuardVSCoverage}{0.945}
\newcommand{\StrongCouplingGuardVSWidth}{0.649}
\newcommand{\StrongCouplingGuardStressCoverage}{0.948--0.978}

\newcommand{\CVMHRawSharedVSRMSE}{0.252}
\newcommand{\CVMHRawSharedVSContraction}{66.5\%}
\newcommand{\CVMHRawSharedVSCoverage}{0.581}
\newcommand{\CVMHRawSharedVSWidth}{0.511}
\newcommand{\CVMHStrongRawSharedVSRMSE}{0.222}
\newcommand{\CVMHStrongRawSharedVSContraction}{75.0\%}
\newcommand{\CVMHStrongRawSharedVSCoverage}{0.483}
\newcommand{\CVMHStrongRawSharedVSWidth}{0.390}
\newcommand{\CVMHIndependentFullVSRMSE}{0.204}
\newcommand{\CVMHIndependentFullVSContraction}{$-0.7$\%}
\newcommand{\CVMHIndependentFullVSCoverage}{0.997}
\newcommand{\CVMHIndependentFullVSWidth}{1.581}
\newcommand{\CVMHCouplingGuardVSRMSE}{0.231}
\newcommand{\CVMHCouplingGuardVSContraction}{47.6\%}
\newcommand{\CVMHCouplingGuardVSCoverage}{0.926}
\newcommand{\CVMHCouplingGuardVSWidth}{0.733}
\newcommand{\CVMHCouplingGuardStressCoverage}{0.912--0.935}
\newcommand{\CVMHStrongCouplingGuardVSRMSE}{0.200}
\newcommand{\CVMHStrongCouplingGuardVSContraction}{53.4\%}
\newcommand{\CVMHStrongCouplingGuardVSCoverage}{0.891}
\newcommand{\CVMHStrongCouplingGuardVSWidth}{0.638}
\newcommand{\CVMHStrongCouplingGuardStressCoverage}{0.864--0.916}

\newcommand{\OpenFWIInverseImageRMSE}{0.083}
\newcommand{\OpenFWIInverseMaternRMSE}{0.085}

\newcommand{\FieldTrendTravelRMSE}{1.195}
\newcommand{\FieldTrendGravityRMSE}{7.033}
\newcommand{\FieldRBFTravelRMSE}{0.703}
\newcommand{\FieldRBFGravityRMSE}{4.006}
\newcommand{\GroupedImageTravelRMSE}{1.014}
\newcommand{\GroupedImageGravityRMSE}{4.313}
\newcommand{\GroupedDCTTravelRMSE}{1.259}
\newcommand{\GroupedDCTGravityRMSE}{4.821}

\newcommand{\VIRankEightVPMeanGap}{0.0263}
\newcommand{\VIRankEightVSMeanGap}{0.00650}
\newcommand{\VIRankEightRhoMeanGap}{0.00578}
\newcommand{\VIRankEightTravelRMSE}{0.715}
\newcommand{\VIRankEightGravityRMSE}{4.362}
\newcommand{\VISecondSeedVPMeanGap}{0.0135}
\newcommand{\VISecondSeedVSMeanGap}{0.00692}
\newcommand{\VISecondSeedRhoMeanGap}{0.00355}
\newcommand{\VISecondSeedTravelRMSE}{0.725}
\newcommand{\VISecondSeedGravityRMSE}{4.287}
\newcommand{\VIRankFortyEightVPMeanGap}{0.0381}
\newcommand{\VIRankFortyEightVSMeanGap}{0.0123}
\newcommand{\VIRankFortyEightRhoMeanGap}{0.00828}
\newcommand{\VIRankFortyEightTravelRMSE}{0.733}
\newcommand{\VIRankFortyEightGravityRMSE}{4.247}

\newcommand{\MeasuredTransferImagePriorRMSE}{0.149}
\newcommand{\MeasuredTransferImagePriorCRPS}{0.098}
\newcommand{\MeasuredTransferImagePriorInclusion}{0.627}
\newcommand{\MeasuredTransferImagePosteriorRMSE}{0.213}
\newcommand{\MeasuredTransferImagePosteriorCRPS}{0.165}
\newcommand{\MeasuredTransferImagePosteriorInclusion}{0.137}
\newcommand{\MeasuredTransferIndependentPriorRMSE}{0.150}
\newcommand{\MeasuredTransferIndependentPriorCRPS}{0.098}
\newcommand{\MeasuredTransferIndependentPriorInclusion}{0.608}
\newcommand{\MeasuredTransferIndependentPosteriorRMSE}{0.150}
\newcommand{\MeasuredTransferIndependentPosteriorCRPS}{0.098}
\newcommand{\MeasuredTransferIndependentPosteriorInclusion}{0.608}
\newcommand{\MeasuredTransferSmoothPriorRMSE}{0.147}
\newcommand{\MeasuredTransferSmoothPriorCRPS}{0.096}
\newcommand{\MeasuredTransferSmoothPriorInclusion}{0.620}
\newcommand{\MeasuredTransferSmoothPosteriorRMSE}{0.097}
\newcommand{\MeasuredTransferSmoothPosteriorCRPS}{0.059}
\newcommand{\MeasuredTransferSmoothPosteriorInclusion}{0.761}

\title{From Image Morphology to Multiphysics Earth Posteriors}
\author[1]{Ziye Yu\thanks{Corresponding author: \href{mailto:yuziye@cea-igp.ac.cn}{yuziye@cea-igp.ac.cn}. ORCID: \href{https://orcid.org/0000-0002-1720-3811}{0000-0002-1720-3811}.}}
\author[2,3]{Xin Liu}
\author[4]{Yuqi Cai}
\affil[1]{Institute of Geophysics, China Earthquake Administration, Beijing 100081, China}
\affil[2]{Laboratory of Seismology and Physics of Earth's Interior, School of Earth and Space Sciences, University of Science and Technology of China, Hefei 230026, China}
\affil[3]{Institute of Advanced Technology, University of Science and Technology of China, Hefei 230088, China}
\affil[4]{University of Chinese Academy of Sciences, Beijing 100029, China}
\date{}
\hypersetup{pdftitle={From Image Morphology to Multiphysics Earth Posteriors},pdfauthor={Ziye Yu, Xin Liu, Yuqi Cai}}

\begin{document}
\maketitle

\begin{abstract}
Structural learning and physical-property inference can draw on different sources of knowledge. We learn morphology from 33,140 natural, satellite, and texture images, without geological training models, then infer Earth properties from physical observations. Identical frozen weights map 128 coordinates into each $256\times256$ compressional-velocity, shear-velocity, or density field. Fixed adapters assign physical meaning; observation-specific likelihoods determine new posteriors without retraining. Across 256 unseen Earth models, image coordinates reduce projection error fourfold relative to a dimension-matched cosine basis. Matched inversions favor image coordinates for the tested curved-velocity targets and a smooth radial basis for the tested faulted targets. A simplified 2.5-dimensional application with 915 measured observations along one 120.9-km southern-California profile constrains a velocity--density contrast. Grouped arrival-time and gravity errors fall by 43\% and 44\% relative to the pretrained image prior, although the matched smooth radial basis predicts better. Direct layered-Rayleigh updating reduces blind phase-velocity error by 28\% at five sites, whereas P-and-gravity-only sharing worsens it, distinguishing reusable morphology from beneficial cross-property coupling. These results establish a practical separation: images supply structural alternatives, and heterogeneous physical observations determine their experiment-specific posterior probabilities.
\end{abstract}


\section*{Significance}
Structural knowledge for Earth inference need not originate in geological training models. We learn spatial patterns from public images, assign physical meaning through explicit adapters, and use observations to infer seismic velocities and density. One frozen generator supports these different properties. Matched tests reveal where this choice helps: image coordinates recover the tested curved interfaces better, whereas smooth coordinates perform better on the tested faulted models and southern-California profile. Sharing coordinates across properties can also impair an unobserved property's prediction. These results separate learning a structural prior from deciding how physics constrains it. Images provide an alternative source of structural knowledge whose usefulness depends on the structures and observations being resolved.

\section*{Introduction}

Must a structural prior for the Earth be learned from Earth models? Images contain boundaries, layers, and textures without labeling seismic velocities or density. We show that these spatial patterns can supply a structural description whose physical interpretation is acquired separately. Fixed adapters define property ranges and background trends; earthquake arrivals, surface waves, and gravity constrain departures from them. Bayesian inference combines this description with the physics of each measurement to produce distributions over possible Earth models \cite{Stuart2010,HaberOldenburg1997,ColomboRovetta2018}.

Generative priors already compress inverse problems, including geological variational-autoencoder (VAE) tomography \cite{BoraEtAl2017,PatelEtAl2022,LopezAlvisEtAl2022,MelesEtAl2024}. Natural-image denoisers support posterior seismic inversion \cite{IzzatullahEtAl2024}, and structural joint inversion permits flexible coupling \cite{HaberOldenburg1997,ColomboRovetta2018,PianaAgostinettiBodin2018}. Our question is whether one image-learned description can serve different Earth properties while keeping physical knowledge explicit. One generator retains identical weights across compressional velocity $V_P$, shear velocity $V_S$, and density. Its coordinates describe spatial alternatives; physical likelihoods update their probabilities.

\begin{figure}[H]
\centering
\includegraphics[width=\linewidth]{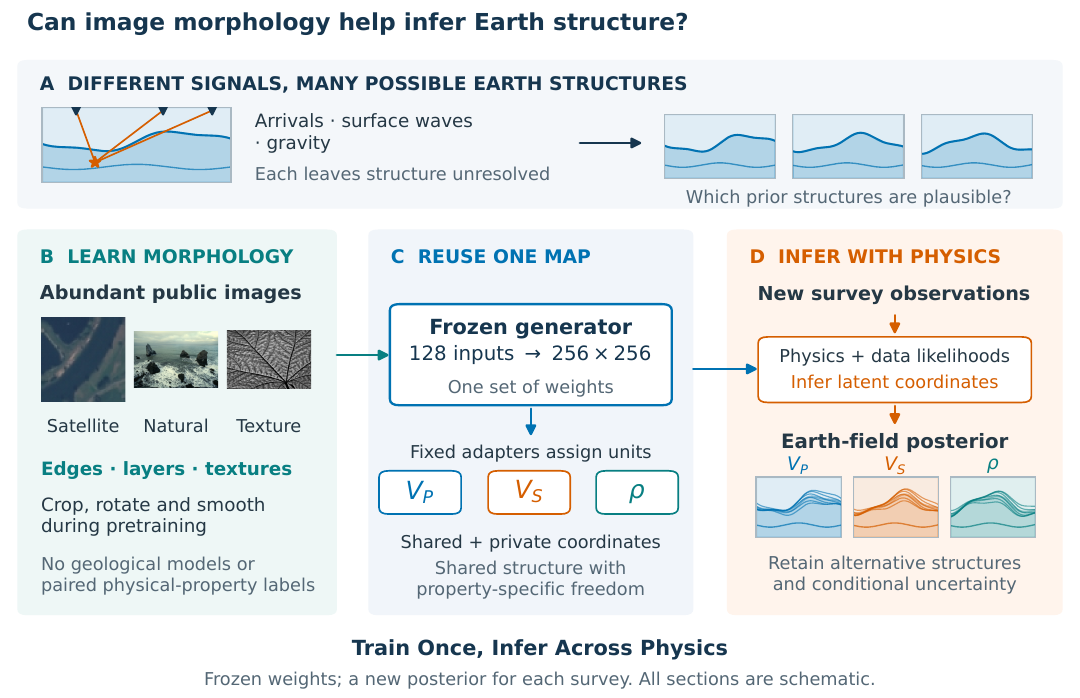}
\caption{\textbf{Can image morphology help infer Earth structure?} (A) Different physical observations leave multiple subsurface structures possible. (B) Public satellite, natural, and texture images supply morphology without geological training models or paired property labels. Examples are from EuroSAT, BSDS500, and DTD. (C) One frozen scalar generator maps 128 coordinates to a $256\times256$ field. Identical weights describe $V_P$, $V_S$, and density through fixed physical adapters; shared and private coordinates specify coupling. (D) Forward physics and data likelihoods update coordinate probabilities to produce distributions over Earth fields. ``Train once'' refers to generator weights; each experiment requires a new posterior. Sections and alternative boundaries are conceptual illustrations, not reconstructions. Quantitative tests follow in Figs.~\ref{fig:coordinates}--\ref{fig:field}.}
\label{fig:scientificquestion}
\end{figure}

The generator is trained once and remains frozen while each experiment determines a new posterior (Fig.~\ref{fig:scientificquestion}). Independent coordinates reuse morphology without cross-property dependence; shared coordinates add a testable coupling prior. We distinguish three questions: can the description represent unseen Earth structures, when does it improve recovery over smooth coordinates, and when does sharing help an unobserved property? Matched structural tests and one measured regional profile answer different parts of this question. Together they test separation of structural learning from physical inference, rather than universal superiority of a single prior.

One frozen decoder implements this separation. Let $\boldsymbol{\zeta}=(\bm z_s,\bm z_{V_P},\bm z_{V_S},\bm z_\rho)$ collect one shared and three property-specific latent blocks, each in $\mathbb R^{64}$. At spatial position $\mathbf{x}=(s,h)$, with profile coordinate $s$ and depth $h$,
\begin{equation}
\begin{aligned}
q_p(\mathbf{x})&=\G([\bm z_s,\bm z_p])(\mathbf{x}),\\
    m_p(\mathbf{x})&=\operatorname{clip}_{[\ell_p,u_p]}\!\left\{t_p(h)+a_p[2q_p(\mathbf{x})-1]\right\},\\
p(\boldsymbol{\zeta}\mid\bm d)&\propto\N(\boldsymbol{\zeta};\bm0,\bm I)\prod_k p\!\left(\bm d_k\mid F_k(\bm m_{V_P},\bm m_{V_S},\bm m_\rho)\right).
\end{aligned}
\label{eq:state}
\end{equation}
Here $p\in\{V_P,V_S,\rho\}$ and $k$ indexes physical datasets. Each coordinate changes a spatial pattern rather than one grid cell, while the pretrained weights $\theta$ remain fixed. Each property receives 128 inputs; the shared/private joint state has 256 coordinates. Shared inputs coordinate changes across properties, whereas private inputs permit departures. This is a structural prior, not a learned velocity--density law or a fixed $V_P/V_S$ ratio.

Spatial correlation is built into this parameterization. Independent latent variables need not produce independent grid cells: the generator makes spatial locations vary together. It therefore supplies a nonlinear structural constraint in place of a prescribed Gaussian model covariance. Generated ensembles reveal the resulting spatial and cross-property correlations; local derivatives connect them to conventional sensitivity and covariance analysis (\emph{SI Appendix}, Section S16).

Fixed adapters supply background $t_p$, residual amplitude $a_p$, and bounds $[\ell_p,u_p]$. Backgrounds are constant in controlled tests and depth-dependent in the field application. Each Gaussian likelihood divides residuals by its effective error SD and sums over inference observations; development data fix the error scales. Thus uncertainty, sampling, and physical sensitivity determine each dataset's contribution, not numerical units alone. The $3\times256\times256$ outputs are generated field values. Their physical resolution depends on the prior and observations, not grid spacing alone.

\begin{figure}[H]
\centering
\includegraphics[width=0.97\linewidth]{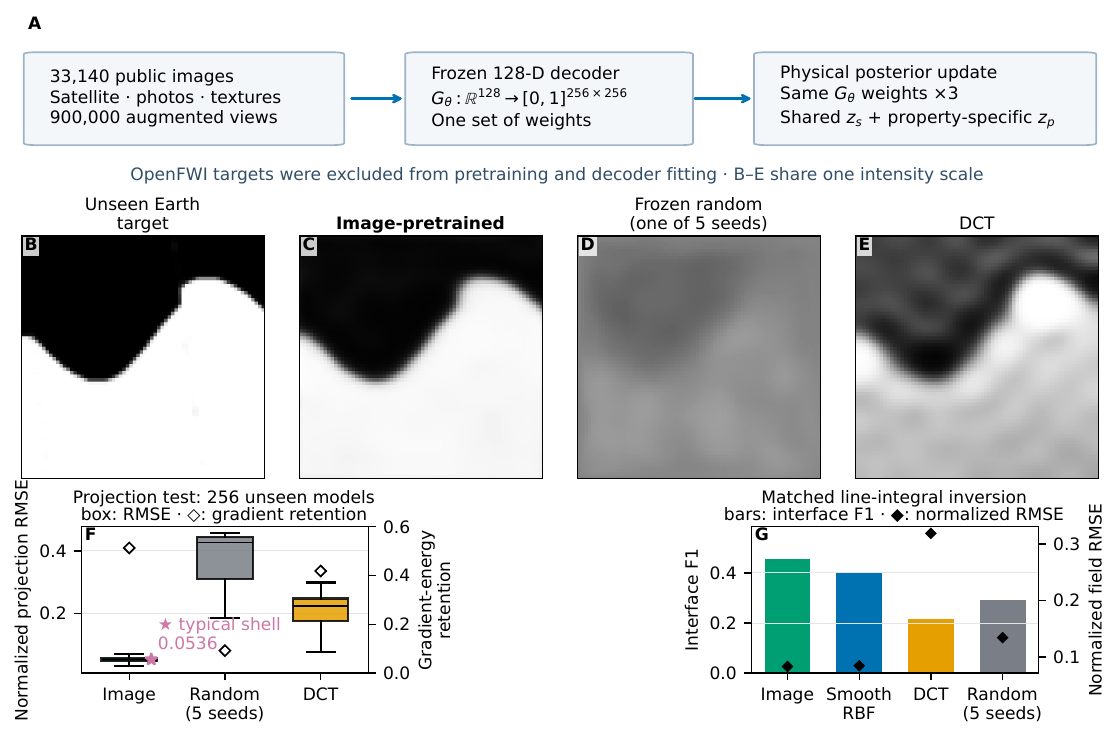}
\caption{\textbf{Generic image pretraining makes unseen Earth morphology accessible.} (A) A 33,140-image corpus yields 900,000 augmented training views for one 128-coordinate decoder. (B--E) A held-out CurveFault target and matched projections share one intensity scale. (F) Projection error over 256 OpenFWI models: boxes show quartiles, black lines medians, and whiskers extend to the most extreme values within 1.5 interquartile ranges. Random-decoder errors average five seeds per target. (G) Field error and interface F1 average eight inverse truths, four per family. Targets entered neither pretraining nor decoder fitting.}
\label{fig:coordinates}
\end{figure}

\section*{Image morphology represents unseen Earth structure}

Public EuroSAT, BSDS500, and Describable Textures images supply the \ImagePopulation-image corpus \cite{HelberEtAl2019,ArbelaezEtAl2011,CimpoiEtAl2014}. Disjoint source identities supply 29,816 training and 2,000 validation images. Cropping, rotation, reflection, contrast inversion, and multiscale smoothing produce \AugmentedViews\ source-balanced training views. No Earth model enters training, normalization, early stopping, or checkpoint selection (\emph{SI Appendix}, Section S1).

Matched representation controls isolate what pretraining contributes. Before inversion, a projection test asks how closely each 128-coordinate parameterization can represent a target under the same optimization budget. Across 256 unseen layered, curved, and faulted OpenFWI models \cite{DengEtAl2022}, normalized root mean square error (RMSE) is \PretrainedOpenFWIRMSE\ for the pretrained decoder, \RandomOpenFWIRMSE\ across five frozen-random decoders, and \DCTOpenFWIRMSE\ for a discrete cosine transform (DCT) basis. Gradient-energy retention is \PretrainedOpenFWIGradient\ for pretraining but \RandomOpenFWIGradient\ for random weights. Restricting coordinates to the 5th--95th percentile standard-normal radial shell changes error only to \TypicalOpenFWIRMSE. These fits therefore do not depend on latent vectors with unusually large magnitudes.

The benefit depends on both structure and observations. On nine smooth regional sections, pretrained and cosine coordinates are nearly tied (\PretrainedRegionalRMSE\ versus \DCTRegionalRMSE), whereas complex OpenFWI projections separate them by a factor of \OpenFWIDCTRatio. A line-integral inversion benchmark then compares eight projection-suite targets under matched dimension, acquisition, noise, prior scale, initialization, and computational budget. Priors match a predeclared field standard deviation (SD) without using target fields. Two development truths set an additive field-discrepancy SD for interval evaluation on the remaining six; structural errors use all eight.

Image coordinates outperform a smooth Gaussian radial-basis (RBF) control on curved-velocity targets in field RMSE (0.078 versus 0.113) and interface F1 (0.404 versus 0.247). On curve-fault targets, the ordering reverses: image versus RBF RMSE is 0.089 versus 0.056, and F1 is 0.510 versus 0.555. The ordering holds for all four truths in each family. Aggregate image and RBF errors are nearly tied (\OpenFWIInverseImageRMSE\ and \OpenFWIInverseMaternRMSE). These are structure-dependent outcomes under one acquisition, not evidence that image priors generally dominate smooth priors or that target morphology alone predicts the winning prior.

\section*{Shared coordinates recover an unmeasured property in known-truth tests}

\begin{figure}[H]
\centering
\includegraphics[width=0.78\linewidth]{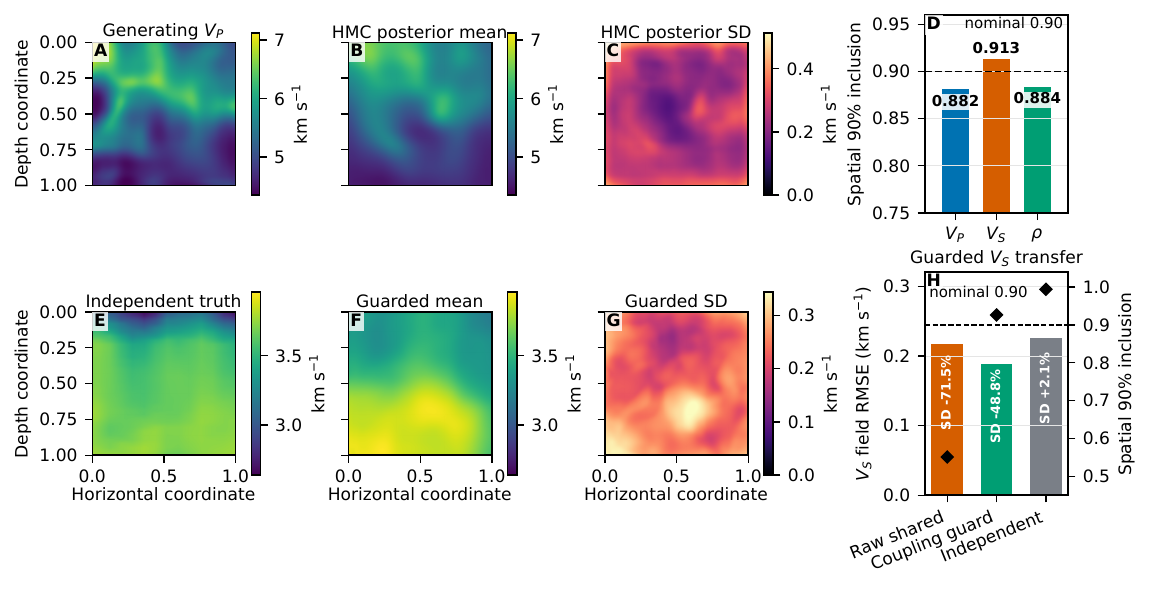}
\caption{\textbf{Explicit shared coordinates propagate posterior information into an unmeasured Earth property.} (A--D) Generating $V_P$, HMC mean and SD from 8,000 draws, and pointwise 90\% spatial inclusion. (E--H) Independent Wang2020 $V_S$ truth, pooled mean and SD with dispersion withheld, and transfer diagnostics. Raw sharing contracts but undercovers; the CVM-H development-selected, frozen-weight pool retains \CouplingGuardVSContraction\ contraction and \CouplingGuardVSCoverage\ inclusion. Wang2020 and mismatch tests are independent of decoder training and guard selection.}
\label{fig:commonstate}
\end{figure}

A decoder-generated known-truth experiment shows how one joint posterior propagates into all three native-grid Earth properties (Fig.~\ref{fig:commonstate}). Traveltime, dispersion, and gravity update the same 256-coordinate state. Four-chain Hamiltonian Monte Carlo (HMC) retains 8,000 draws. For this fixed truth, pointwise 90\% intervals include \HMCCoverageVP, \HMCCoverageVS, and \HMCCoverageRho\ of the $V_P$, $V_S$, and density fields ($\widehat R\leq\HMCRhatMax$; no divergences) \cite{VehtariEtAl2021}. Posterior SD measures the spread of decoded structures, conditional on the generator, coupling, and likelihood. Spatial inclusion describes this one generating field, not calibration across repeated truth--data pairs (\emph{SI Appendix}, Section S6).

Generic images learn the morphological map, not a geophysical relation among $V_P$, $V_S$, and density. We remove dispersion, leaving no direct $V_S$ likelihood term, and test against an independent Wang2020 section \cite{Wang2020Model}. Raw sharing lowers $V_S$ error and contracts its posterior SD. A separate 256-dimensional factorial supports transfer at fixed total dimension but changes each property's marginal prior by reducing independent active inputs (\emph{SI Appendix}, Section S11). It complements the full 128-input independent comparison below, without isolating coupling at identical marginals.

To retain useful transfer under structural mismatch, we combine decoded field distributions as $q_{\rm pool}(\bm m)=wq_{\rm shared}(\bm m\mid\bm d)+(1-w)q_{\rm independent}(\bm m\mid\bm d)$. The largest admissible shared weight, \CouplingGuardSharedWeight, is selected on five CVM-H-derived development cases \cite{ShawEtAl2015,CVMHData2020} under a predefined 0.88--0.95 spatial-inclusion criterion, then frozen before every Wang2020 test. This linear pool is a development-selected combination of approximate posteriors, separate from the single-model posterior in Eq.~\ref{eq:state}. With dispersion withheld, it reduces Wang $V_S$ error from \IndependentFullVSRMSE\ to \CouplingGuardVSRMSE\ km s$^{-1}$, contracts mean SD by \CouplingGuardVSContraction, and includes \CouplingGuardVSCoverage\ of the field in nominal 90\% intervals; four Wang-derived mismatch tests give \CouplingGuardStressCoverage. Heterogeneous observations therefore communicate through explicit shared coordinates acting through an image-pretrained morphological map, rather than through a fixed petrophysical equation.

\section*{Measured observations turn the structural prior into Earth posteriors}

\begin{figure}[H]
\centering
\includegraphics[width=0.90\linewidth]{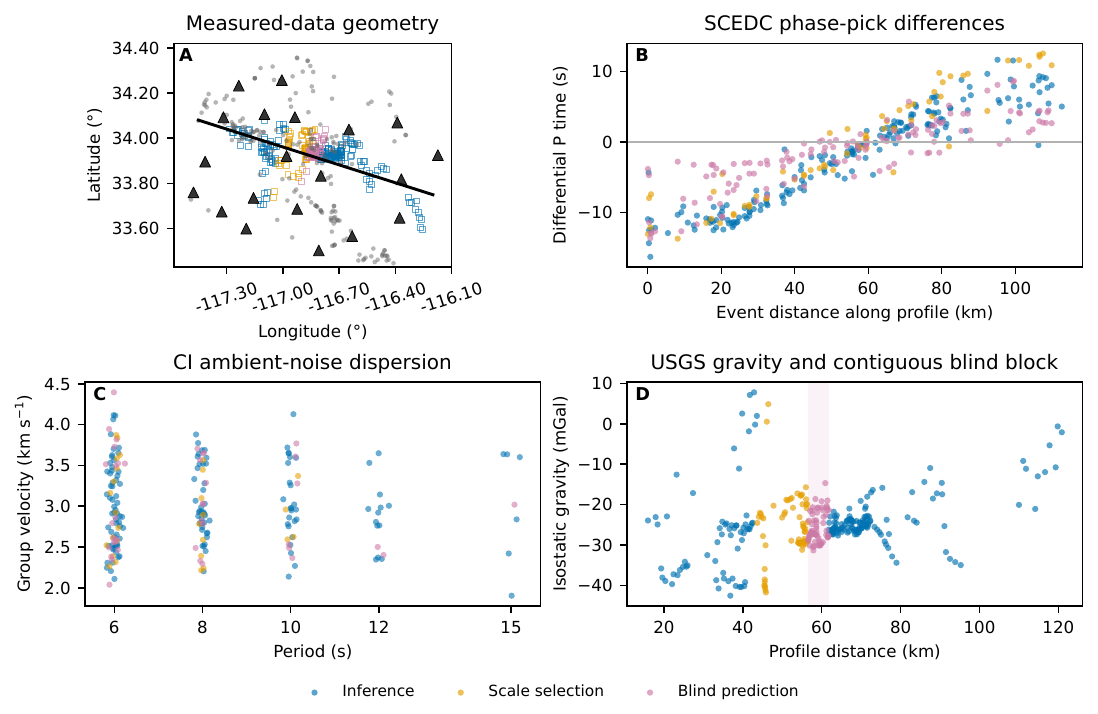}
\caption{\textbf{Measured southern-California observations and fixed partitions.} (A) SCEDC events (gray circles), CI stations (dark triangles), USGS gravity sites (colored squares), and the profile (black line). Events and stations may contribute to several original partitions. (B--D) Differential P arrivals, ambient-noise dispersion, and isostatic gravity with a contiguous blind block. Colors show original observation partitions. A separate grouped challenge enforces zero event and station-pair-family overlap.}
\label{fig:geometry}
\end{figure}

We apply the same frozen decoder to measured data along one 120.9-km San Gorgonio Pass--San Jacinto profile (Fig.~\ref{fig:geometry}). This 2.5-dimensional inversion uses 915 observations: differential P arrivals from Southern California Earthquake Data Center phase picks, group velocities measured from 30 days of CI-network ambient-noise records, and USGS isostatic gravity \cite{SCEDC2013,CINetwork,LangenheimEtAl2025}. Straight-ray slowness integrals, depth-averaged surface-wave kernels, and finite-strike gravity define the simplified forward physics. The experiment tests structural recovery and predictive differences within these approximations, not a full-wave or three-dimensional Earth reconstruction.

Fixed partitions assign 527 observations to inference, 156 to modality-level discrepancy selection, and 232 to blind prediction. A second evaluation holds out entire earthquake events and surface-wave station-pair families, together with a contiguous gravity block. All three property-sensitive likelihoods are active in this field fit. These are measured-data tests within one region; partitioning observations does not provide an independent regional replication.

The frozen decoder produces posterior means and conditional spatial SDs for all three properties (Fig.~\ref{fig:field}). On the original blind split, differential-arrival RMSE falls from \FieldTravelPriorRMSE\ to \FieldTravelPosteriorRMSE\ s and gravity RMSE from \FieldGravityPriorRMSE\ to \FieldGravityPosteriorRMSE\ mGal; matched RBF coordinates attain lower errors of \FieldRBFTravelRMSE\ s and \FieldRBFGravityRMSE\ mGal. The continuous ranked probability score (CRPS), which evaluates the full predictive distribution, decreases by \FieldTravelCRPSReduction\ and \FieldGravityCRPSReduction\ relative to the image prior. Predictive 90\% intervals include \FieldTravelPredictiveInclusion\ and \FieldGravityPredictiveInclusion\ of observations after accounting for development-selected observation and discrepancy uncertainty.

Under grouped holdouts, arrival-time RMSE falls from 1.776 to \GroupedImageTravelRMSE\ s and gravity RMSE from 7.635 to \GroupedImageGravityRMSE\ mGal, reductions of 43\% and 44\% relative to the pretrained image prior before observation updating. RBF again gives lower errors (0.915 s and 3.994 mGal), whereas DCT errors are higher (\GroupedDCTTravelRMSE\ s and \GroupedDCTGravityRMSE\ mGal; \emph{SI Appendix}, Section S10). The reductions quantify information gained from observations within the image representation, not an advantage over the strongest alternative prior.

The posterior adds laterally coherent structure to the fixed crustal trends. It places a co-located velocity--density contrast beneath San Gorgonio Pass, coincident with independently mapped basement and sedimentary wedges, and higher velocities toward the Coachella margin \cite{LangenheimEtAl2005,LangenheimFuis2022}. Prior-to-posterior difference maps separate the inferred update from the imposed trends; depth profiles show the remaining alternatives as pointwise 90\% intervals (\emph{SI Appendix}, Section S9). CVM-H \cite{ShawEtAl2015,CVMHData2020}, excluded from the likelihood, correlates with the posterior at \FieldCVMHCorrelationVP, \FieldCVMHCorrelationVS, and \FieldCVMHCorrelationRho. These whole-field correlations include depth trends and provide context, not local-resolution validation. Blind predictions directly test the $V_P$ and density components; the blocked layered-Rayleigh experiment below tests the reusable $V_S$ coordinates with measured data.

\section*{Image morphology and smoothness favor different structures}

\begin{figure}[H]
\centering
\includegraphics[width=0.82\linewidth]{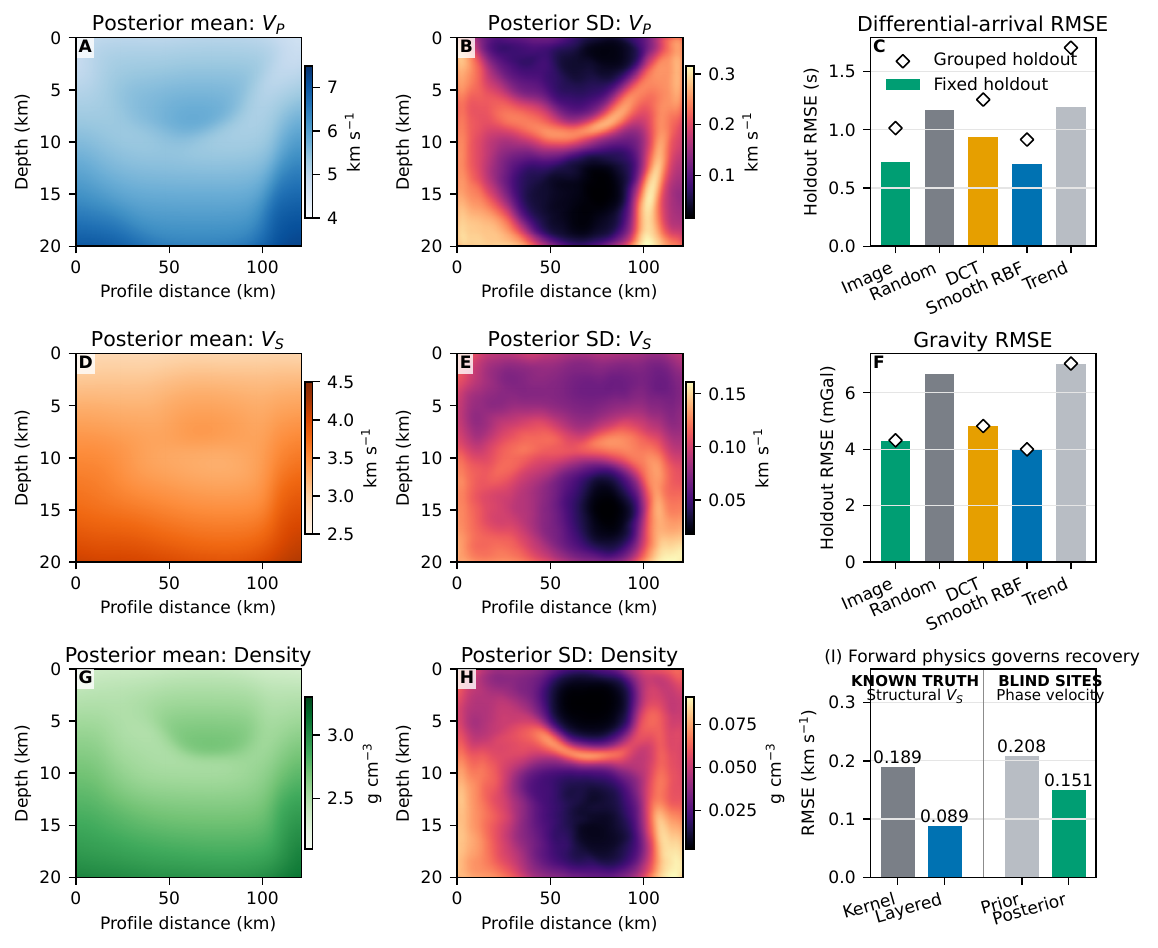}
\caption{\textbf{Measured posterior, controlled comparisons, and forward-physics test.} (A,B,D,E,G,H) Means and conditional SDs for $V_P$, $V_S$, and density under the fixed image-pretrained state. (C,F) Original-holdout RMSE; diamonds mark event/path-grouped holdouts and random bars average five seeds. Image coordinates beat random, DCT, and trend, while smooth RBF is strongest on this profile. (I) Left: kernel versus layered structural $V_S$ error for 40 ten-node known-truth profiles. Right: phase-velocity RMSE across 85 period--site values at five blind sites, before and after the generator update.}
\label{fig:field}
\end{figure}

Nine controlled inversions compare parameterizations under fixed observations, physical adapters, discrepancy scales, computational budget, and rank-24-plus-diagonal posterior family. Every nontrivial model has 256 coordinates; the trend-only model has no inferred spatial coordinates. DCT and RBF amplitudes match the image decoder's prior field SD. Random decoders retain identical architecture and adapters, without induced-field SD matching. Image-pretrained coordinates outperform these random, DCT, and trend controls for both differential arrivals and gravity (Fig.~\ref{fig:field}). Trend-only errors of \FieldTrendTravelRMSE\ s and \FieldTrendGravityRMSE\ mGal show that the imposed depth dependence alone does not explain the predictive gains.

The original and grouped field comparisons favor smooth RBF coordinates. Together with the inverse benchmark, they provide an empirical map of the tested conditions: image coordinates recover the four CurveVel targets better, while RBF recovers the four CurveFault targets better and predicts this regional profile better (\emph{SI Appendix}, Section S17). These outcomes depend jointly on structural representation and observational sensitivity. They do not assign all curved geology to image priors or all faults to smooth priors.

Variational inference (VI) with covariance ranks 8--48 and independent starts gives similar field means and blind predictive scores. Covariance rank affects SD more than means (\emph{SI Appendix}, Section S13). Field SDs are conditional VI estimates, whereas predictive intervals also include the fixed discrepancy scales.

A retrospective P-and-gravity-only intervention tests cross-property transfer at five measured Rayleigh sites while omitting the dispersion likelihood. Holding $V_P$ and density fixed during predictive evaluation isolates changes mediated by $V_S$. Across three optimizer seeds, shared-image phase RMSE increases from \MeasuredTransferImagePriorRMSE\ to \MeasuredTransferImagePosteriorRMSE\ km s$^{-1}$, whereas shared smooth-basis error decreases from \MeasuredTransferSmoothPriorRMSE\ to \MeasuredTransferSmoothPosteriorRMSE. Independent-image $V_S$ remains exactly at its prior (\emph{SI Appendix}, Section S11). The fixed image coupling changes the unobserved property but moves its predictions away from the measured Rayleigh constraints. Reusing the decoder and transferring information between properties are therefore distinct claims; direct shear-sensitive observations test the former without relying on the latter.

\section*{Forward physics determines which structures observations recover}

Surface waves distinguish representational capacity from depth sensitivity. The simplified five-period kernel in the 915-observation inversion gives a near-null blind update (0.625 to 0.624 km s$^{-1}$), motivating a paired operator test. With the same ten-node depth parameterization, layered Rayleigh physics reduces structural error by 53\% across 40 known $V_S$ profiles relative to a Gaussian period--depth kernel. Both operators nevertheless fit phase velocity at approximately 0.016 km s$^{-1}$ RMSE. This local comparison precedes the generator update and attributes the structural gain to forward physics. On the separate Qiu/Cai phase-velocity profile, both operators again fit the 3--16-s curves at about 0.014 km s$^{-1}$ \cite{QiuEtAl2019,CaiEtAl2022}. Similar data fit can thus accompany a twofold difference in structural error.

The Qiu/Cai release contains 4,076 paired local phase- and group-velocity curves derived from ambient-noise tomography. We use phase velocity. Thirty-one distinct source curves near the profile are aggregated to 25 sites, partitioned into contiguous blocks of 15 inference, five development, and five blind sites with no source-curve overlap. Local layered inversions supply depth-profile means and marginal SDs to a separate $V_S$-only update of the frozen decoder. This modular experiment is not a replacement joint fit of the 915 observations. All uncertainty scales are fixed before blind evaluation.

Across 85 period--site values at the five blind sites, RMSE falls from \TwoStepPriorRMSE\ to \TwoStepPosteriorRMSE\ km s$^{-1}$. Nominal 90\% predictive intervals include \TwoStepPredictiveInclusion\ of observations. These are correlated period measurements at five locations, not 85 independent validation cases. The result establishes a second observation-specific use of the frozen generator within the same regional setting: uncertain local profiles update a 2.5-dimensional section and improve predictions at withheld locations.

\section*{Structural learning and physical inference draw on different knowledge}

The central result is a separation of knowledge sources. Spatial patterns learned without geological training models supply a prior description for several Earth properties. Physical units, trends, bounds, coupling, and observation operators remain explicit geophysical choices. Observations then select among the represented alternatives. The same frozen weights implement this separation in controlled inversions and measured-data updates, without learning a joint velocity--density law.

The comparisons make this separation scientifically useful. Representability across 256 OpenFWI and nine regional sections is broader than the inverse evidence. The eight inverse targets reveal opposite image--RBF rankings across two structural families; the measured profile favors RBF. Direct Rayleigh updating improves prediction where fixed cross-property sharing worsens it. The resulting map of tested conditions distinguishes a useful representation from a useful coupling and from adequate forward physics (\emph{SI Appendix}, Section S17).

Pretraining improves on frozen random weights, demonstrating that learned information contributes beyond architecture alone. The experiments do not isolate whether spectra, edge statistics, or layered textures explain that contribution. Ensemble covariance and local sensitivity describe how the trained map constrains fields, not which image features caused transfer. Nor is there a matched generator trained on geological models; avoiding geological pretraining establishes an alternative source of structural knowledge, not superiority to domain-trained generators.

Weight reuse is demonstrated across properties and likelihoods, while regional generalization remains untested. The field conclusions concern recoverable differences under simplified 2.5-dimensional physics and the separate layered-Rayleigh update; uncertainty remains conditional on those operators, coupling choices, and posterior approximations.

The same underdetermination is familiar in medical reconstruction, where structural regularization can stabilize sparse observations, but the admissible structures and uncertainty model must be validated within the medical domain. Images supply reusable morphology; physical observations determine a new Earth posterior for each experiment. Earth inference can thus combine structural knowledge learned from abundant images with physical knowledge supplied independently by each experiment.

\section*{Materials and methods}
A single-channel variational autoencoder learns from source-balanced public images, with Earth fields excluded from training and checkpoint selection. The frozen decoder and fixed adapters map shared/private coordinates to physical fields (Eq.~\ref{eq:state}). Likelihoods use Gaussian residual errors. Field nuisance offsets/trends are profiled on inference rows; separate development rows fix modality error multipliers before posterior fitting.

Four-chain HMC supplies 8,000 controlled posterior draws, with matched diagonal-VI comparisons. The measured 2.5-dimensional system uses straight-ray slowness integrals, depth-averaged surface sensitivity, and finite-strike gravity. Rank-24-plus-diagonal VI supplies 512 draws, with covariance ranks and starts tested separately. Predictive intervals include fixed observation/discrepancy uncertainty. The P-and-gravity-only intervention preserves likelihood-inactive private coordinates exactly at their prior.

Controls fix observations and optimization budgets. Nontrivial OpenFWI priors and field DCT/RBF controls also match dimension and prior field SD; field random decoders match architecture and adapters. Known-truth transfer tests a development-selected linear pool. The two-step experiment passes \texttt{disba} \cite{LuuDisba} profile means and marginal SDs to a Gaussian generator likelihood with a frozen development scale. Full algorithms, splits, and bounds are in the \emph{SI Appendix}.

OpenAI Codex assisted software development, reproducibility checks, plotting code, and manuscript drafting and revision. The authors direct the research and are responsible for verifying all code, results, and claims.

\section*{Data and code availability}
Code, trained weights, posterior chains, processed public observations, split manifests, and figure inputs are publicly available in the \path{image_morphology_earth_posteriors} directory at \url{https://huggingface.co/cangyeone/relational-geophysics}. The release includes per-file checksums, source provenance, provider terms, and verification scripts; the immutable release revision is recorded in the \emph{SI Appendix}. Original image archives and regional reference models are retrieved from their cited providers rather than mirrored where redistribution terms are restricted or unspecified. Raw phase picks are available from SCEDC (doi:10.7909/C3WD3XH1), gravity from USGS (doi:10.5066/P13B2RWF), and local Rayleigh curves from the cited Qiu/Cai sources.

\section*{Acknowledgements}
This work was supported by the National Natural Science Foundation of China (No. 42674122) and the Fundamental Research Funds of the China Earthquake Administration (No. DQJB25B43). This study used public and simulated data; no proprietary or confidential data were exposed to the AI service. Numerical figures were rendered from recorded results, and the introductory diagram is explicitly schematic.

\section*{Competing interests}
The authors declare no competing interests.

\FloatBarrier
\bibliographystyle{unsrtnat}
\bibliography{references}

\clearpage
\suppressfloats[t]
\setcounter{figure}{0}
\setcounter{table}{0}
\setcounter{equation}{0}
\setcounter{section}{0}
\renewcommand{\theHfigure}{S.\arabic{figure}}
\renewcommand{\theHtable}{S.\arabic{table}}
\renewcommand{\theHequation}{S.\arabic{equation}}
\renewcommand{\theHsection}{S.\arabic{section}}
\section*{Supplementary Information}
\newcommand{\MedicalPatients}{5}
\newcommand{\MedicalFullPSNRMean}{33.11}
\newcommand{\MedicalFullPSNRSD}{1.27}
\newcommand{\MedicalFullSSIMMean}{0.828}
\newcommand{\MedicalFullSSIMSD}{0.101}
\newcommand{\MedicalFullResidualMean}{0.0461}
\newcommand{\MedicalFullResidualSD}{0.0063}
\newcommand{\MedicalSparsePSNRMean}{29.81}
\newcommand{\MedicalSparsePSNRSD}{2.87}
\newcommand{\MedicalSparseSSIMMean}{0.657}
\newcommand{\MedicalSparseSSIMSD}{0.195}
\newcommand{\MedicalSparseResidualMean}{0.0448}
\newcommand{\MedicalSparseResidualSD}{0.0056}
\newcommand{\MedicalTVPSNRMean}{32.26}
\newcommand{\MedicalTVPSNRSD}{1.48}
\newcommand{\MedicalTVSSIMMean}{0.826}
\newcommand{\MedicalTVSSIMSD}{0.055}
\newcommand{\MedicalTVResidualMean}{0.0509}
\newcommand{\MedicalTVResidualSD}{0.0048}
\newcommand{\MedicalPSNRGainMean}{2.45}
\newcommand{\MedicalPSNRGainSD}{4.12}
\newcommand{\MedicalSSIMGainMean}{0.169}
\newcommand{\MedicalSSIMGainSD}{0.244}
\newcommand{\MedicalImprovedCount}{4}

\renewcommand{\thefigure}{S\arabic{figure}}
\renewcommand{\thetable}{S\arabic{table}}
\renewcommand{\theequation}{S\arabic{equation}}

\section*{SI Appendix overview}

The manuscript tests whether structural learning and physical-property inference can draw on different sources of knowledge. A single generator learns spatial patterns from images and remains frozen as explicit adapters and physical likelihoods determine Earth posteriors. Four complementary evaluations trace this separation: image validation measures reconstruction within the pretraining domain; optimized projection measures representability of held-out Earth fields; a typical-shell constraint tests radial accessibility under the latent prior; and posterior experiments measure inference from physical observations. Reuse of the same weights is distinct from the cross-property dependence imposed by shared coordinates. Section S17 maps the tested conditions and their evidential scope; no single metric substitutes for another.

Public images train the generator, and image validation selects an epoch within each state dimension. The 128-dimensional deployment architecture was trained before geophysical projection; regional and OpenFWI evaluations followed, and 12- and 64-dimensional runs were added afterward as a retrospective paired capacity analysis. No Earth field updates decoder weights. An independently generated in-prior Earth and a public Wang regional model provide compact known-truth tests. The main manuscript prioritizes the southern-California inversion of measured seismic and gravity observations; the SI retains the fuller synthetic diagnostics needed to interpret that field result. A deterministic medical CT illustration is not evidence for the geophysical generator or posterior.

\section*{S1. Public-image population and data separation}

\subsection*{Image sources}

The pretraining population contains \ImagePopulation\ images from three public collections (Table~\ref{si:tab:imagesources}). EuroSAT contributes remote-sensing organization \cite{HelberEtAl2019}; BSDS500 contributes natural boundaries and multiscale scenes \cite{ArbelaezEtAl2011}; and DTD contributes directional, repeated, and irregular textures \cite{CimpoiEtAl2014}. Labels are never used.

\begin{table}[H]
\centering
\caption{Public image sources used for structural pretraining.}
\label{si:tab:imagesources}
\begin{adjustbox}{max width=\linewidth}\begin{tabular}{lrrp{7.6cm}}
\toprule
Source & Images & Sampling weight & Structural contribution \\
\midrule
EuroSAT & 27,000 & 0.45 & Regional patches, edges, and multiscale spatial organization. \\
BSDS500 & 500 & 0.20 & Natural boundaries and scene-scale variation. \\
DTD & 5,640 & 0.35 & Repeated, anisotropic, cracked, banded, and irregular texture. \\
\bottomrule
\end{tabular}\end{adjustbox}
\end{table}

Files are indexed by a stable hash of source name and relative path. Sorting these hashes gives \TrainIdentities\ training and \ValidationIdentities\ validation identities with zero overlap. The remaining source identities are unused. At each draw, a square crop covering 55--100\% of the shorter side is resized to $256\times256$, converted to grayscale, and robustly normalized by its 1st and 99th percentiles. Independent right-angle rotation, horizontal and vertical reflection, and contrast inversion remove preferred orientation and polarity. A fine Gaussian scale (1--5 pixels) is blended with a coarse scale (6--18 pixels) at a random fraction from 0.20 to 0.45. Thirty epochs with 30,000 newly augmented views per epoch produce \AugmentedViews\ training views.

OpenFWI, CVM-H, Wang2020, Berg2018, and Berg2021 are blocked by configuration from training sources. They do not enter image normalization, validation, early stopping, or a gradient update. The deployed compact checkpoint has SHA-256
\begin{quote}
\ttfamily\small \CheckpointHash.
\end{quote}

\section*{S2. Native-grid variational generator and latent capacity}

The encoder contains six stride-2 convolutional blocks with channels 24, 48, 96, 128, 192, and 256, group normalization, and SiLU activation. A $4\times4\times256$ representation maps to latent mean and log variance. The decoder reverses this construction through six bilinear-upsample--convolution blocks with channels 192, 128, 96, 64, 40, and 24. A final convolution and sigmoid emit one $256\times256$ scalar field. No smaller raster is enlarged after inference.

For input $\bm x$, training minimizes
\begin{equation}
\mathcal L(\bm x)=\|\bm x-\G(\bm u)\|_2^2/N
+0.1\|\nabla\bm x-\nabla\G(\bm u)\|_1/N
+\beta D_{\mathrm{KL}}[q_\phi(\bm u\mid\bm x)\,\|\,\N(\bm0,\bm I)],
\end{equation}
where $N=256^2$. The KL weight increases linearly to $5\times10^{-5}$ over eight epochs with 0.02 free bits per coordinate. AdamW uses learning rate $3\times10^{-4}$, weight decay $10^{-5}$, batch size 8, and gradient-norm clipping at 5.0. The epoch-29 checkpoint minimizes validation loss. The 128-dimensional model contains 3,911,217 parameters and attains validation normalized RMSE \ImageValidationRMSE.

\begin{figure}[H]
\centering
\includegraphics[width=0.90\linewidth]{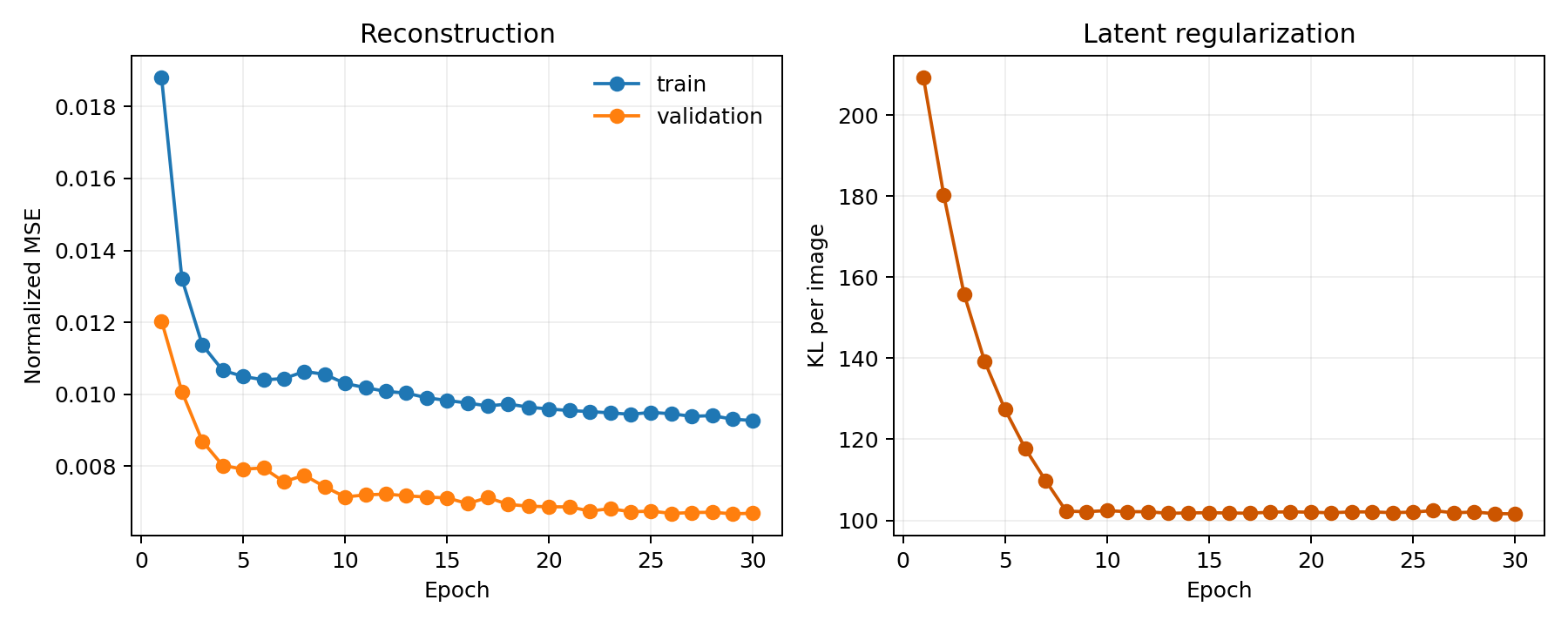}
\caption{Training history for the deployed 128-dimensional generator. Reconstruction terms are shown at left and the KL contribution after scheduled warmup at right. The held-out image loss, not an Earth model, selects the checkpoint.}
\label{si:fig:traininghistory}
\end{figure}

The paired ablation reveals a monotonic capacity--transfer relation: from 12 to 128 dimensions, image validation error falls by 45\%, regional projection error by 39\%, OpenFWI error by 45\%, and OpenFWI gradient retention rises from 0.343 to 0.513. The experiment changes only latent dimension and the associated latent projection matrices. Image identities, augmentations, encoder and decoder channels, optimizer, epoch count, views, and seed are shared (Table~\ref{si:tab:ablation}). The chronology is material: the 128-dimensional run completed first, followed by its regional and OpenFWI projections; the 12- and 64-dimensional runs completed later. The table therefore characterizes capacity retrospectively. It did not select the deployed dimension among three candidates before Earth evaluation. Within every dimension, the best epoch is selected only by held-out image normalized MSE, and no Earth field enters a weight update.

\begin{table}[H]
\centering
\caption{Paired latent-capacity ablation. Lower RMSE and higher gradient-energy retention are better.}
\label{si:tab:ablation}
\begin{adjustbox}{max width=\linewidth}\begin{tabular}{rrrrrr}
\toprule
Latent dim. & Parameters & Image RMSE & Regional RMSE & OpenFWI RMSE & OpenFWI gradient retention \\
\midrule
12  & 2,485,577 & 0.1476 & 0.0792 & 0.0953 & 0.343 \\
64  & 3,124,657 & 0.0960 & 0.0530 & 0.0596 & 0.473 \\
128 & 3,911,217 & 0.0817 & 0.0480 & 0.0524 & 0.513 \\
\bottomrule
\end{tabular}\end{adjustbox}
\end{table}

Across 256 standard-normal draws from the deployed model, field mean is 0.4953, field SD is 0.2395, and 0.831\% of pixels lie within 0.01 of either sigmoid boundary. Automated checks verify output shape, finite gradients and histories, bounded outputs, nonzero prior variation, checkpoint reload identity, image-identity separation, and absence of geophysical training sources.

\section*{S3. Matched pretraining controls and prior accessibility}

\subsection*{Relation to geological, plug-and-play, and structural priors}

Lopez-Alvis et al. trained a 40-dimensional convolutional VAE on $10^7$ crops drawn from 18 related geological training images derived from two outcrop-conditioned base models, then optimized latent coordinates and two velocity-mixing parameters by stochastic gradient descent for synthetic and field cross-borehole GPR traveltimes \cite{LopezAlvisEtAl2022}. Repeated initializations assessed optimization variability; the paper reports repeated SGD endpoint reconstructions rather than posterior draws. The present study adopts the established frozen-decoder/latent-inversion construction but changes both its source and inferential target: images without geophysical-property fields, subsurface models, or survey labels train a 128-dimensional single-channel decoder; identical weights parameterize three physical properties; three likelihoods update a joint state; and VI or HMC returns posterior field distributions. The contribution is therefore not the use of a VAE in an inverse problem. It is the conversion of generic visual morphology into a reusable multiphysics posterior coordinate system. Locally assembled geological priors and generic pretraining remain complementary routes to regional specificity and cross-experiment reuse, respectively.

Meles et al. combine a geological VAE with MCMC and a PCA--polynomial-chaos surrogate for Bayesian GPR tomography \cite{MelesEtAl2024}. Natural-image-trained convolutional denoisers have also entered deterministic and posterior poststack seismic inversion through plug-and-play regularization \cite{RomeroEtAl2022,IzzatullahEtAl2024}. Those studies establish geological DGM posteriors and generic-image regularization. They do not test whether one low-dimensional frozen generator can be reused with identical weights across $V_P$, $V_S$, and density and updated by heterogeneous likelihoods without retraining.

Structural coupling itself also predates this work. Joint inversion has imposed common interfaces or gradients, and Bayesian structure-decoupling formulations allow properties to separate where observations do not support common geometry \cite{HaberOldenburg1997,ColomboRovetta2018,PianaAgostinettiBodin2018}. Our shared/private construction does not learn a joint $V_P$--$V_S$--density law from image pairs. It makes cross-property coupling explicit, ablatable, and falsifiable: shared coordinates transmit morphology, private coordinates retain departures, and a development-frozen guard limits transfer under mismatch.

The capacity ablation cannot identify why the 128-dimensional model transfers. We therefore compare four matched parameterizations: the trained decoder; the same architecture with frozen random weights; a 128-dimensional low-frequency two-dimensional DCT basis; and the trained decoder with optimized coordinates constrained to the standard-normal radial shell. The frozen-random calculation is repeated for decoder-weight seeds 20260814--20260818. Every seed uses the same normalized targets, latent-coordinate initialization seeds, steps, learning rate, and restarts. Regional targets use 600 steps and four restarts. OpenFWI targets use 400 steps and two restarts. No method updates decoder or basis parameters.

\begin{table}[H]
\centering
\caption{Matched geometric-representability controls. Values are means over 9 regional or 256 OpenFWI targets. Frozen-random RMSE is averaged over five decoder-weight seeds; the parenthetical value is the SD of the five target-averaged seed means.}
\label{si:tab:controls}
\begin{adjustbox}{max width=\linewidth}\begin{tabular}{llrrr}
\toprule
Parameterization & Target set & Normalized RMSE & Correlation & Gradient retention \\
\midrule
Image-pretrained decoder & Regional & 0.0470 & 0.9666 & 0.5605 \\
Frozen-random decoder & Regional & \RandomRegionalRMSE\ (\RandomRegionalSeedSD) & \RandomRegionalCorrelation\ & \RandomRegionalGradient\ \\
128-D DCT & Regional & 0.0496 & 0.9669 & 0.4906 \\
Image-pretrained, typical shell & Regional & 0.0470 & 0.9666 & 0.5609 \\
\midrule
Image-pretrained decoder & OpenFWI & 0.0524 & 0.9917 & 0.5132 \\
Frozen-random decoder & OpenFWI & \RandomOpenFWIRMSE\ (\RandomOpenFWISeedSD) & \RandomOpenFWICorrelation\ & \RandomOpenFWIGradient\ \\
128-D DCT & OpenFWI & 0.2130 & 0.9456 & 0.4186 \\
Image-pretrained, typical shell & OpenFWI & 0.0536 & 0.9914 & 0.5069 \\
\bottomrule
\end{tabular}\end{adjustbox}
\end{table}

Smooth regional profiles make DCT a genuine positive control: it nearly matches the trained decoder, localizing the gain from image learning to harder morphology. On every one of the 256 OpenFWI targets, the pretrained decoder outperforms both random-weight and DCT controls. After averaging the frozen-random error for each target over five weight seeds, median paired frozen-random/pretrained and DCT/pretrained RMSE ratios are \OpenFWIRandomRatio\ and \OpenFWIDCTRatio. The five frozen-random seed means span \RandomOpenFWISeedMin--\RandomOpenFWISeedMax. This control isolates information stored in pretrained weights while preserving architecture and state dimension. It is not a deep image prior in which network weights are optimized independently for every target; the cited untrained-network result motivates architectural bias but does not make the methods identical \cite{HeckelSoltanolkotabi2020}.

\begin{figure}[H]
\centering
\includegraphics[width=0.98\linewidth]{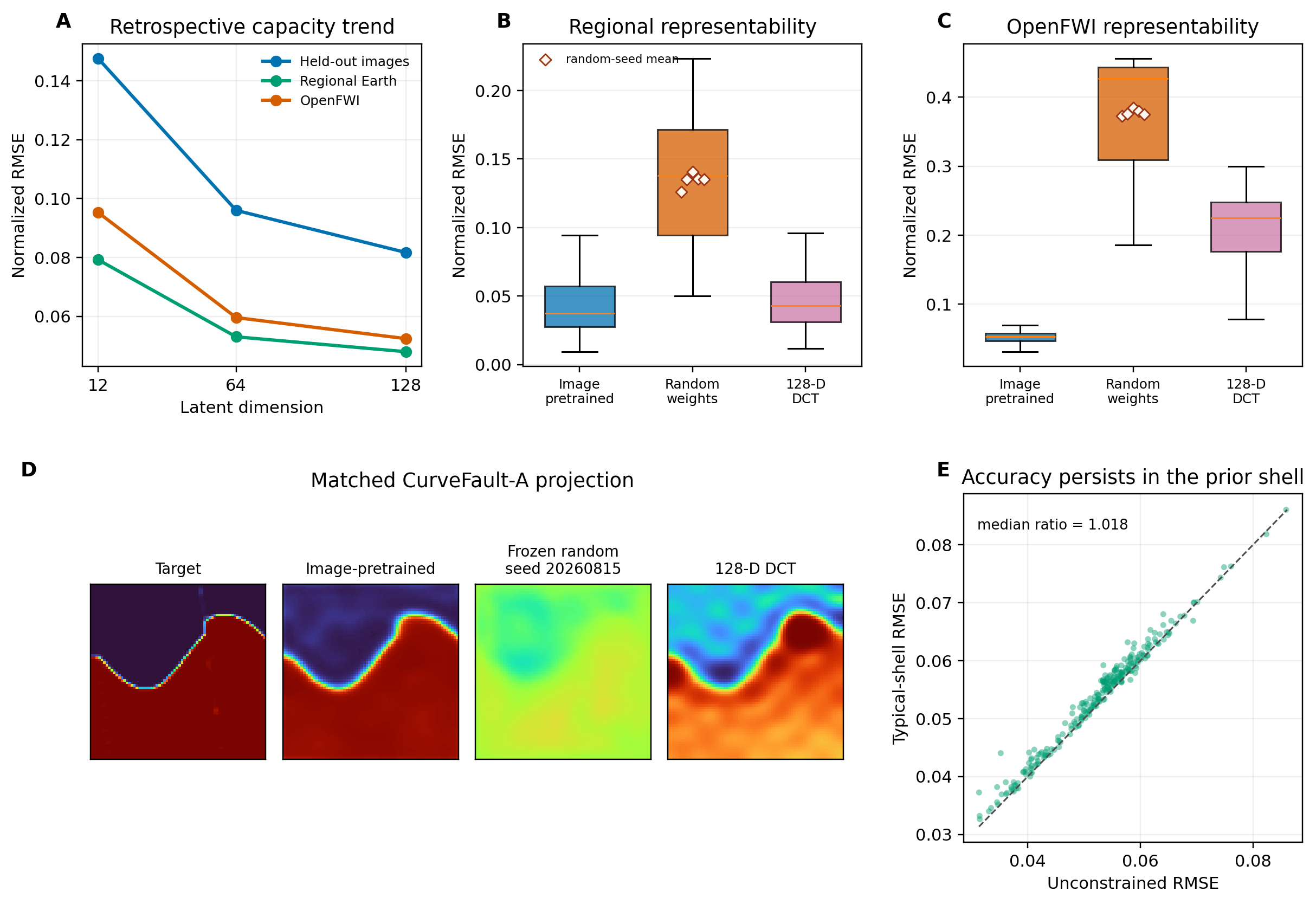}
\caption{Learning from generic images contributes beyond architecture and low-dimensional smoothness. (A) Retrospective paired 12/64/128-dimensional training characterizes capacity. (B and C) Matched regional and OpenFWI projections compare the deployed decoder, five frozen-random initializations, and a 128-dimensional DCT basis. Boxes summarize targetwise errors; each frozen-random target is averaged over seeds, and overlaid points show the five seed means. (D) A CurveFault-A example uses the seed nearest the frozen-random ensemble mean. (E) Constraining optimized coordinates to the 5th--95th percentile standard-normal radial shell preserves OpenFWI accuracy; the line denotes equality.}
\label{si:fig:transfer}
\end{figure}

An optimized projection measures distance to the decoder range. It does not by itself measure probability mass. For the single-field coordinate $\bm u\in\mathbb R^{128}$, $\|\bm u\|^2$ has a $\chi^2_{128}$ distribution under the declared prior. Unconstrained trained-decoder projections have median energies 31.95 for regional targets and 48.19 for OpenFWI, both at very low radial percentiles. They lie closer to the origin than the high-dimensional typical shell. We therefore constrain $\|\bm u\|$ to 10.142--12.466, the 5th--95th percentile radial interval. Median constrained energies are 140.35 for regional fields and 155.40 for OpenFWI. Projection accuracy changes negligibly (Table~\ref{si:tab:controls}). Thus the morphology remains reachable at typical radii. This radial calculation does not claim complete characterization of the induced field density \cite{AsimEtAl2020}.

\section*{S4. Regional and OpenFWI representational transfer}

The regional profile runs from longitude $-117.45^{\circ}$, latitude $34.08^{\circ}$ to longitude $-116.20^{\circ}$, latitude $33.75^{\circ}$ and depth 0--20 km. CVM-H contributes $V_P$, $V_S$, and density \cite{ShawEtAl2015,CVMHData2020}. Wang2020 contributes released Voigt-averaged $V_P$, $V_S$, and density. Berg2018 contributes $V_S$; Berg2021 contributes $V_S$ and $V_P/V_S$, from which $V_P$ is calculated \cite{BergEtAl2018,BergEtAl2021,Wang2020Model,Berg2018Model,Berg2021Model}. Fields are sampled or resized to $256\times256$ and normalized to the declared property ranges.

\begin{table}[H]
\centering
\caption{Regional-Earth projections through the frozen 128-dimensional decoder.}
\label{si:tab:regionalmetrics}
\begin{adjustbox}{max width=\linewidth}\begin{tabular}{llrrr}
\toprule
Model & Property & Physical RMSE & Normalized RMSE & Correlation \\
\midrule
CVM-H & $V_P$ & 0.2074 km s$^{-1}$ & 0.0593 & 0.9469 \\
CVM-H & $V_S$ & 0.1926 km s$^{-1}$ & 0.0963 & 0.9035 \\
CVM-H & density & 0.0343 g cm$^{-3}$ & 0.0286 & 0.9641 \\
Wang2020 & $V_P$ & 0.0734 km s$^{-1}$ & 0.0210 & 0.9891 \\
Wang2020 & $V_S$ & 0.0648 km s$^{-1}$ & 0.0324 & 0.9844 \\
Wang2020 & density & 0.0114 g cm$^{-3}$ & 0.0095 & 0.9947 \\
Berg2018 & $V_S$ & 0.1786 km s$^{-1}$ & 0.0893 & 0.9519 \\
Berg2021 & $V_S$ & 0.1142 km s$^{-1}$ & 0.0571 & 0.9753 \\
Berg2021 & $V_P$ & 0.1346 km s$^{-1}$ & 0.0385 & 0.9892 \\
\bottomrule
\end{tabular}\end{adjustbox}
\end{table}

\begin{figure}[H]
\centering
\includegraphics[height=0.85\textheight]{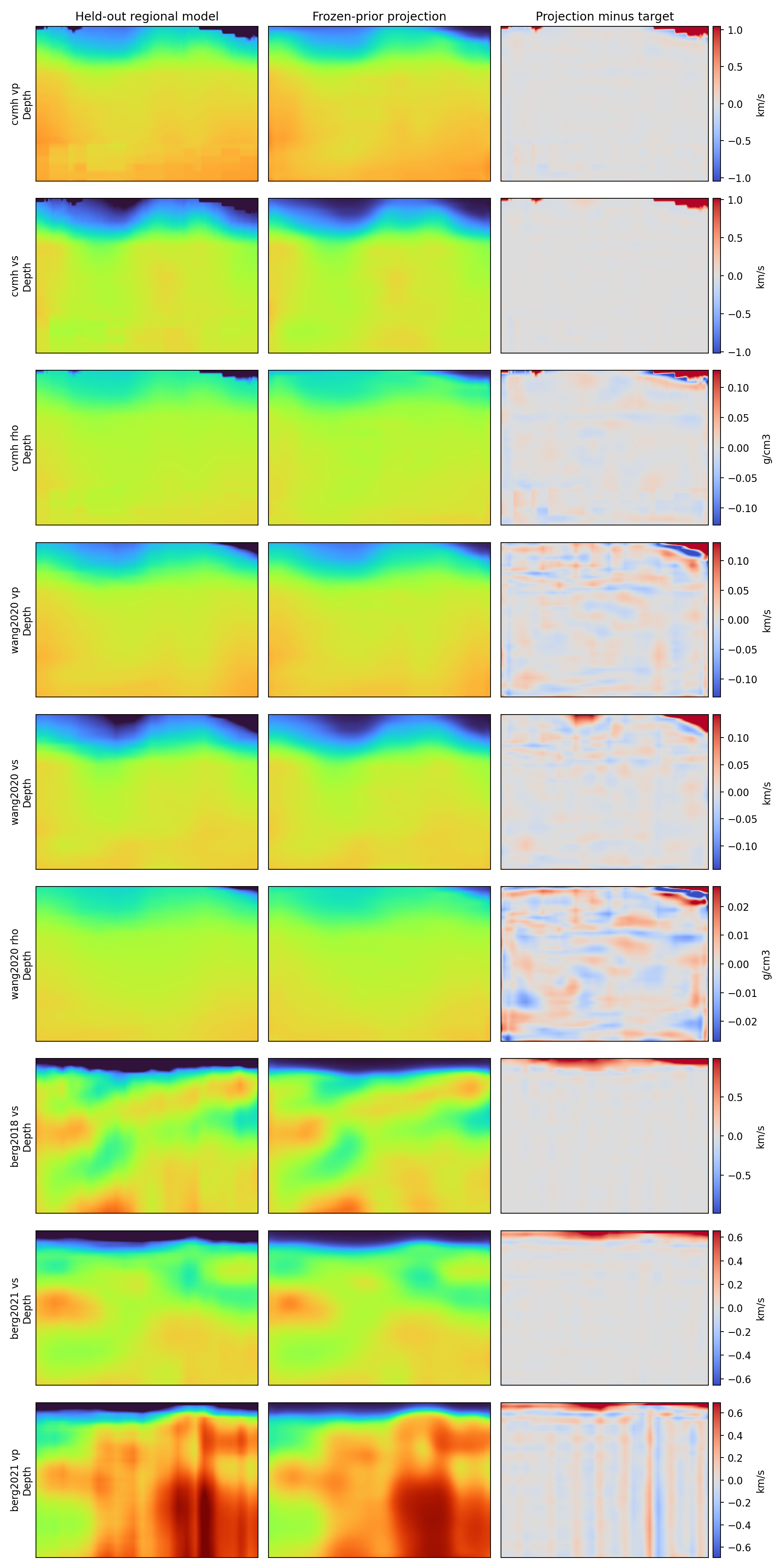}
\caption{All nine regional representational-transfer tests. Columns show public-model target, best frozen-decoder projection, and residual. CVM-H masks, fine streaks, and sharp transitions account for the largest residuals.}
\label{si:fig:regionalprojections}
\end{figure}

OpenFWI supplies abrupt layers and faults \cite{DengEtAl2022}. A seeded rule selects 64 models each from FlatVel-A, CurveVel-A, FlatFault-A, and CurveFault-A. Each is resized to $256\times256$ and normalized within its family range. Only coordinates are optimized.

\begin{table}[H]
\centering
\caption{OpenFWI projection metrics, averaged over 64 fields per family.}
\label{si:tab:openfwimetrics}
\begin{adjustbox}{max width=\linewidth}\begin{tabular}{lrrrr}
\toprule
Family & Normalized RMSE & Correlation & Gradient RMSE & Gradient retention \\
\midrule
FlatVel-A & 0.0420 & 0.9923 & 0.0417 & 0.5027 \\
CurveVel-A & 0.0554 & 0.9848 & 0.0605 & 0.4067 \\
FlatFault-A & 0.0548 & 0.9947 & 0.0425 & 0.5782 \\
CurveFault-A & 0.0574 & 0.9949 & 0.0460 & 0.5652 \\
\bottomrule
\end{tabular}\end{adjustbox}
\end{table}

High correlation records preserved layer and fault geometry. Gradient retention of 0.41--0.58 records the complementary boundary: the decoder smooths the sharpest contrasts. Residual concentration at discontinuities is a quantitative gradient-domain diagnostic, not a Fourier or spectral analysis. It identifies discontinuity-aware training and residual refinement as specific extensions. Both metrics are required to describe the transferred representation.

\subsection*{Matched complex-geometry inverse benchmark}

Projection alone does not prove inverse performance. We therefore invert eight OpenFWI CurveVel-A and CurveFault-A fields using image-pretrained, 128-dimensional RBF, 128-dimensional DCT, five 128-dimensional frozen-random decoders, and a fixed trend. These targets are part of the preceding 256-model projection suite but entered neither pretraining nor decoder fitting. Every nontrivial prior is matched to the predeclared 0.2395 normalized field SD using 48 standard-normal prior draws, without target fields. The first truth in each family is used only to freeze each method's additive field-discrepancy SD; structural summaries use all eight and interval inclusion scores the remaining six. Every nontrivial state uses the same 768 sparse training line integrals, 256 held-out line integrals, 0.0075 observation-noise SD, two 500-step MAP starts, 500 diagonal-VI steps, initialization, posterior draws, and compute device.

The aggregate image and RBF field errors are nearly identical, but the family ordering is consistent across all four truths in each family. On CurveVel-A, image coordinates lower field RMSE from 0.113 to 0.078 and raise interface F1 from 0.247 to 0.404. On CurveFault-A, RBF lowers field RMSE from 0.089 to 0.056 and raises interface F1 from 0.510 to 0.555. RBF also gives the best held-out line-integral prediction and lowest error inside the least-sampled sensitivity quartile. DCT, random, and trend controls are substantially worse. The result distinguishes prior performance under the tested line-integral acquisition. Without varying acquisition and structural family independently, it cannot attribute the ordering to morphology alone or establish a general curved-interface/fault boundary.

\begin{table}[H]
\centering
\caption{Matched line-integral inversion of eight complex OpenFWI truths. Random values average five frozen-weight seeds. Development-adjusted inclusion uses two family-first development truths to freeze one additive field-discrepancy SD, then scores the remaining six truths.}
\label{si:tab:openfwiinverse}
\small
\setlength{\tabcolsep}{3.3pt}
\begin{adjustbox}{max width=\linewidth}\begin{tabular}{lrrrrrr}
\toprule
State & Field RMSE & Poor-region RMSE & Line RMSE & Gradient corr. & Interface F1 & Test 90\% inclusion \\
\midrule
Image pretrained & 0.083 & 0.111 & 0.026 & 0.607 & 0.457 & 0.850 \\
RBF--128 & 0.085 & 0.083 & 0.019 & 0.512 & 0.401 & 0.937 \\
Frozen random & 0.134 & 0.187 & 0.038 & 0.195 & 0.290 & 0.877 \\
DCT--128 & 0.319 & 0.448 & 0.096 & 0.134 & 0.216 & 0.884 \\
Trend only & 0.277 & 0.225 & 0.193 & $-0.001$ & 0.083 & --- \\
\bottomrule
\end{tabular}\end{adjustbox}
\end{table}

Raw diagonal-VI field intervals under-include for every parameterization and are not presented as calibrated uncertainty. A field-error SD selected on two development truths raises conditional spatial inclusion on six test truths to 0.850 for image coordinates, 0.937 for RBF, 0.877 for the five-random mean, and 0.884 for DCT. In the lowest sensitivity quartile, corresponding inclusion is 0.751, 0.938, 0.765, and 0.771. The image posterior SD is 1.305 times larger in poorly than well sampled cells; RBF's ratio is 1.013. These results reinforce the representation regime: image coordinates carry the strongest CurveVel interface signal, whereas RBF supplies wider development-adjusted intervals and is strongest for CurveFault under this line-integral acquisition.

\section*{S5. Multiphysics construction and variational comparison}

For each property, a 64-dimensional shared block is concatenated with a 64-dimensional property block. The joint state is
\begin{equation}
\boldsymbol{\zeta}=(\bm z_s,\bm z_{V_P},\bm z_{V_S},\bm z_\rho)\in\mathbb R^{256},
\qquad \bm z_s,\bm z_p\sim\N(\bm0,\bm I_{64}).
\end{equation}
The same decoder is evaluated three times. Affine transforms map $[0,1]$ to 4.0--7.5 km s$^{-1}$ for $V_P$, 2.5--4.5 km s$^{-1}$ for $V_S$, and 2.1--3.3 g cm$^{-3}$ for density. The shared block is imposed by the inversion; no cross-property relation is learned from generic images. A fixed Birch-type or other empirical relation would bind the reusable parameterization to one regional and scale-dependent assumption. Shared coordinates instead supply flexible coupling, property-specific coordinates retain departures, and the likelihood weights both. A regionally justified petrophysical relation can be added later as a hierarchical prior or likelihood term without retraining the decoder.

All operators act on native-grid outputs. Fifty-two curved paths, sampled at 160 segments, form a bilinear path-length matrix $\bm A_t$ and predict $\bm f_t=\bm A_t\operatorname{vec}(1/V_P)$. Twelve horizontal sites and four characteristic depths give 48 separable Gaussian kernels that average $V_S$. Twenty-four surface sites use two-dimensional kernels proportional to $h[(s-s_j)^2+h^2]^{-1.25}$ on density contrast from 2.65 g cm$^{-3}$. These differentiable operators expose ray, localized-depth, and nonlocal sensitivity; they are not full wave or modal solvers.

Observation-noise SD is 10\% of coordinatewise prior-predictive SD from 384 joint prior draws, with a floor at 20\% of the modality median. Mean noise SD is 0.001889 for traveltime, 0.03726 for dispersion, and 0.01701 for gravity in operator units.

The matched single-modality/all-modality factorial uses diagonal Gaussian VI,
\begin{equation}
q(\boldsymbol{\zeta})=\N[\bm\mu,\operatorname{diag}\{\exp(2\bm\ell)\}],
\end{equation}
optimized for 1,200 Adam steps at learning rate 0.012, four Monte Carlo samples per step, and gradient clipping at 50. Field summaries use 128 draws. This approximation supports the complete modality factorial but omits posterior correlations \cite{BleiEtAl2017}.

\begin{figure}[H]
\centering
\includegraphics[width=0.80\linewidth]{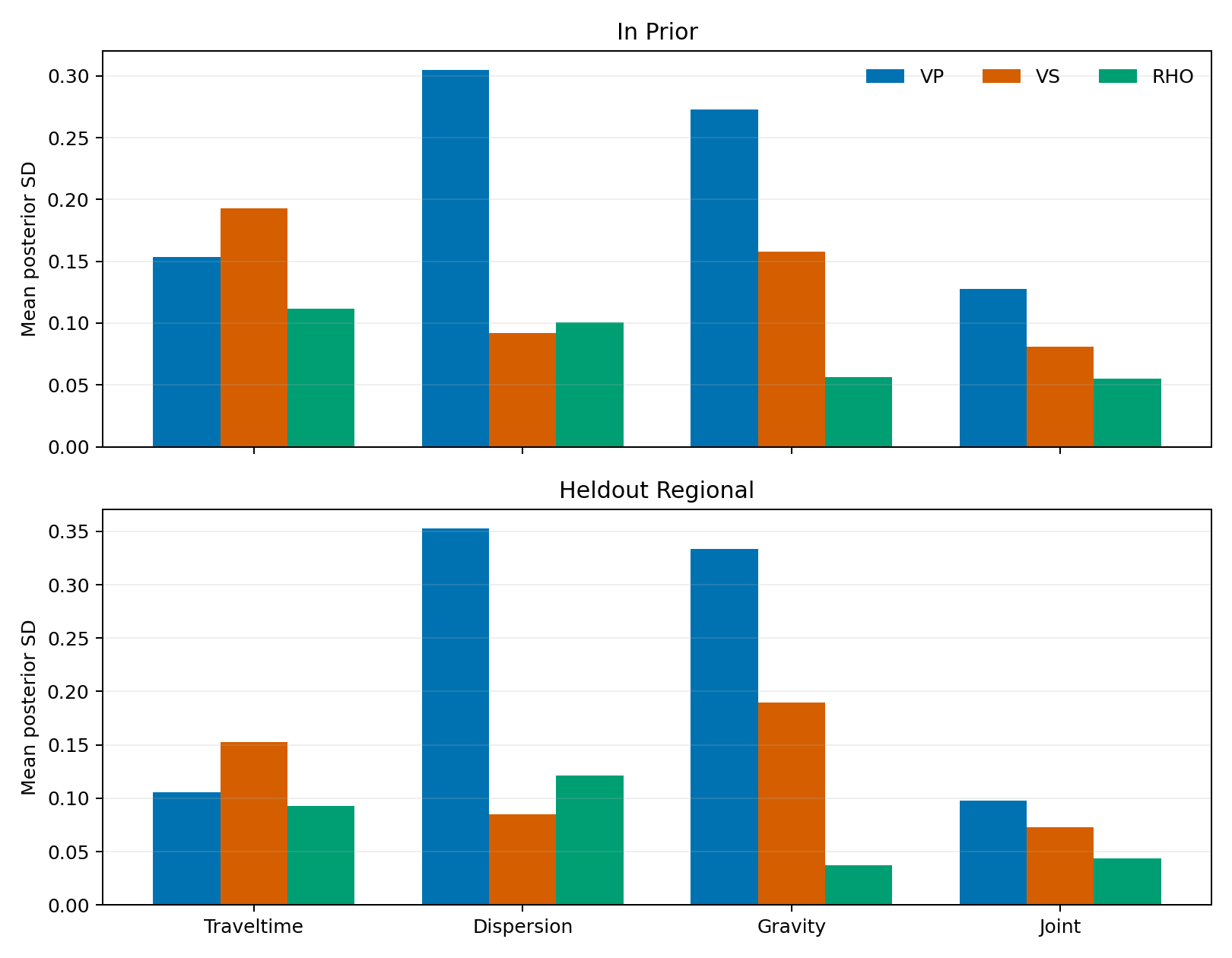}
\caption{Mean posterior SD under matched single-physics and joint diagonal VI. Joint observations contract all properties in the in-prior and held-out experiments. Physical units differ, so comparisons are made within a property.}
\label{si:fig:contraction}
\end{figure}

\begin{table}[H]
\centering
\caption{In-prior mean posterior SD under diagonal VI.}
\label{si:tab:contraction}
\begin{adjustbox}{max width=\linewidth}\begin{tabular}{lrrr}
\toprule
Posterior & $V_P$ (km s$^{-1}$) & $V_S$ (km s$^{-1}$) & Density (g cm$^{-3}$) \\
\midrule
Matched single physics & 0.1536 & 0.0921 & 0.0566 \\
Joint physics & 0.1278 & 0.0808 & 0.0550 \\
\bottomrule
\end{tabular}\end{adjustbox}
\end{table}

\begin{figure}[H]
\centering
\includegraphics[width=0.94\linewidth]{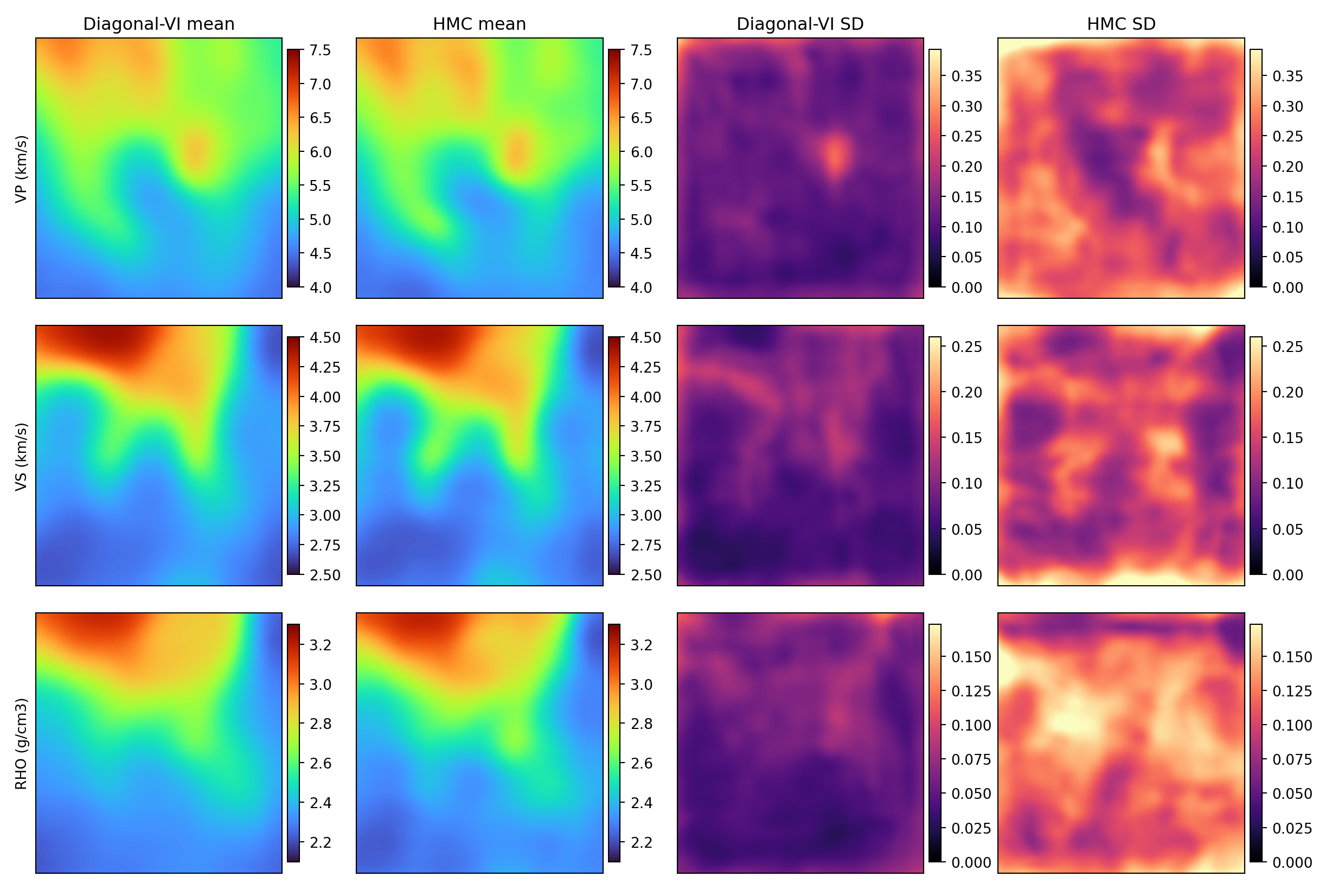}
\caption{Diagonal VI and HMC use the same generator, observations, and joint likelihood. Their central fields are similar, whereas diagonal VI produces smaller pixelwise SD. The post-hoc latent-space decomposition below attributes this gap primarily to contracted marginal scales in influential latent coordinates, not to omission of covariance alone.}
\label{si:fig:vihmc}
\end{figure}

The absolute VI--HMC uncertainty difference is substantial and algorithmic. HMC mean field SD is \VIHMCSDRatioVP, \VIHMCSDRatioVS, and \VIHMCSDRatioRho\ times the diagonal-VI value for $V_P$, $V_S$, and density. The variational family minimizes $D_{\mathrm{KL}}(q\|p)$ over diagonal Gaussians \cite{BleiEtAl2017}; it cannot represent skewness or posterior ridges and can contract marginals whose wider support would enter low-density regions. In latent space, mean marginal SD is \VILatentMeanSD\ under VI and \HMCLatentMeanSD\ under HMC. The mean coordinatewise HMC/VI ratio is \HMCLatentMeanSDRatio, although its median is only \HMCLatentMedianSDRatio, showing that the contraction is concentrated in a subset of coordinates rather than spread uniformly.

We tested whether discarded covariance alone explains the factor-of-two field-space gap. The HMC posterior contains some strong coordinate pairs (maximum $|r|=\HMCLatentMaximumCorrelation$), but a Gaussian carrying the empirical HMC mean and full covariance reproduces the three HMC mean field SDs to within \HMCFullGaussianMaximumSDGap. A diagonal Gaussian carrying the same HMC marginal variances remains within \HMCDiagonalGaussianMaximumSDGap\ and is slightly wider, not narrower. Independently shuffling every HMC coordinate across draws gives the same conclusion. These post-hoc ensembles are diagnostics, not posterior fits. Contracted marginal scales in the fitted VI solution account for most of the field-space SD gap; omitting covariance alone does not reproduce it. This decomposition does not establish a global variational optimum. We consequently use diagonal VI for matched modality and sensitivity comparisons, with HMC providing the absolute-uncertainty reference for this controlled posterior.

\section*{S6. Full-preconditioned HMC and field-level stability}

HMC targets the declared 256-dimensional posterior up to numerical integration error \cite{Betancourt2017}. A previous 4,000-draw run is used only as a pilot covariance estimate and is excluded from inference. The covariance receives 5\% diagonal shrinkage and $9.09\times10^{-7}$ jitter. Four fresh chains use 750 warmup iterations, 2,000 retained draws, six leapfrog steps, and step size 0.199979. Sampling acceptance is 0.9039.

Rank-normalized diagnostics follow Vehtari et al. \cite{VehtariEtAl2021}. All 256 coordinate $\widehat R$ values are below 1.01; median and maximum are 1.00125 and 1.00731. Bulk ESS has median 2,243 and minimum 1,288; tail ESS has median 3,992 and minimum 1,993. There are zero nonfinite transitions and zero divergences under $|\Delta H|>1000$. The 99th percentile of $|\Delta H|$ is 0.741. Chain E-BFMI values are 0.874, 0.854, 0.918, and 0.969.

\begin{figure}[H]
\centering
\includegraphics[width=0.96\linewidth]{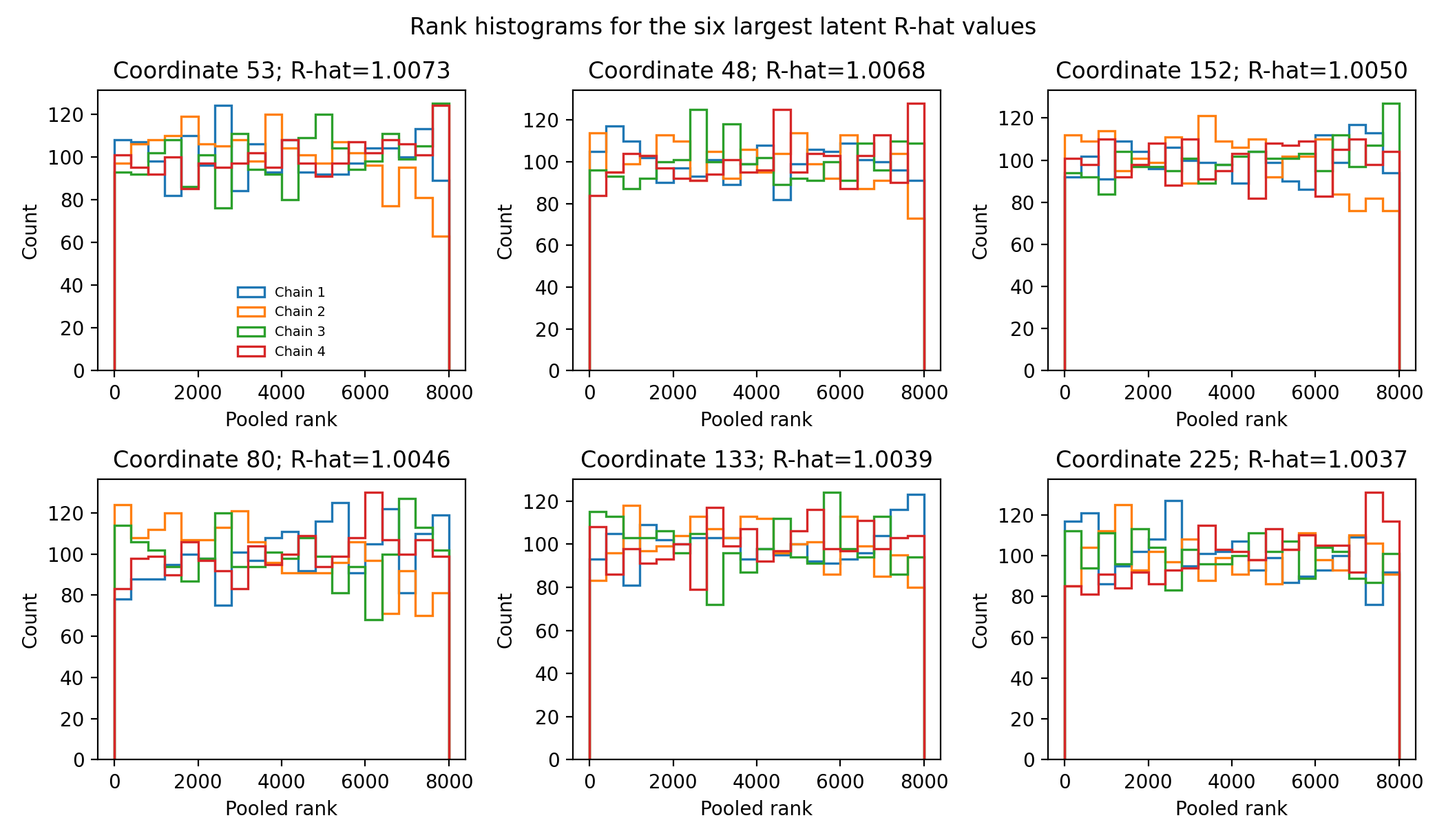}
\caption{Rank-normalized HMC diagnostics. Coordinatewise $\widehat R$, bulk ESS, and tail ESS satisfy the prespecified final thresholds. Representative rank histograms overlap across chains, and energy behavior is stable.}
\label{si:fig:hmcrank}
\end{figure}

\begin{table}[H]
\centering
\caption{Field-map stability across the four fresh HMC chains. The maximum compares each chain with the pooled result.}
\label{si:tab:hmcstability}
\begin{adjustbox}{max width=\linewidth}\begin{tabular}{lrrrr}
\toprule
Property & Mean-map RMSE & SD-map relative $L_2$ & Lower-90\% RMSE & Upper-90\% RMSE \\
\midrule
$V_P$ & 0.01090 & 0.0229 & 0.02670 & 0.04165 \\
$V_S$ & 0.00746 & 0.0353 & 0.01973 & 0.03109 \\
Density & 0.00550 & 0.0287 & 0.01153 & 0.02183 \\
\bottomrule
\end{tabular}\end{adjustbox}
\end{table}

Field-level scalar $\widehat R$ values are at most 1.0053. MCSE for spatial mean is 0.000687 km s$^{-1}$ for $V_P$, 0.000262 km s$^{-1}$ for $V_S$, and 0.000438 g cm$^{-3}$ for density. Using all 8,000 draws, HMC posterior-mean RMSEs are 0.2860, 0.1478, and 0.1439 in physical units. Mean SDs are \HMCSDVP, \HMCSDVS, and \HMCSDRho. Spatial 90\% interval-inclusion fractions are \HMCCoverageVP, \HMCCoverageVS, and \HMCCoverageRho.

We examine more than the 90\% interval by computing a spatial posterior rank at each cell $j$,
\begin{equation}
\widehat u_j=\frac{\#\{s:m_j^{(s)}<m_j^\ast\}+0.5\,\#\{s:m_j^{(s)}=m_j^\ast\}+0.5}{S+1},
\qquad S=8{,}000.
\end{equation}
The mean ranks are \HMCSpatialPITMeanVP, \HMCSpatialPITMeanVS, and \HMCSpatialPITMeanRho\ for $V_P$, $V_S$, and density. Equal-tailed 95\% intervals include \HMCSpatialCoverageNinetyFiveVP, \HMCSpatialCoverageNinetyFiveVS, and \HMCSpatialCoverageNinetyFiveRho\ of their respective grids. Across nominal levels 50, 60, 70, 80, 90, and 95\%, the largest absolute spatial-inclusion gaps are \HMCSpatialCoverageMaxGapVP, \HMCSpatialCoverageMaxGapVS, and \HMCSpatialCoverageMaxGapRho. The $V_P$ curve stays close to the identity, $V_S$ is modestly conservative at central levels, and the density PIT is left-skewed with underinclusion at central levels. This last pattern exposes spatially coherent conditional bias for this generating field even though its 90--95\% inclusion is close to nominal.

These curves are intentionally labeled \emph{spatial} rather than calibrated coverage. All 65,536 locations are coupled through 256 coordinates, and only one truth and one noise realization were generated. Treating cells as independent replicates would manufacture a misleadingly small binomial error bar. Formal simulation-based calibration requires repeated truth--data pairs; the present PIT and inclusion curves are a conditional diagnostic and no repeated-sampling calibration claim is made.

\begin{figure}[H]
\centering
\includegraphics[width=0.98\linewidth]{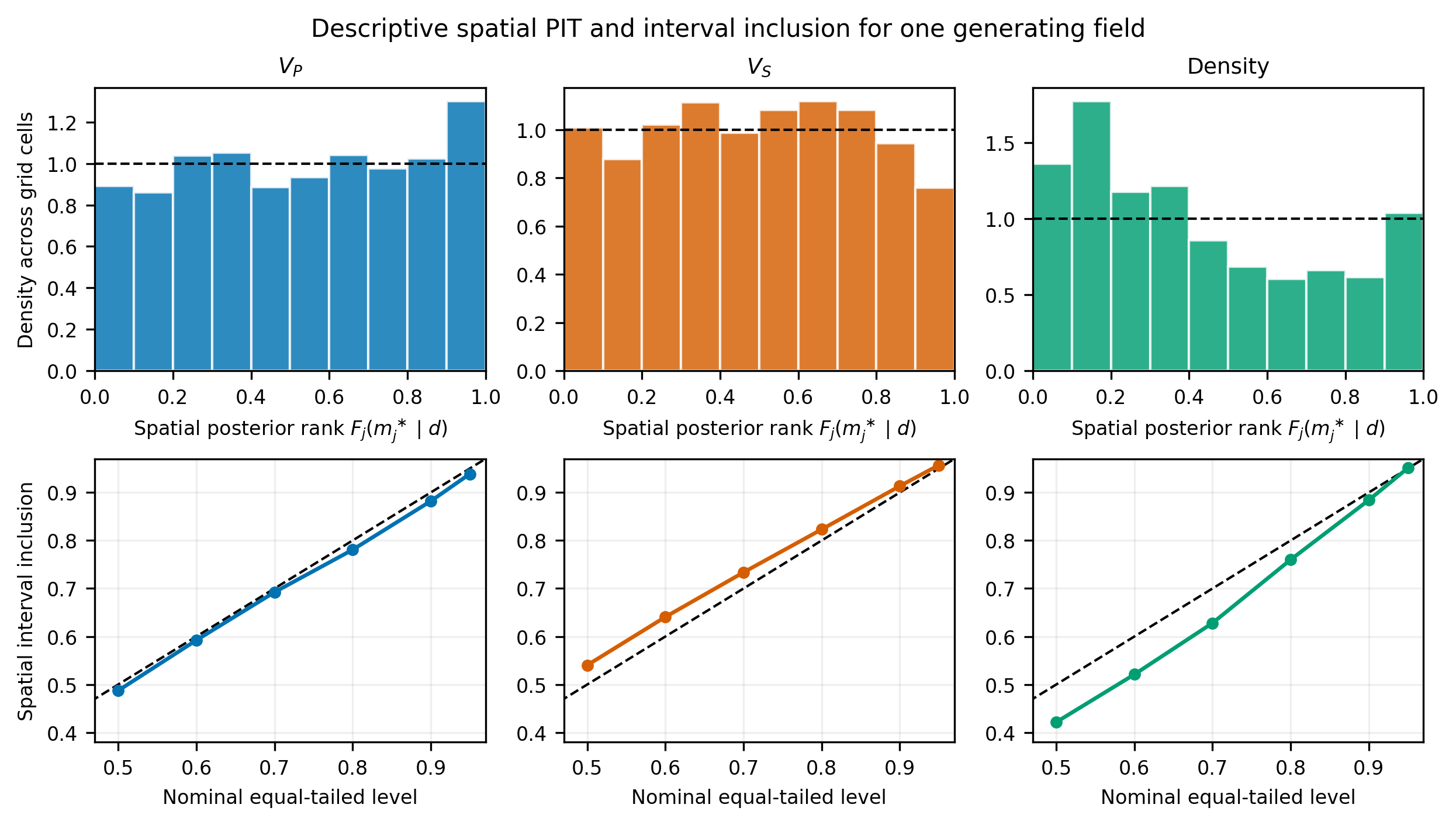}
\caption{Descriptive spatial PIT and equal-tailed interval inclusion from all 8,000 HMC draws. Dashed lines mark uniform PIT density (top) and identity between nominal level and spatial inclusion (bottom). Grid cells share latent coordinates, so these panels diagnose one generating field and must not be interpreted as independent simulation-based calibration replicates.}
\label{si:fig:hmcspatialpit}
\end{figure}

Primary physical-field summaries decode all 8,000 retained draws. Balanced subsets take equal numbers of evenly spaced draws from each chain. Table~\ref{si:tab:hmcdrawstability} reports the largest map RMSE across properties as a fraction of each property range, together with the largest change in spatial inclusion and low/high-sensitivity SD ratio relative to all draws.

\begin{table}[H]
\centering
\caption{Decoded HMC field-summary stability relative to all 8,000 retained draws. Values in the four map columns are the maximum propertywise RMSE divided by the corresponding physical range.}
\label{si:tab:hmcdrawstability}
\small
\begin{adjustbox}{max width=\linewidth}\begin{tabular}{rrrrrrr}
\toprule
Draws & Mean & SD & 5th pct. & 95th pct. & Max. inclusion change & Max. SD-ratio change \\
\midrule
256 & \HMCStabMeanTwoFiveSix & \HMCStabSDTwoFiveSix & \HMCStabLowTwoFiveSix & \HMCStabHighTwoFiveSix & \HMCStabInclusionTwoFiveSix & \HMCStabRatioTwoFiveSix \\
1,024 & \HMCStabMeanOneZeroTwoFour & \HMCStabSDOneZeroTwoFour & \HMCStabLowOneZeroTwoFour & \HMCStabHighOneZeroTwoFour & \HMCStabInclusionOneZeroTwoFour & \HMCStabRatioOneZeroTwoFour \\
2,048 & \HMCStabMeanTwoZeroFourEight & \HMCStabSDTwoZeroFourEight & \HMCStabLowTwoZeroFourEight & \HMCStabHighTwoZeroFourEight & \HMCStabInclusionTwoZeroFourEight & \HMCStabRatioTwoZeroFourEight \\
\bottomrule
\end{tabular}\end{adjustbox}
\end{table}

Mean absolute normalized residuals of the HMC posterior mean are 0.683 for traveltime, 0.562 for dispersion, and 0.532 for gravity. Diagonal-VI spatial interval-inclusion fractions on the same truth are 0.588, 0.680, and 0.419. The difference isolates posterior approximation within a fixed generative parameterization.

Pixelwise posterior SD is computed over all decoded physical draws. The per-pixel $L_2$ norm of each acquisition matrix defines a normalized sensitivity image for the corresponding property. Comparing the lowest and highest sensitivity quintiles gives SD ratios \SensitivityRatioVP, \SensitivityRatioVS, and \SensitivityRatioRho; sensitivity--SD correlations are \SensitivityCorrelationVP, \SensitivityCorrelationVS, and \SensitivityCorrelationRho. The 256 coordinates couple locations, but weak acquisition remains visible in the field uncertainty.

\begin{figure}[H]
\centering
\includegraphics[width=0.96\linewidth]{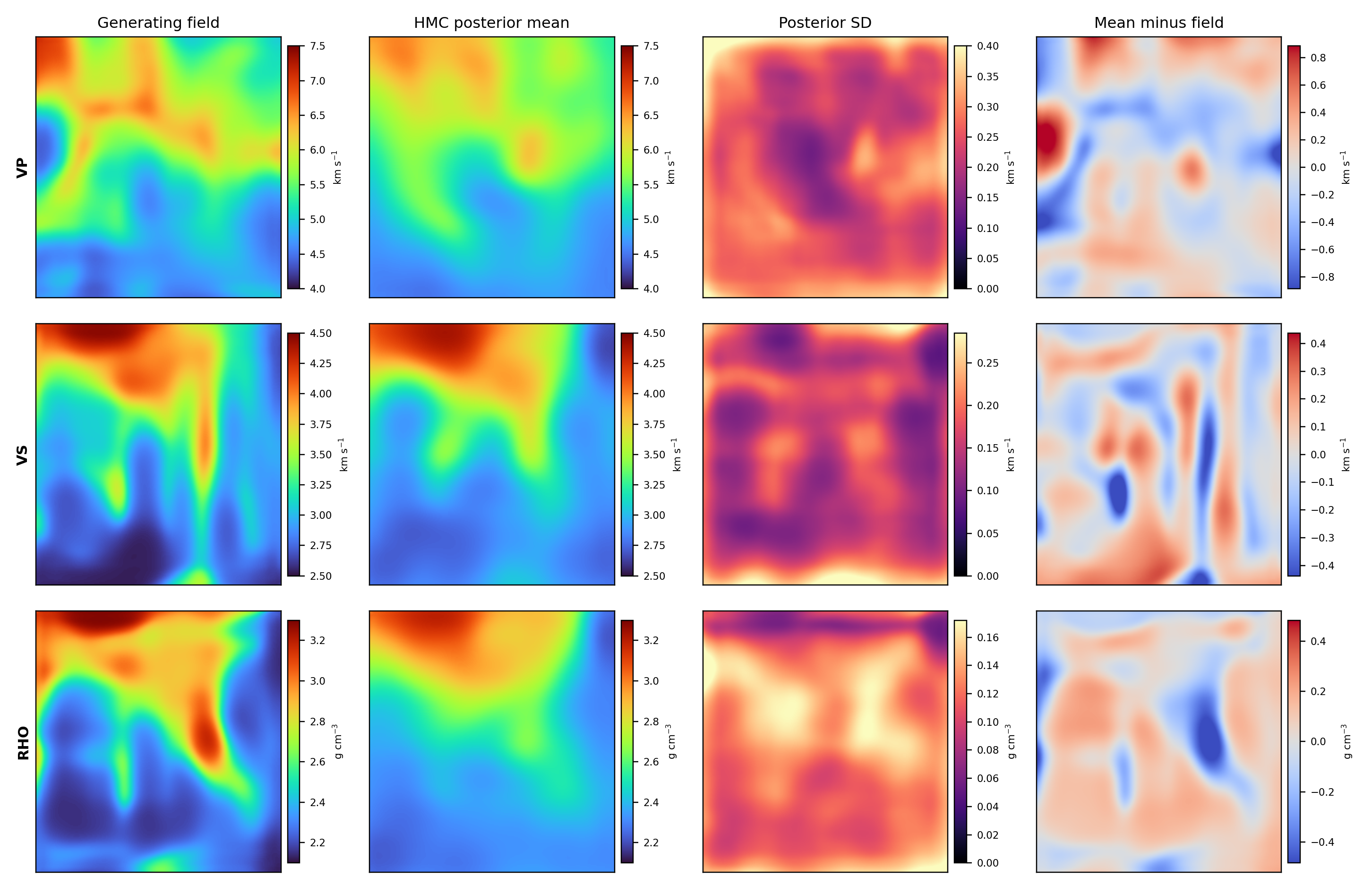}
\caption{Controlled known-truth multiphysics posterior within the frozen image-pretrained coordinates. Rows show $V_P$, $V_S$, and density. Common physical color limits are used for each generating field and its HMC posterior mean; posterior SD and mean error use separate labeled scales. This figure establishes recovery and uncertainty behavior under known truth and is not a field-data result.}
\label{si:fig:hmcposterior}
\end{figure}

\section*{S7. Shared-coordinate sensitivity}

A conventional VAE does not assign privileged semantics to the first coordinates. We therefore rerun held-out regional VI with 0, 32, 64, 96, and 128 shared coordinates and two seeded permutations of a 64-coordinate shared block. Every configuration uses 1,200 VI steps, the same 124 observations, and 128 posterior draws. Each property retains 128 coordinates; explicit overlap therefore reduces the joint state from 384 dimensions under zero sharing to 128 under full sharing. Every fit minimizes the Monte Carlo expected negative log likelihood plus one analytic $D_{\mathrm{KL}}[q\|\mathcal N(0,I)]$. Table~\ref{si:tab:shared} reports posterior-mean RMSE, the maximum of the three mean absolute normalized residuals, and cross-property gradient coherence.

\begin{table}[H]
\centering
\caption{Held-out regional sensitivity to the shared-coordinate construction.}
\label{si:tab:shared}
\scriptsize
\setlength{\tabcolsep}{3.2pt}
\begin{adjustbox}{max width=\linewidth}\begin{tabular}{lrrrrrr}
\toprule
Shared construction & Joint dim. & $V_P$ RMSE & $V_S$ RMSE & $\rho$ RMSE & Max. residual & Gradient coherence \\
 & & (km s$^{-1}$) & (km s$^{-1}$) & (g cm$^{-3}$) & (noise SD) & \\
\midrule
0 shared & 384 & 0.2037 & 0.0734 & 0.0807 & 0.742 & 0.330 \\
32 shared & 320 & 0.1844 & 0.0837 & 0.0812 & 0.717 & 0.402 \\
64 shared, leading & 256 & 0.2127 & 0.0992 & 0.0407 & 0.748 & 0.433 \\
96 shared & 192 & 0.1748 & 0.1164 & 0.0328 & 0.803 & 0.729 \\
128 shared & 128 & 0.1796 & 0.1225 & 0.0506 & 0.849 & 1.000 \\
64 shared, permutation A & 256 & 0.1148 & 0.0971 & 0.0794 & 0.752 & 0.551 \\
64 shared, permutation B & 256 & 0.1207 & 0.0878 & 0.0735 & 0.740 & 0.623 \\
\bottomrule
\end{tabular}\end{adjustbox}
\end{table}

\begin{figure}[H]
\centering
\includegraphics[width=0.96\linewidth]{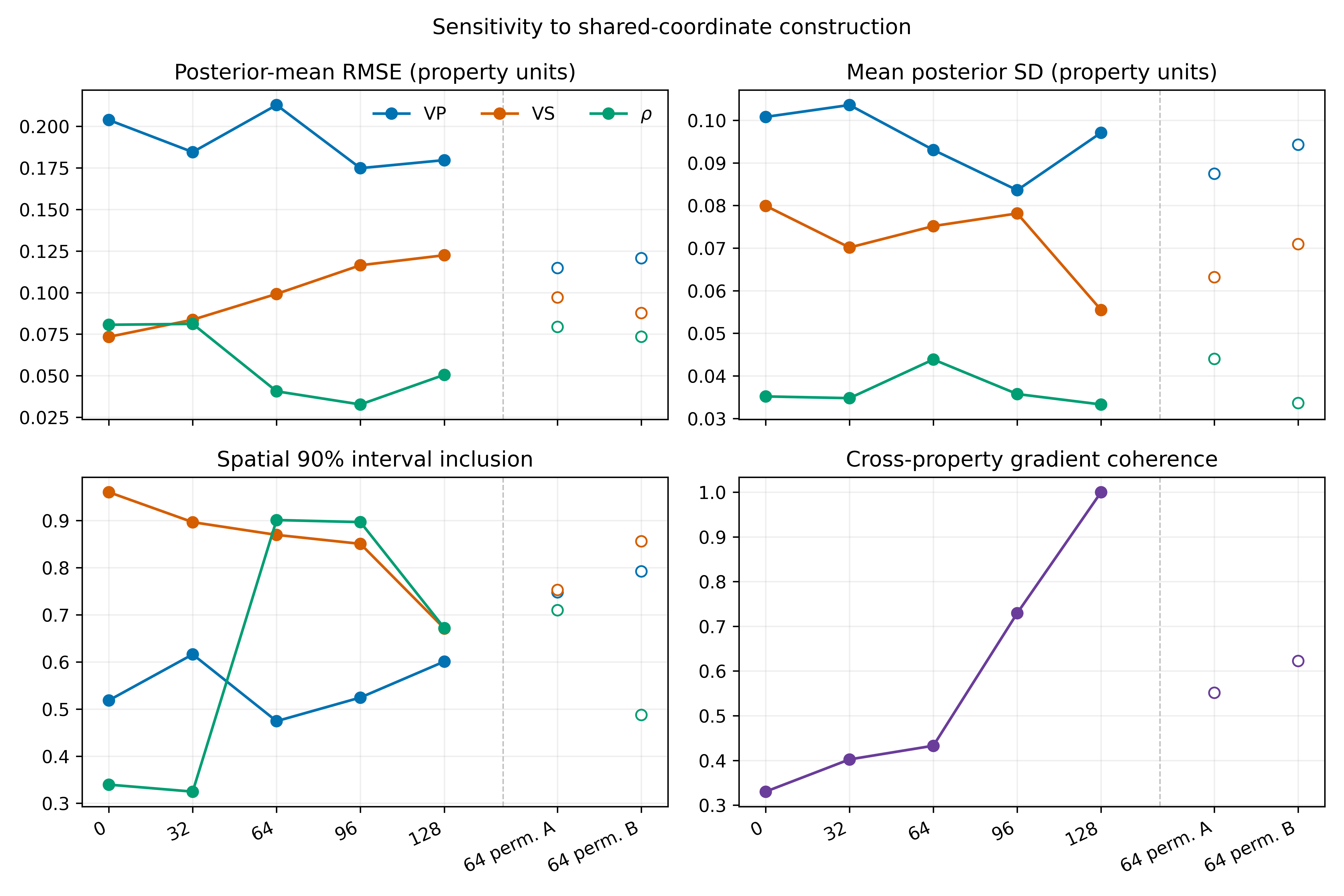}
\caption{Held-out regional shared-coordinate sensitivity. All constructions fit the observations within one noise SD in mean absolute residual. More sharing generally increases cross-property gradient coherence, but full sharing forces identical normalized geometry and reduces property-specific flexibility. The two permuted 64-coordinate blocks show that the leading split is not uniquely privileged.}
\label{si:fig:shared}
\end{figure}

The sweep exposes a coherence--flexibility frontier. Increasing overlap compresses the joint state and raises cross-property gradient coherence from 0.330 to 1.000, whereas property-specific blocks preserve departures that full sharing suppresses. Every construction fits the observations within 0.85 noise SD, so the frontier is not an optimization-failure artifact. The two permuted 64-coordinate blocks show that explicit overlap, rather than privileged semantics of the leading coordinates, creates the coupling. The 64/64 construction occupies the interior of this frontier; Section S11 supplies an intervention test of its scientific value by showing that it transfers information into an unmeasured property on an independent Earth truth, including against a dimension-matched independent state.

\section*{S8. Held-out regional representation and inference}

The held-out truth uses Wang2020 sections stored by the external test. Rows 32--256 are resized vertically to 256 cells without clipping. Observations are simulated with the same acquisition and noise construction as the in-prior experiment. This is a semi-synthetic test based on a public Earth model, not an inversion of the original observations.

Two Wang projection summaries appear in the evidence archive and refer to different target arrays. Table~\ref{si:tab:regionalmetrics} uses the released full 0--20 km profile and gives 0.0734, 0.0648, and 0.0114 in physical units. The posterior experiment removes the upper 32 rows and bilinearly resizes the remaining subsurface section; the resulting array is exactly the stored posterior truth (maximum absolute difference 0). Projection of that exact truth gives the smaller values below. Both calculations use the same checkpoint, 600 optimization steps, four restarts, physical-unit RMSE after affine restoration, and unconstrained coordinates. The difference is therefore target identity and preprocessing, not a favorable replacement of one result by another.

Four-restart, 600-step projection through the frozen decoder gives RMSE 0.01491 km s$^{-1}$ for $V_P$, 0.01154 km s$^{-1}$ for $V_S$, and 0.00626 g cm$^{-3}$ for density; correlations are 0.99858, 0.99846, and 0.99620. The corresponding joint VI posterior-mean errors are much larger (Table~\ref{si:tab:heldout}). The experiment therefore distinguishes an accurately represented truth from incomplete recovery under sparse observations.

\begin{table}[H]
\centering
\caption{Joint diagonal-VI posterior for the held-out Wang regional truth. ``Inclusion'' is the fraction of grid locations whose pointwise 90\% interval contains the truth.}
\label{si:tab:heldout}
\begin{adjustbox}{max width=\linewidth}\begin{tabular}{lrrrr}
\toprule
Property & Mean MAE & Mean RMSE & Mean posterior SD & Spatial inclusion \\
\midrule
$V_P$ & 0.1696 km s$^{-1}$ & 0.2156 km s$^{-1}$ & 0.0975 km s$^{-1}$ & 0.4880 \\
$V_S$ & 0.0680 km s$^{-1}$ & 0.1003 km s$^{-1}$ & 0.0727 km s$^{-1}$ & 0.8428 \\
Density & 0.0331 g cm$^{-3}$ & 0.0419 g cm$^{-3}$ & 0.0437 g cm$^{-3}$ & 0.8841 \\
\bottomrule
\end{tabular}\end{adjustbox}
\end{table}

Mean absolute normalized residuals are 0.752 for traveltime, 0.654 for dispersion, and 0.653 for gravity. The small best-projection error and larger posterior error separate representational capacity from sparse-data recovery. Together with the in-prior VI--HMC comparison, property-dependent inclusion locates the limiting layer in sparse-data recovery and posterior approximation rather than decoder access.

\begin{figure}[H]
\centering
\includegraphics[width=0.98\linewidth]{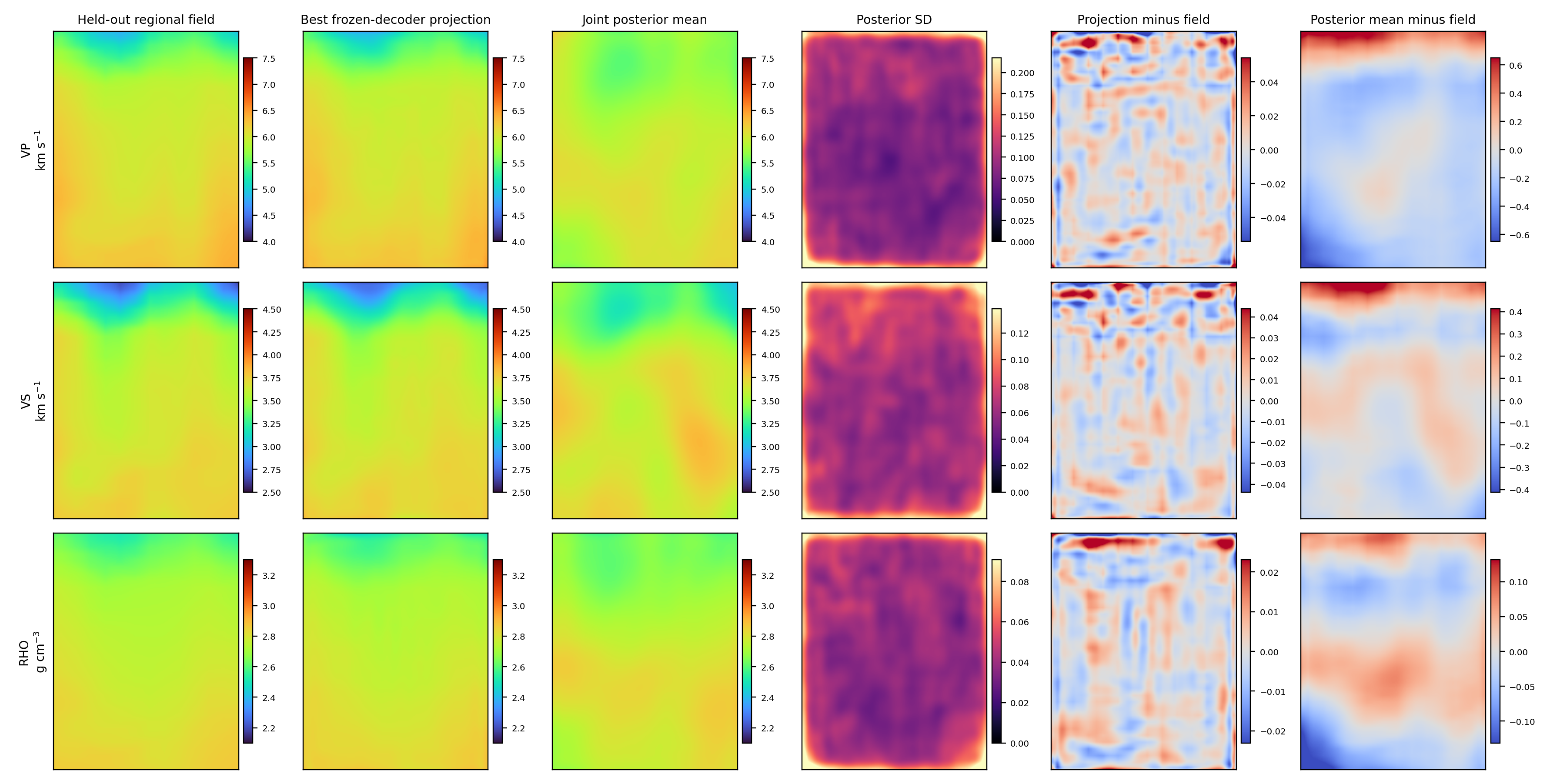}
\caption{Representation and inference in the held-out semi-synthetic regional experiment. Columns separate the public Wang-model truth, its best projection through the frozen decoder, the joint VI posterior mean, posterior SD, projection residual, and inference residual. The known truth makes the gap between decoder access and sparse-data recovery directly measurable and complements the measured-data field experiment in Section S9.}
\label{si:fig:regionalposterior}
\end{figure}

\section*{S9. Measured southern-California field inversion}

\subsection*{Observations and fixed splits}

The field profile extends from longitude $-117.45^{\circ}$, latitude $34.08^{\circ}$ to longitude $-116.20^{\circ}$, latitude $33.75^{\circ}$ and to 20 km depth. Main-text Figure 4 shows the measurement geometry, raw observations, and fixed partitions; Figure 5 shows the multiphysics posterior, matched field controls, and the surface-physics experiment. The seismic observations are not sampled from Wang, CVM-H, or another Earth model. Differential P arrivals are constructed from 2,111 event-phase files in the SCEDC AWS archive and 97 WebSTP fallback files \cite{SCEDC2013}. For each event, differencing picks at two stations cancels the catalog origin time exactly. The gravity values are USGS isostatic anomalies measured in 2000--2004 and released with principal facts for the San Gorgonio Pass structural knot \cite{LangenheimEtAl2025}. Surface-wave group velocities were measured from 30 days of archived CI-network LHZ waveforms at 6, 8, 10, 12, and 15 s periods \cite{CINetwork}.

Stable station hashes define the seismic partitions. The gravity holdout is a contiguous central profile block rather than an interleaved random sample. Training rows enter posterior fitting. Development rows fix one modality-level forward-discrepancy multiplier before inversion: 5.938 for differential arrivals, 1.471 for surface waves, and 6.262 for gravity. Holdout rows do not set a model coordinate, nuisance coefficient, optimizer state, or uncertainty scale (Table~\ref{si:tab:fieldcounts}). One period-specific surface-wave offset per period and a gravity intercept and linear trend are profiled from training rows only.

\begin{table}[H]
\centering
\caption{Measured observations in the southern-California field experiment.}
\label{si:tab:fieldcounts}
\begin{adjustbox}{max width=\linewidth}\begin{tabular}{lrrrp{5.2cm}}
\toprule
Modality & Training & Development & Holdout & Public source and measured quantity \\
\midrule
Differential P arrivals & \FieldTravelTrain & \FieldTravelDevelopment & \FieldTravelHoldout & SCEDC phase picks; station-pair differential arrival time. \\
Surface-wave dispersion & \FieldSurfaceTrain & \FieldSurfaceDevelopment & \FieldSurfaceHoldout & CI LHZ waveforms; 30-day ambient-noise group velocity. \\
Gravity & \FieldGravityTrain & \FieldGravityDevelopment & \FieldGravityHoldout & USGS principal facts; isostatic gravity anomaly. \\
\bottomrule
\end{tabular}\end{adjustbox}
\end{table}

\subsection*{Field parameterization and likelihood}

The decoder checkpoint is identical to that used in all manuscript experiments. Its SHA-256 is
\begin{quote}
\small\path{37d9494a4b9ccffb6f3c7ca31aefc207576dcbc6a03a28ec8ae814da6b2daf13}.
\end{quote}
For this field application, a deterministic monotonic depth trend is declared before inversion and the frozen generator contributes a bounded texture residual. With $H=20$ km, $\mathbf{x}=(s,h)$, and property $p\in\{V_P,V_S,\rho\}$, the implemented mapping is
\begin{equation}
q_p(s,h)=G_\theta([\bm z_s,\bm z_p])_{i(s,h)},\qquad
m_p(s,h)=\operatorname{clip}_{[\ell_p,u_p]}\!\left[\alpha_p+(\beta_p-\alpha_p)\frac{h}{H}+a_p\{2q_p(s,h)-1\}\right].
\label{si:eq:fieldadapter}
\end{equation}
The fixed tuples $(\alpha_p,\beta_p,a_p,\ell_p,u_p)$ are $(4.5,6.9,0.55,4.0,7.5)$ km s$^{-1}$ for $V_P$, $(2.7,4.2,0.35,2.5,4.5)$ km s$^{-1}$ for $V_S$, and $(2.25,3.10,0.17,2.1,3.3)$ g cm$^{-3}$ for density. Bilinear grid lookup $i(s,h)$ maps physical position to the native $256\times256$ decoder raster. Thus $\theta$, $H$, $i$, and all five adapter constants per property are fixed; $\boldsymbol{\zeta}=(\bm z_s,\bm z_{V_P},\bm z_{V_S},\bm z_\rho)$ is the only inferred spatial state. One period-specific surface-wave offset per period and the gravity intercept and slope are profiled from training observations only. This adapter supplies the direction of depth that is absent from orientation-augmented image pretraining. It is fixed independently of the observations and CVM-H, and no decoder weight is changed.

The forward calculation is a differentiable 2.5-dimensional approximation on the $256\times256$ profile. Differential P times use differences of straight-ray slowness integrals over 96 segments, retaining along-profile, cross-profile, and depth contributions to segment length. At period $T$, the surface operator path-averages 60 along-path samples with Gaussian depth weights centered at $\min(0.78T,0.82H)$, width $\max(1.2,0.36T)$, and fixed contributions $0.90V_S+0.08V_P+0.02(\rho-2.65)$. Gravity uses an 18-km finite-strike 2.5-dimensional point-cell density kernel with cross-profile offsets. A rank-24-plus-diagonal Gaussian variational family is optimized in the 256-dimensional joint coordinate state after three MAP restarts; 512 posterior coordinate draws are decoded into physical fields.

\subsection*{Untouched predictive evidence and field uncertainty}

Table~\ref{si:tab:fieldpredictive} reports the prespecified holdout scores. Differential-arrival RMSE decreases by \FieldTravelReduction\ and CRPS by \FieldTravelCRPSReduction. Gravity RMSE decreases by \FieldGravityReduction\ and CRPS by \FieldGravityCRPSReduction. Surface-wave RMSE changes by less than 0.1\%, isolating forward physics as the next testable bottleneck. Section S12 replaces the approximate kernel with layered Rayleigh physics while keeping the reusable generator fixed. Predictive interval inclusion incorporates the development-set discrepancy scale and should be read as data-space compatibility under the declared approximate likelihood, not as field-space calibration.

\begin{table}[H]
\centering
\caption{Held-out predictive performance for measured southern-California observations. CRPS is in the unit of each observation.}
\label{si:tab:fieldpredictive}
\small
\setlength{\tabcolsep}{3pt}
\begin{adjustbox}{max width=\linewidth}\begin{tabular}{lrrrrr}
\toprule
Modality & Prior RMSE & Posterior RMSE & Prior CRPS & Posterior CRPS & Posterior 90\% inclusion \\
\midrule
Differential P time & 1.255 s & 0.726 s & 0.700 s & 0.502 s & \FieldTravelPredictiveInclusion \\
Surface-wave velocity & 0.625 km s$^{-1}$ & 0.624 km s$^{-1}$ & 0.357 km s$^{-1}$ & 0.357 km s$^{-1}$ & \FieldSurfacePredictiveInclusion \\
Isostatic gravity & 7.635 mGal & 4.278 mGal & 4.236 mGal & 2.437 mGal & \FieldGravityPredictiveInclusion \\
\bottomrule
\end{tabular}\end{adjustbox}
\end{table}

\subsection*{Physical two-step Rayleigh-wave experiment}

The simplified-kernel null result creates a mechanistic test among representation, data, and forward physics. We therefore added a separate fixed-split experiment with published phase-dispersion data and layered Rayleigh calculations while retaining the image-pretrained decoder and its native $256\times256$ output. The public array contains \TwoStepSourceCurves\ local Rayleigh phase- and group-dispersion curves at \TwoStepPeriods\ periods from 3 to 16 s \cite{QiuEtAl2019,CaiEtAl2022}. Qiu et al. estimated these curves from one year of ambient-noise correlations recorded across southern California and Eikonal tomography; Cai et al. distribute the resulting data-derived dispersion array in their reproducibility repository. We use phase velocity alone because the distributed group velocities were derived from smoothed phase dispersion and are not statistically independent measurements. This dataset and likelihood are distinct from the five-period CI group velocities in the 915-row field experiment. Section S12 directly compares the simplified and layered operators and shows why prediction fit alone does not identify structural fidelity.

Kernel interpolation uses \TwoStepUsedSourceCurves\ distinct curves near the profile to form \TwoStepProfileSites\ sites along 121.2 km; the maximum distance to the nearest source cell is 3.17 km. Contiguous profile blocks assign \TwoStepTrainSites\ sites to inference, \TwoStepDevelopmentSites\ to uncertainty selection, and \TwoStepHoldoutSites\ to blind evaluation. Exact coordinate and source-curve audits find zero inference--development, inference--blind, or development--blind source overlap. A layered fundamental-mode Rayleigh solver replaces the former Gaussian period--depth kernel. Independent local inversions fit the 25 observed phase curves with mean RMSE \TwoStepLocalMeanRMSE\ km s$^{-1}$ and maximum RMSE \TwoStepLocalMaxRMSE\ km s$^{-1}$; only the 15 inference-site means and Laplace SDs enter the second-stage generator likelihood. The five development sites select its frozen uncertainty scale, and the five blind sites enter evaluation only.

Writing the local Laplace mean and SD at site $i$ and target depth $h_j$ as $\widehat v_{ij}$ and $\widehat\sigma_{ij}$, the generator update minimizes the negative log likelihood, up to an additive constant,
\begin{equation}
-\log L_2(\bm u)=\frac{1}{2}\sum_{i\in\mathcal T}\sum_j
\left[
\frac{m_{V_S}(s_i,h_j;\bm u)-\widehat v_{ij}}
{\gamma\sqrt{\widehat\sigma_{ij}^{2}+\sigma_0^2}}
\right]^2,
\label{si:eq:twosteplikelihood}
\end{equation}
MAP adds $\|\bm u\|^2/2$, and VI minimizes the Monte Carlo expected negative log likelihood plus $D_{\mathrm{KL}}[q(\bm u)\|\N(0,I)]$. The modular interface transmits six local Laplace means and marginal SDs per site into a diagonal site--depth likelihood; within-site Laplace covariance is not propagated. Here $\bm u\in\mathbb R^{128}$ is the single-field coordinate, $\mathcal T$ contains only the 15 inference sites, $h_j\in\{3,5,8,12,16,20\}$ km, and $\sigma_0=0.10$ km s$^{-1}$. The five development sites set $\gamma=\TwoStepModularScale$ and then select the smallest additive phase discrepancy achieving at least 90\% predictive inclusion, \TwoStepDiscrepancy\ km s$^{-1}$. Neither choice is refitted on the five blind sites. Training sites alone set the lower-boundary continuation. Both uncertainty choices are frozen before blind evaluation.

At the \TwoStepHoldoutSites\ blind sites, comprising 85 period--site values, phase-velocity RMSE decreases from \TwoStepPriorRMSE\ to \TwoStepPosteriorRMSE\ km s$^{-1}$, a \TwoStepRMSEDecrease\ reduction (Table~\ref{si:tab:twostepresult}; main-text Fig.~5). After the \TwoStepModularScale\ local-SD multiplier is fixed, development selects zero additional phase discrepancy and the resulting nominal 90\% intervals include \TwoStepPredictiveInclusion\ of blind observations. This is data-space predictive inclusion for one fixed geographic split, not repeated-truth simulation-based calibration. It shows that measured Rayleigh observations can update the frozen image coordinate system through layered local physics and improve predictions never used in inversion; the paired operator test in Section S12 determines which structural statement that fit supports.

\begin{table}[H]
\centering
\caption{Blind predictive performance and external structural comparisons for the physical two-step surface-wave inversion. Historical regional models are likelihood-excluded reference sections, not pointwise truth. Comparisons use depths of at least 3 km.}
\label{si:tab:twostepresult}
\small
\begin{adjustbox}{max width=\linewidth}\begin{tabular}{lrrrr}
\toprule
Quantity or reference & Prior RMSE & Posterior RMSE & Prior correlation & Posterior correlation \\
\midrule
Blind phase velocity & \TwoStepPriorRMSE\ km s$^{-1}$ & \TwoStepPosteriorRMSE\ km s$^{-1}$ & --- & --- \\
CVM-H $V_S$ & 0.286 km s$^{-1}$ & 0.196 km s$^{-1}$ & 0.615 & 0.719 \\
Wang 2020 $V_S$ & 0.247 km s$^{-1}$ & 0.166 km s$^{-1}$ & 0.778 & 0.855 \\
Berg 2021 $V_S$ & 0.295 km s$^{-1}$ & 0.209 km s$^{-1}$ & 0.592 & 0.690 \\
Berg 2018 $V_S$ & 0.329 km s$^{-1}$ & 0.264 km s$^{-1}$ & 0.398 & 0.409 \\
\bottomrule
\end{tabular}\end{adjustbox}
\end{table}

The posterior decreases RMSE and increases correlation relative to all four likelihood-excluded regional reference sections. Because these models differ in construction and resolution, this agreement is supportive structural context rather than an accuracy score. In Cai et al.'s released inversion code, each distributed uncertainty value $u$ is a dimensionless multiplier and the perturbation scale is $\sigma_v=0.0712335495u$ km s$^{-1}$, the source expression \texttt{var\_rms * var\_full}. The phase multipliers span 0.7019--2.0784, giving source-defined SDs of 0.0500--0.1481 km s$^{-1}$. Profile aggregation combines weighted measurement variance and local interpolation variance in quadrature; no extra velocity floor is added. In stage 2, the interpolated local Laplace SD is combined with the separate 0.10 km s$^{-1}$ modular floor before applying the development-selected multiplier. The calculation is modular: local Laplace means and marginal SDs, rather than raw correlations or full local posterior samples, define the stage-2 Gaussian likelihood. It is not the joint posterior conditional on raw ambient-noise cross-correlations, and the measured Rayleigh data constrain $V_S$ rather than independently validating $V_P$ or density.

\begin{figure}[H]
\centering
\includegraphics[width=0.98\linewidth]{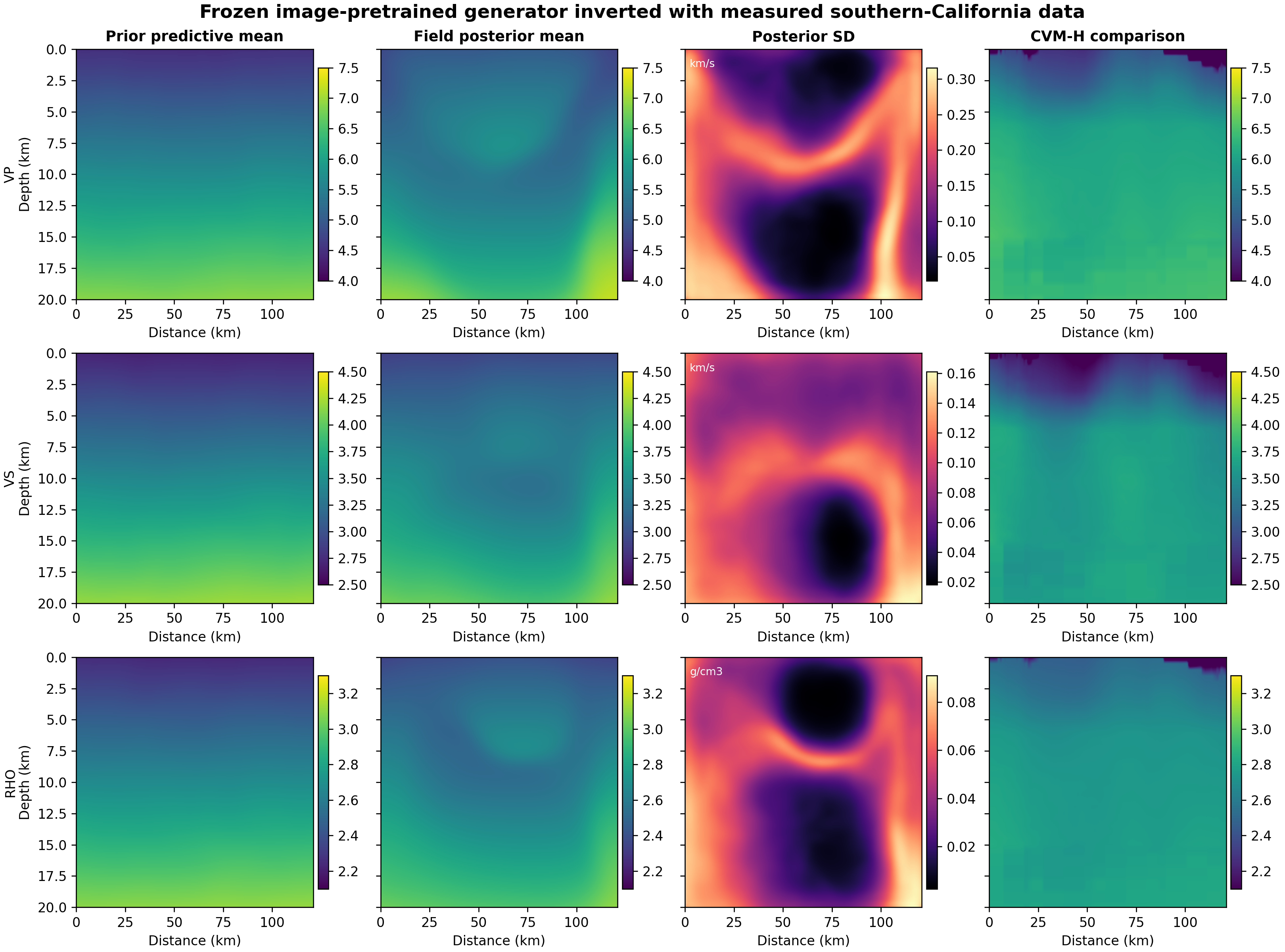}
\caption{Measured-data posterior and likelihood-excluded regional context. Rows show $V_P$, $V_S$, and density; columns show prior mean, posterior mean, posterior SD, and the CVM-H section. CVM-H is excluded from the likelihood and is not labeled as ground truth.}
\label{si:fig:fieldexternal}
\end{figure}

Figure~\ref{si:fig:fieldexternal} compares the prior mean, posterior mean, posterior SD, and external CVM-H section for all three properties. CVM-H is never used in inference. Posterior/CVM-H correlations are \FieldCVMHCorrelationVP, \FieldCVMHCorrelationVS, and \FieldCVMHCorrelationRho\ for $V_P$, $V_S$, and density; these are structural comparisons rather than pointwise real-Earth accuracy scores.

\begin{figure}[H]
\centering
\includegraphics[width=0.98\linewidth]{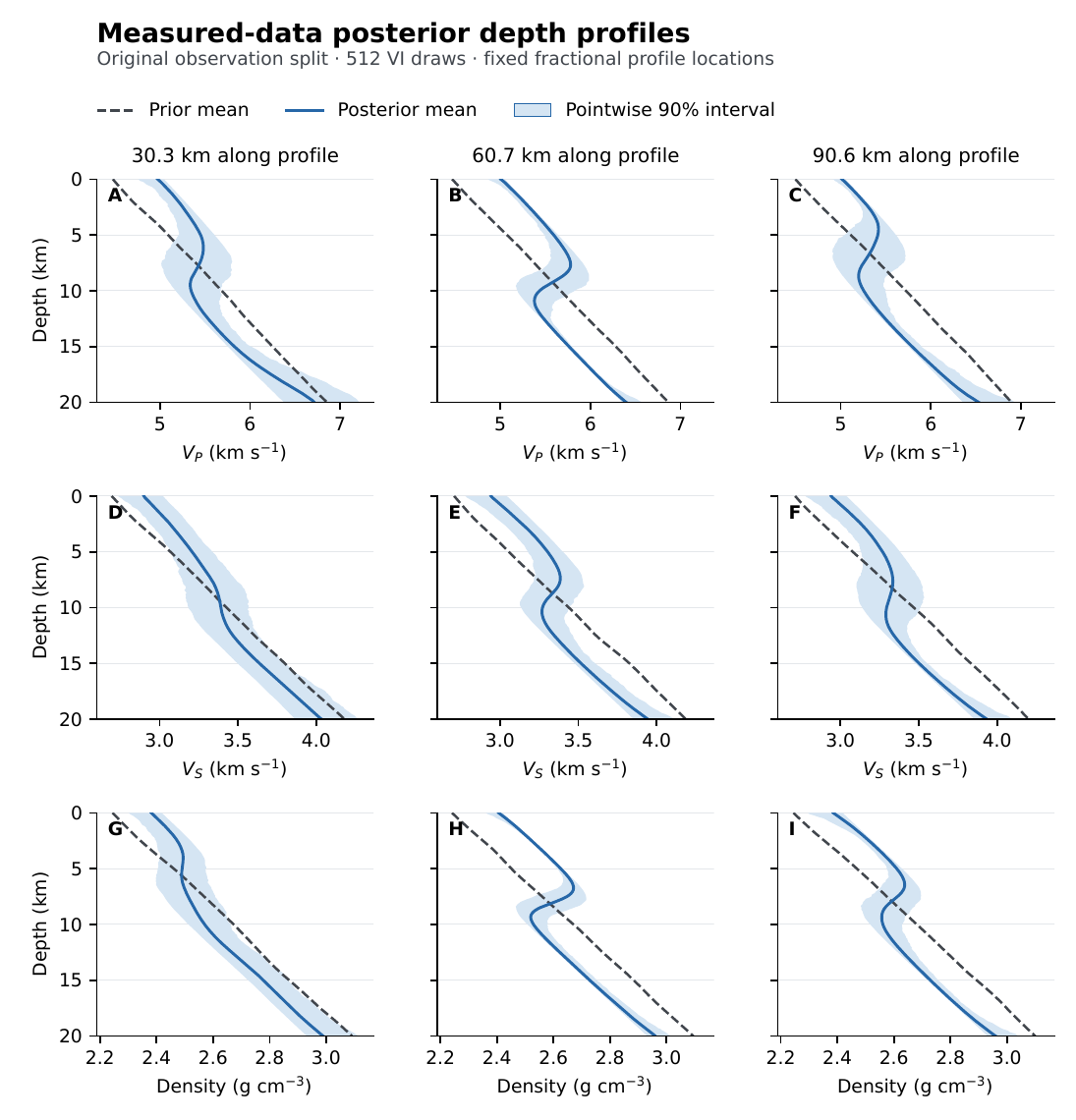}
\caption{Measured-data posterior depth profiles on the original observation split. Rows show $V_P$, $V_S$, and density; columns show fixed fractional positions along the 120.9-km profile. Dashed lines are prior means, solid lines posterior means, and shaded bands pointwise equal-tailed 90\% variational intervals. Horizontal scales match within each row. These locations are not independent wells or validation truths. The $V_S$ panels belong to the joint fit with simplified surface sensitivity, not the separate layered-Rayleigh update.}
\label{si:fig:fielddepthprofiles}
\end{figure}

Figure~\ref{si:fig:fielddepthprofiles} presents the saved pointwise intervals in physical depth coordinates. Locations are the nearest grid columns to one-quarter, one-half, and three-quarters of the full profile, selected by a fixed geometric rule rather than fit quality. The dashed prior mean distinguishes inherited depth dependence from the solid posterior update. Bands are the saved 5th and 95th percentiles of 512 variational field draws, not a Gaussian approximation from SD. They describe field uncertainty conditional on the model; they neither include observation noise as field uncertainty nor define simultaneous 90\% bands for entire profiles.

The posterior mean field SDs are 0.1484 km s$^{-1}$ for $V_P$, 0.0877 km s$^{-1}$ for $V_S$, and 0.0412 g cm$^{-3}$ for density. A second variational optimization with an independent seed changes the three posterior means by RMSE 0.0140 km s$^{-1}$, 0.0060 km s$^{-1}$, and 0.0030 g cm$^{-3}$; mean-field correlations exceed \FieldRepeatMinimumCorrelation\ and SD-map correlations range from 0.980 to 0.993. The independent rerun reproduces the posterior fields under the tested initialization, with the reported mean- and SD-map agreement.

\begin{figure}[H]
\centering
\includegraphics[width=0.96\linewidth]{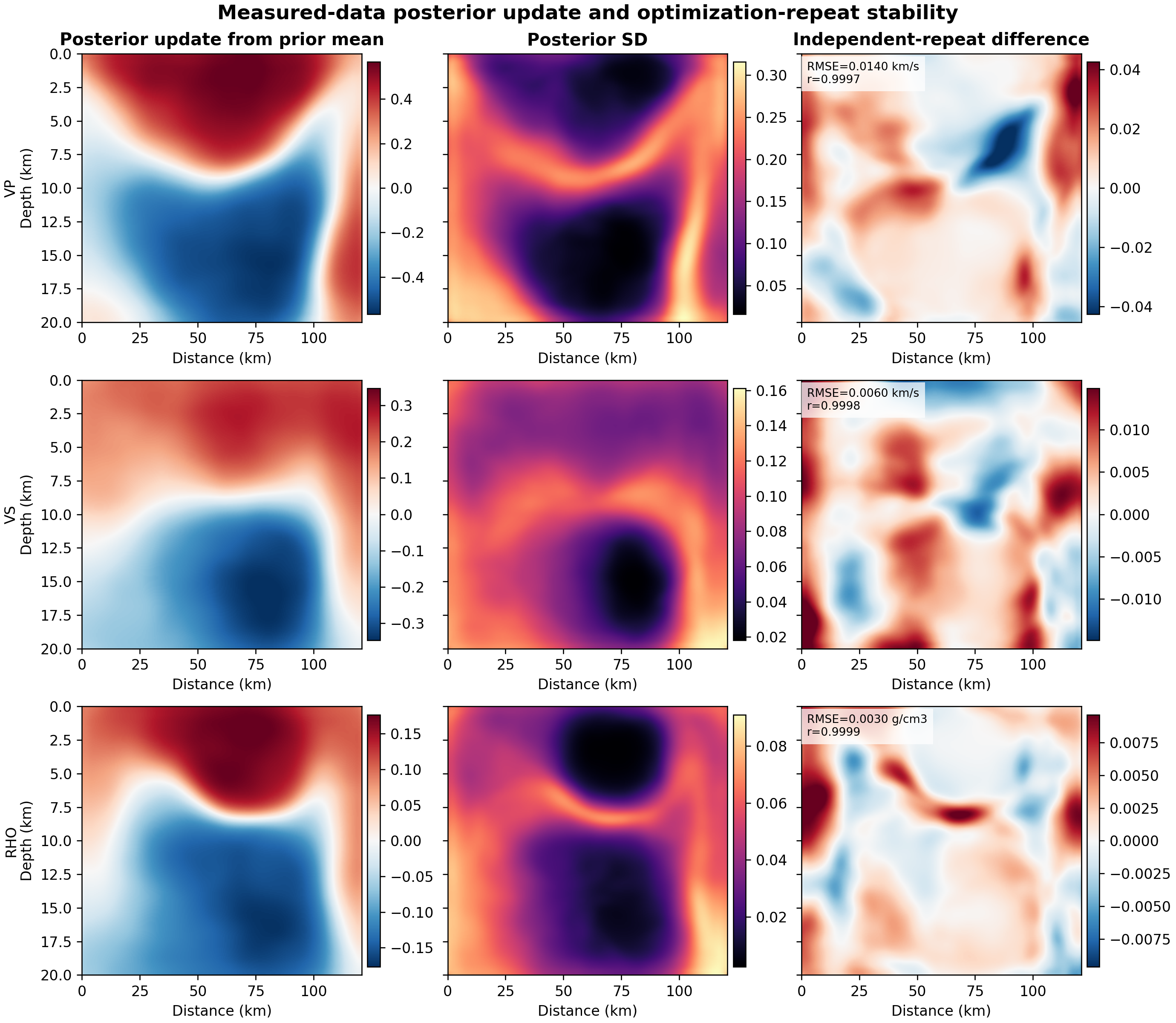}
\caption{Measured-data posterior update and independent-optimization stability. Rows show $V_P$, $V_S$, and density; columns show posterior-minus-prior mean, posterior SD, and the signed difference between independently optimized posterior means.}
\label{si:fig:fieldstability}
\end{figure}

\section*{S10. Matched measured-data parameterization controls}

The field counterfactual fixes $V_P$/$V_S$/density trends and residual amplitudes, inference/\allowbreak development/\allowbreak holdout rows, discrepancy multipliers, three MAP starts with 700 steps each, rank-24-plus-diagonal VI with 1,000 steps, 512 posterior draws, and nuisance projections. The controls are the image-pretrained decoder, five independently initialized frozen-random decoders of identical architecture, DCT and smooth Gaussian radial-basis (RBF) states, and the fixed depth trends. Every nontrivial model has 256 inferred coordinates, with 128 inputs per property; trend only has none. DCT and RBF amplitudes are scaled to the pretrained decoder's standard-normal prior field SD before inversion. Random decoders retain the same adapters without this induced-field SD matching. Their comparison tests pretrained weights against random weights under fixed adapters, not equal induced prior variance. Runtime is wall time on the same Apple MPS device.

The smooth control uses Gaussian basis functions proportional to $\exp\{-[(s-s_j)^2/\ell_s^2+\allowbreak(h-h_j)^2/\ell_h^2]/2\}$ on normalized profile coordinates. Each basis function is demeaned and normalized to unit root mean square; a scaled hyperbolic tangent and the fixed adapter produce bounded fields. Each property uses 128 basis functions on eight depth rows and 16 horizontal columns. This finite basis is not a Mat\'ern covariance model or eigendecomposition. Historical archive identifiers \texttt{matern\_rbf\_128}, \texttt{matern\_rbf\_256}, and \texttt{matern\_shared}, and the associated numerical macro names, are retained for compatibility. Throughout this manuscript they denote the smooth Gaussian RBF control. The scalar OpenFWI benchmark uses the same basis construction but its own all-model prior-SD matching described in Section S4.

\begin{table}[H]
\centering
\caption{Matched fixed-holdout field counterfactuals. Random-decoder entries are means across five seeds; the reported SD applies to RMSE. Width is mean total predictive SD.}
\label{si:tab:fieldcounterfactual}
\small
\setlength{\tabcolsep}{3pt}
\begin{adjustbox}{max width=\linewidth}\begin{tabular}{llrrrrr}
\toprule
Modality & Parameterization & RMSE & Seed SD & CRPS & 90\% inclusion & Width \\
\midrule
Differential P time & Image pretrained & 0.727 & --- & 0.502 & 1.000 & 1.640 \\
 & Frozen random & 1.172 & 0.023 & 0.665 & 0.930 & 1.635 \\
 & DCT--256 & 0.938 & --- & 0.575 & 0.992 & 1.652 \\
 & Smooth RBF--256 & 0.703 & --- & 0.495 & 1.000 & 1.643 \\
 & Trend only & 1.195 & --- & 0.675 & 0.930 & 1.635 \\
\addlinespace
Isostatic gravity & Image pretrained & 4.278 & --- & 2.437 & 0.922 & 4.933 \\
 & Frozen random & 6.666 & 0.336 & 3.975 & 0.772 & 4.761 \\
 & DCT--256 & 4.830 & --- & 2.792 & 0.922 & 5.026 \\
 & Smooth RBF--256 & 4.006 & --- & 2.325 & 0.984 & 5.490 \\
 & Trend only & 7.033 & --- & 4.226 & 0.750 & 4.726 \\
\bottomrule
\end{tabular}\end{adjustbox}
\end{table}

The image coordinates beat every frozen-random realization, DCT, and trend only for both informative field modalities. The RBF basis is slightly stronger on this smooth profile. The field and OpenFWI counterfactuals play complementary causal roles: measured data show that pretrained weights outperform architecture, cosine, and trend controls, while the curved/faulted benchmark distinguishes image morphology from a dimension-matched geophysical smooth basis.

\subsection*{Grouped and blocked predictive challenge}

The original station-hash split can share earthquakes across differential-time partitions. The grouped challenge instead hashes event identifiers, and forms at most one station pair per selected event. Train/development/holdout event overlap is exactly zero. Surface rows are grouped by the unordered station pair; path-family overlap is exactly zero. Gravity retains a contiguous holdout interval from 56.56 to 61.77 km along profile. Counts are 192/64/128 for differential arrivals, 127/30/54 for surface dispersion, and 192/64/64 for gravity.

\begin{table}[H]
\centering
\caption{Posterior RMSE under event-disjoint, station-pair-family-disjoint, and contiguous spatial holdouts.}
\label{si:tab:groupedfield}
\small
\begin{adjustbox}{max width=\linewidth}\begin{tabular}{lrrr}
\toprule
Parameterization & Differential P time (s) & Surface velocity (km s$^{-1}$) & Gravity (mGal) \\
\midrule
Image pretrained & 1.014 & 0.548 & 4.313 \\
DCT--256 & 1.259 & 0.544 & 4.821 \\
Smooth RBF--256 & 0.915 & 0.538 & 3.994 \\
Trend only & 1.705 & 0.550 & 7.033 \\
\bottomrule
\end{tabular}\end{adjustbox}
\end{table}

Image coordinates retain their advantage over DCT and trend-only predictions after correlated events and station-pair paths are removed. The RBF ordering also persists. Predictive inclusion for the image model is 0.984 for differential arrivals, 0.907 for surface dispersion, and 0.906 for gravity. The complete split audit is stored in \texttt{grouped\_field\_holdout/split\_overlap\_audit.json}.

\section*{S11. Cross-property transfer through explicit shared coordinates}

The shared-state ablation uses an independent Wang2020 section and identical frozen decoder weights. Wang2020 entered neither image pretraining nor checkpoint selection; the decoder remains frozen in every posterior fit. Three state constructions are compared: 64 shared plus 64 property-specific coordinates (256 joint dimensions); 96 shared plus 32 property-specific coordinates (192 joint dimensions); and independent 128-dimensional coordinates for every property (384 joint dimensions). Within each truth, all constructions use identical fields, observations, noise realizations, optimization budgets, and draw counts. Their variational objective is the Monte Carlo expected negative log likelihood plus the analytic $D_{\mathrm{KL}}[q(\boldsymbol{\zeta})\|\N(\bm0,\bm I)]$; the standard-normal latent prior is counted exactly once.

The decisive intervention withholds dispersion. Traveltime depends only on $V_P$ and gravity only on density; forward predictions are numerically invariant to every controlled $V_S$ modification. Any $V_S$ update must therefore travel through shared coordinates rather than a direct likelihood term. That transfer is positive and measurable: relative to the full-independent Wang posterior, the raw 64/64 and 96/32 states reduce $V_S$ mean error from \IndependentFullVSRMSE\ to \SharedVSRMSE\ and \StrongSharedVSRMSE\ km s$^{-1}$. Relative to each construction's own prior, their posterior SDs contract by \SharedVSContraction\ and \StrongSharedVSContraction. A separate joint-dimension-matched factorial gives 0.219 versus 0.235 km s$^{-1}$ RMSE and 71.9\% contraction versus 5.0\% expansion for the shared64 and dimension-matched independent states. The independent control has 85, 85, and 86 active inputs per property, with remaining inputs fixed to zero. Equal total dimension therefore does not imply equal marginal property priors. This is a complementary dimension-budget comparison, not an isolated coupling effect under identical marginals. This factorial is archived in \path{experiments/pnas_strengthening_20260825/shared_state_transfer/shared_state_transfer.json} and uses a distinct frozen noise realization from the paired guard below. The raw intervals nevertheless under-include the guard's fixed truth. Table~\ref{si:tab:sharedtransfer} shows that the coupling guard retains the transferred accuracy and substantial contraction while restoring near-nominal conditional spatial inclusion across held-out structural-mismatch tests.

\begin{table}[H]
\centering
\caption{Paired $V_S$ transfer with dispersion withheld. Five CVM-H-derived cases are used only for development; Wang2020 is the untouched test. Inclusion and width describe pointwise equal-tailed 90\% intervals for each fixed truth. Negative contraction denotes slight prior-relative expansion.}
\label{si:tab:sharedtransfer}
\small
\setlength{\tabcolsep}{3.5pt}
\begin{adjustbox}{max width=\linewidth}\begin{tabular}{llrrrr}
\toprule
Truth & Posterior & Mean RMSE (km s$^{-1}$) & SD contraction & 90\% inclusion & Width (km s$^{-1}$) \\
\midrule
CVM-H development & Raw shared 64/64 & \CVMHRawSharedVSRMSE & \CVMHRawSharedVSContraction & \CVMHRawSharedVSCoverage & \CVMHRawSharedVSWidth \\
 & Raw shared 96/32 & \CVMHStrongRawSharedVSRMSE & \CVMHStrongRawSharedVSContraction & \CVMHStrongRawSharedVSCoverage & \CVMHStrongRawSharedVSWidth \\
 & Full independent & \CVMHIndependentFullVSRMSE & \CVMHIndependentFullVSContraction & \CVMHIndependentFullVSCoverage & \CVMHIndependentFullVSWidth \\
 & Guarded 64/64 & \CVMHCouplingGuardVSRMSE & \CVMHCouplingGuardVSContraction & \CVMHCouplingGuardVSCoverage & \CVMHCouplingGuardVSWidth \\
 & Guarded 96/32 & \CVMHStrongCouplingGuardVSRMSE & \CVMHStrongCouplingGuardVSContraction & \CVMHStrongCouplingGuardVSCoverage & \CVMHStrongCouplingGuardVSWidth \\
\addlinespace
Wang2020 test & Raw shared 64/64 & \SharedVSRMSE & \SharedVSContraction & \RawSharedVSCoverage & \RawSharedVSWidth \\
 & Raw shared 96/32 & \StrongSharedVSRMSE & \StrongSharedVSContraction & \StrongRawSharedVSCoverage & \StrongRawSharedVSWidth \\
 & Full independent & \IndependentFullVSRMSE & \IndependentFullVSContraction & \IndependentFullVSCoverage & \IndependentFullVSWidth \\
 & Guarded 64/64 & \CouplingGuardVSRMSE & \CouplingGuardVSContraction & \CouplingGuardVSCoverage & \CouplingGuardVSWidth \\
 & Guarded 96/32 & \StrongCouplingGuardVSRMSE & \StrongCouplingGuardVSContraction & \StrongCouplingGuardVSCoverage & \StrongCouplingGuardVSWidth \\
\bottomrule
\end{tabular}\end{adjustbox}
\end{table}

We construct the guard as a linear pool of shared-state and full-independent posterior draws. Candidate mixture weights are evaluated only on five CVM-H-derived development cases: the public section, aligned morphology, a shifted boundary, a $V_S$-specific anomaly, and deliberate decoupling. The rule selects the largest shared weight for which all five spatial-inclusion fractions lie in the predefined 0.88--0.95 interval; if none is admissible, it minimizes the worst absolute deviation from 0.90. The selected weight is a development-set safeguard, not a posterior probability that cross-property sharing is physically correct. For 64/64 sharing, weight \CouplingGuardSharedWeight\ satisfies the criterion: inclusion is \CVMHCouplingGuardVSCoverage\ on the public CVM-H section and \CVMHCouplingGuardStressCoverage\ across its four structural stress cases. The weight is then frozen. On independent Wang2020, the guard raises inclusion from \RawSharedVSCoverage\ to \CouplingGuardVSCoverage, retains \CouplingGuardVSContraction\ SD contraction, and lowers mean error from \SharedVSRMSE\ to \CouplingGuardVSRMSE\ km s$^{-1}$. Its four untouched Wang-derived mismatch tests span \CouplingGuardStressCoverage. Thus the shared coordinate carries useful cross-property information, and the prespecified guard converts that transfer into intervals that remain informative under structural mismatch.

The same frozen-weight calculation is repeated for 96/32 sharing. No candidate places all five CVM-H-derived development-case inclusions inside the target band, so the prespecified minimax fallback selects weight \StrongCouplingGuardSharedWeight. It gives \CVMHStrongCouplingGuardVSCoverage\ inclusion on the public CVM-H section and \CVMHStrongCouplingGuardStressCoverage\ across the four CVM-H-derived stress cases. After freezing, independent-Wang inclusion is \StrongCouplingGuardVSCoverage, contraction remains \StrongCouplingGuardVSContraction, and mean error is \StrongCouplingGuardVSRMSE\ km s$^{-1}$; the four Wang-derived mismatch cases span \StrongCouplingGuardStressCoverage. Descriptive spatial PIT histograms and tail fractions are archived for every development and test case.

The property-specific blocks preserve deviations inside each shared model. When the active likelihood is exactly invariant to $V_S$, the observations cannot identify a $V_S$-only anomaly; any $V_S$ update is therefore the declared shared-coordinate prediction. The linear pool retains a development-selected fraction of independent-state draws, limiting cross-property transfer under mismatch. Its components are combined after decoding into a common physical field space, rather than concatenating coordinates of different dimensions. A $V_S$-sensitive dispersion likelihood supplies direct information about such departures. Generic images learn the morphological map, not a $V_P$--$V_S$--density law; heterogeneous observations communicate through explicit shared coordinates acting through that learned map rather than a fixed petrophysical equation.

\subsection*{Measured cross-observable intervention}

A separate intervention tests transfer against measured Rayleigh phase velocities. The pretrained decoder, 256-by-256 grid, physical adapters, P-arrival event partitions, gravity partitions, and development error scales remain fixed. Only the 192 inference P-arrival pairs and 192 inference gravity observations enter the likelihood. All surface-wave terms are removed, including the simplified kernel and the local layered-profile targets. The test therefore asks whether P and gravity data improve shear-sensitive predictions through shared coordinates, rather than through a direct dispersion constraint.

The targets are the five Qiu/Cai sites numbered 20--24 in the existing blocked split, with 17 periods per site. We retain its archived 3.5-km spatial aggregation and propagated source-curve/interpolation uncertainty. Those curves entered neither likelihood nor scale selection in this intervention. They were examined in the earlier surface-wave analysis, however, and the original 10-km interpolation neighborhood had already used nearby source curves. This is a fixed retrospective cross-observable test, not a new prospective blind experiment. A protocol and input hashes were locked before the new fits and scoring. No sites, periods, seeds, checkpoints, or mixture weights were selected using the new scores.

Three conditions use 256 total coordinates: the shared image decoder with 64 shared and three 64-private blocks; the independent image decoder with 85, 85, and 86 nonzero-capable inputs for $V_P$, $V_S$, and density; and the same shared smooth radial-basis control used in Section S10. Unused independent decoder inputs are zero. The independent model matches total dimension but not each property's marginal prior. Accordingly, the full 128-input shared-image prior also supplies a marginal-matched no-transfer reference. Physical adapters, P/gravity discrepancy scales, optimizer budgets, and seeds are identical across conditions.

The absence of a $V_S$ likelihood imposes an exact computational check. With independent standard-normal coordinate priors and likelihood-active and inactive blocks $\bm z_a$ and $\bm z_u$, the posterior factorizes as $p(\bm z_a,\bm z_u\mid\bm d)=p(\bm z_a\mid\bm d)p(\bm z_u)$. We fit only the active block and draw the inactive block exactly from its prior, independently of the fitted coordinates. The shared models have 192 active and 64 inactive coordinates; the matched independent model has 171 active and 85 inactive coordinates. The independent $V_S$ posterior must therefore equal its prior. Paired prior and posterior draws reuse identical inactive coordinates, making this invariance test exact rather than subject to Monte Carlo error.

Each of three optimizer seeds (20260905--20260907) uses three 700-step MAP starts, 1,000 rank-24-plus-diagonal VI steps with four Monte Carlo samples per step, and 512 posterior draws. A MAP start is selected by the training posterior objective only. All seeds are retained. We forward-project a fixed evenly spaced subset of 128 draws, interpolating the decoded section at depths 0, 1.5, 3, 5, 8, 12, 16, and 20 km. A common extension uses $V_S=4.27$ and 4.48 km s$^{-1}$ at 30 and 45 km, $V_P=1.73V_S$, and Brocher density \cite{Brocher2005}. These deep values are prescribed, not inferred by the 0--20-km decoder. Layered Rayleigh predictions use 0.5-km layers and \texttt{disba} \cite{LuuDisba}.

The primary diagnostic varies $V_S$ while holding $V_P$ and density at their common fixed depth trends. It isolates the shear-mediated transfer channel by preventing updated $V_P$ or density from directly changing Rayleigh predictions. This intervention distribution is distinct from the secondary full three-property posterior predictive distribution. Scores comprise site-balanced phase RMSE, empirical-ensemble CRPS with the released measurement uncertainty, nominal 90\% predictive inclusion, and interval width. No new discrepancy scale is fitted to the Rayleigh targets. Periods within a site and neighboring sites are correlated; optimizer seeds assess numerical sensitivity, not independent geological replication or simulation-based calibration. Solver failures are recorded and cause scoring to stop instead of silently dropping samples.

\begin{figure}[H]
\centering
\includegraphics[width=0.98\linewidth]{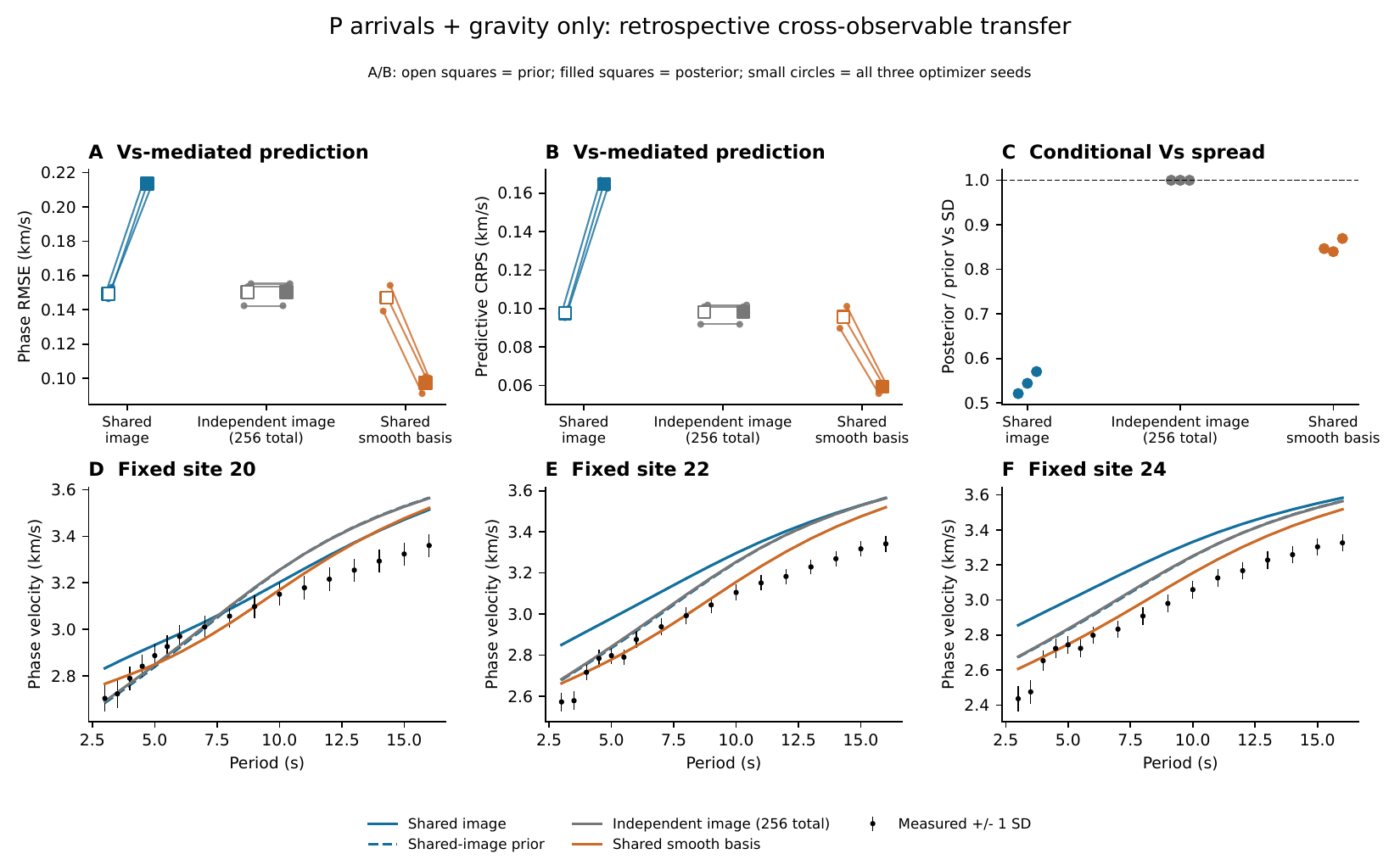}
\caption{\textbf{Measured cross-observable transfer distinguishes reusable morphology from regional coupling.} (A,B) Primary shear-mediated phase RMSE and CRPS over all five fixed sites and 17 periods, with all three optimizer seeds retained. Open and filled squares denote mean prior and posterior scores; small circles show individual seeds. (C) Posterior/prior $V_S$ SD ratios over eight depths at the five sites. The independent condition remains exactly one. (D--F) Predictive means at the preselected first, middle, and last sites; all five sites contribute to A--C. Error bars show the propagated source-observation SD, not uncertainty in an independent geological truth. All curves use fixed $V_P$ and density trends to isolate the $V_S$ channel.}
\label{si:fig:measuredtransfer}
\end{figure}

The measured intervention separates information transfer from predictive benefit (Fig.~\ref{si:fig:measuredtransfer}). Shared-image $V_S$ changes by 0.191 km s$^{-1}$ RMS in the posterior mean over the scored depth/site grid, while its mean SD contracts by 45.5\%. Nevertheless, its shear-mediated phase RMSE increases from \MeasuredTransferImagePriorRMSE\ to \MeasuredTransferImagePosteriorRMSE\ km s$^{-1}$. All three seeds give the same direction, with posterior RMSE 0.211--0.216 km s$^{-1}$. CRPS worsens from \MeasuredTransferImagePriorCRPS\ to \MeasuredTransferImagePosteriorCRPS\ km s$^{-1}$ and predictive inclusion decreases from \MeasuredTransferImagePriorInclusion\ to \MeasuredTransferImagePosteriorInclusion. Independent-image $V_S$, its primary predictions, and all primary scores are exactly unchanged, as required by the likelihood factorization. The shared smooth basis improves primary RMSE from \MeasuredTransferSmoothPriorRMSE\ to \MeasuredTransferSmoothPosteriorRMSE\ and CRPS from \MeasuredTransferSmoothPriorCRPS\ to \MeasuredTransferSmoothPosteriorCRPS\ km s$^{-1}$. Table~\ref{si:tab:measuredtransfer} reports seed-averaged scores, not uncertainty bounds on population performance.

\begin{table}[H]
\centering
\caption{Retrospective shear-mediated prediction from P arrivals and gravity only. Entries average three optimizer seeds on the same five sites. Prior/posterior columns use the same physical prediction rule and released observation uncertainty; no phase discrepancy is fitted.}
\label{si:tab:measuredtransfer}
\small
\begin{adjustbox}{max width=\linewidth}\begin{tabular}{lrrrrrr}
\toprule
 & \multicolumn{2}{c}{RMSE (km s$^{-1}$)} & \multicolumn{2}{c}{CRPS (km s$^{-1}$)} & \multicolumn{2}{c}{90\% inclusion} \\
Condition & Prior & Posterior & Prior & Posterior & Prior & Posterior \\
\midrule
Shared image & \MeasuredTransferImagePriorRMSE & \MeasuredTransferImagePosteriorRMSE & \MeasuredTransferImagePriorCRPS & \MeasuredTransferImagePosteriorCRPS & \MeasuredTransferImagePriorInclusion & \MeasuredTransferImagePosteriorInclusion \\
Independent image & \MeasuredTransferIndependentPriorRMSE & \MeasuredTransferIndependentPosteriorRMSE & \MeasuredTransferIndependentPriorCRPS & \MeasuredTransferIndependentPosteriorCRPS & \MeasuredTransferIndependentPriorInclusion & \MeasuredTransferIndependentPosteriorInclusion \\
Shared smooth basis & \MeasuredTransferSmoothPriorRMSE & \MeasuredTransferSmoothPosteriorRMSE & \MeasuredTransferSmoothPriorCRPS & \MeasuredTransferSmoothPosteriorCRPS & \MeasuredTransferSmoothPriorInclusion & \MeasuredTransferSmoothPosteriorInclusion \\
\bottomrule
\end{tabular}\end{adjustbox}
\end{table}

The secondary full three-property predictions retain the ordering: shared-image RMSE changes from 0.153 to 0.227 km s$^{-1}$, independent-image from 0.144 to 0.173, and shared smooth-basis from 0.143 to 0.071. The independent model can change full Rayleigh predictions through updated $V_P$ and density despite an unchanged $V_S$ distribution, demonstrating why the primary intervention is needed. No dispersion solve fails. Meanwhile, the shared-image P-arrival and gravity holdout RMSEs remain 1.007--1.018 s and 4.238--4.360 mGal. Successful prediction of the assimilated modalities therefore does not by itself determine the direction of cross-property transfer.

These results concern this fixed shared-coordinate prior, 2.5-dimensional section, and prescribed deep continuation. They do not identify a unique cause among structural coupling, depth trends, deep continuation, and physical approximation. Nor do they test the independently selected known-truth mixture as a field-calibrated guard. They establish that this image coupling should not substitute for measured shear-sensitive constraints on the present profile. Section S9 supplies the complementary positive result: adding direct layered-Rayleigh information updates the same frozen decoder and improves spatially withheld phase predictions. The physical prediction operators differ between the interventions, so their RMSE magnitudes are not a matched algorithm ranking.

\section*{S12. Direct surface-wave operator comparison}

The direct comparison uses the same 17 periods from 3 to 16 s. Forty synthetic profiles are represented at 10 depth nodes. Each profile adds a Gaussian perturbation with covariance $(0.12\ {\rm km\,s}^{-1})^2\exp[-|h_i-h_j|/(10\ {\rm km})]$ and a random linear trend to the inversion prior, then enforces 2.0--4.8 km s$^{-1}$ bounds. Noisy fundamental-mode Rayleigh phase velocities are generated with \texttt{disba} \cite{LuuDisba} at 0.02 km s$^{-1}$ noise. Each profile is inverted twice: once with the layered eigenproblem and once with the simplified Gaussian period--depth averaging rule. The two fits use identical depth nodes, prior, smoothing penalty, bounds, four restarts, and conversions $V_P=1.73V_S$ and Brocher density; only the forward operator changes. The measured-data comparison repeats this paired design at all 25 blocked Qiu profile sites.

\begin{table}[H]
\centering
\caption{Paired simplified-kernel and layered-Rayleigh operator results. Measured Qiu sites lack a structural truth, so only curve-space scores are reported.}
\label{si:tab:surfaceoperator}
\footnotesize
\setlength{\tabcolsep}{3.8pt}
\begin{adjustbox}{max width=\linewidth}\begin{tabular}{llrrr}
\toprule
Setting & Operator & Phase RMSE & Structural $V_S$ RMSE & Curve $r$ \\
 & & (km s$^{-1}$) & (km s$^{-1}$) & \\
\midrule
Known truth ($n=40$) & Simplified kernel & 0.01570 & 0.18917 & --- \\
 & Layered Rayleigh & 0.01583 & 0.08852 & --- \\
Measured Qiu ($n=25$) & Simplified kernel & 0.01491 & --- & 0.99734 \\
 & Layered Rayleigh & 0.01441 & --- & 0.99746 \\
\bottomrule
\end{tabular}\end{adjustbox}
\end{table}

Phase-curve RMSE differs by only 0.00013 km s$^{-1}$ under known truth and 0.00049 km s$^{-1}$ at measured sites. Known truth nevertheless shows a 53.2\% reduction in structural $V_S$ error for layered physics. Combined with the blind-site improvement of the two-step generator update, the result establishes a precise boundary: physical dispersion makes represented $V_S$ informative, whereas an approximate kernel can fit the data and still distort the recovered depth profile.

\section*{S13. Field variational-posterior sensitivity}

The measured inversion uses a rank-24-plus-diagonal Gaussian posterior. Formal reruns use ranks 8 and 48, a second rank-24 initialization, and a common 512-draw summary. Each run retains the field checkpoint, observations, discrepancy scales, MAP restart budget, VI steps, and random-number manifest. A nested draw-count audit recomputes predictive scores and field summaries from 128, 256, and 512 draws of the rank-24 posterior.

\begin{table}[H]
\centering
\caption{Sensitivity of measured-data posterior summaries to VI rank and initialization. Field-mean RMSE is relative to the reference rank-24 run after decoding.}
\label{si:tab:visensitivity}
\scriptsize
\setlength{\tabcolsep}{2pt}
\begin{adjustbox}{max width=\linewidth}\begin{tabular}{lrrrrr}
\toprule
Run & $V_P$ mean gap & $V_S$ mean gap & Density mean gap & Travel RMSE & Gravity RMSE \\
\midrule
Rank 8 & \VIRankEightVPMeanGap & \VIRankEightVSMeanGap & \VIRankEightRhoMeanGap & \VIRankEightTravelRMSE & \VIRankEightGravityRMSE \\
Rank 24, second seed & \VISecondSeedVPMeanGap & \VISecondSeedVSMeanGap & \VISecondSeedRhoMeanGap & \VISecondSeedTravelRMSE & \VISecondSeedGravityRMSE \\
Rank 48 & \VIRankFortyEightVPMeanGap & \VIRankFortyEightVSMeanGap & \VIRankFortyEightRhoMeanGap & \VIRankFortyEightTravelRMSE & \VIRankFortyEightGravityRMSE \\
\bottomrule
\end{tabular}\end{adjustbox}
\end{table}

Posterior means and untouched predictive scores remain stable across these changes. SD maps vary more than means because factor rank changes covariance expressivity; the result is reported as approximation sensitivity rather than field calibration. The controlled HMC experiment supplies the separate known-truth probability check.

\section*{S14. External medical CT mechanism illustration}

LoDoPaB-CT is an open low-dose parallel-beam benchmark derived from LIDC/IDRI chest CT data \cite{LeuschnerEtAl2021}. Five patient-distinct official test slices are reconstructed on a $256\times256$ grid from 1,000 views and a deterministic 200-view subset. Filtered backprojection is compared with nonnegative total-variation primal--dual hybrid gradient. No generative decoder and no posterior distribution are used.

\begin{figure}[H]
\centering
\includegraphics[width=0.98\linewidth]{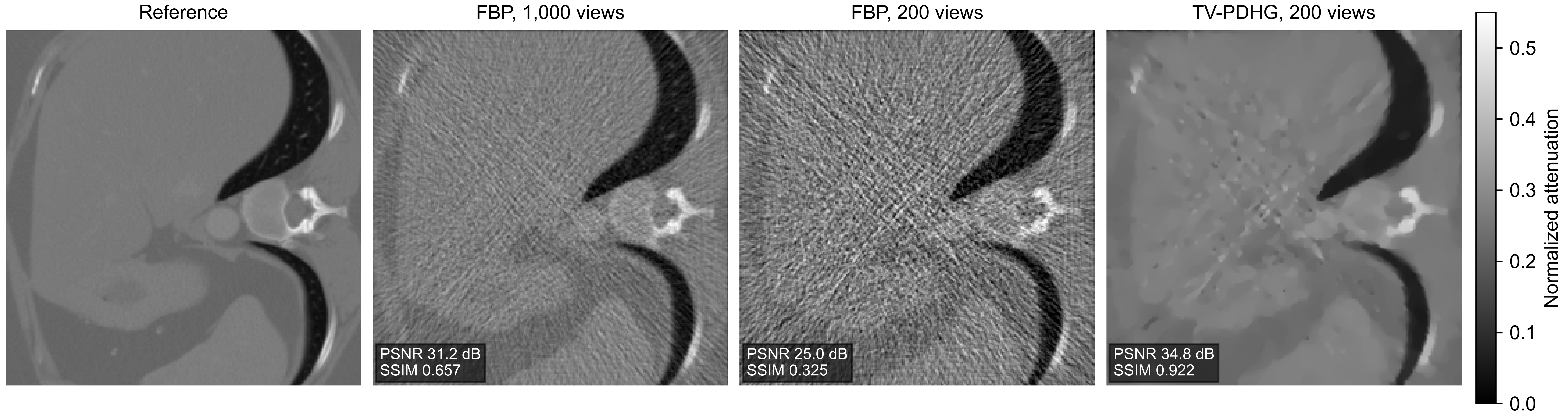}
\caption{Separate medical CT underdetermination illustration. From left: reference, 1,000-view filtered backprojection, 200-view filtered backprojection, and TV-PDHG from 200 views. This deterministic calculation is not validation of the Earth generator or a medical posterior.}
\label{si:fig:medicalct}
\end{figure}

Across five patients, 200-view FBP gives PSNR $\MedicalSparsePSNRMean\pm\MedicalSparsePSNRSD$ dB and SSIM $\MedicalSparseSSIMMean\pm\MedicalSparseSSIMSD$. TV-PDHG gives $\MedicalTVPSNRMean\pm\MedicalTVPSNRSD$ dB and SSIM $\MedicalTVSSIMMean\pm\MedicalTVSSIMSD$. Four of five slices improve under TV. The calculation illustrates structural regularization only; admissible anatomy and medical uncertainty require domain-specific validation. These numbers are absent from the main manuscript.

\section*{S15. Reproducibility, provenance, and boundaries}

The public reproducibility release is available at \url{https://huggingface.co/cangyeone/relational-geophysics/tree/c07aee68a6b5653247a054738fb2667c80c81f04/image_morphology_earth_posteriors}. Its \path{PUBLIC_RELEASE_MANIFEST.json} records file-level hashes and public-source families. The release contains the 12-, 64-, and 128-coordinate weights, retained posterior artifacts, processed public observations, fixed native-grid operators, and numerical verification scripts. No domestic Chinese observational datasets or coordinates are distributed. Third-party data retain their original terms; full image archives and original EMC model files are retrieved from the providers rather than mirrored. Run \texttt{bash reproduce.sh audit} from the project root to verify the pristine release. Before a full recomputation, use \texttt{python fetch\_public\_models.py} to retrieve the four hash-pinned regional reference models and follow the package README for image downloads and environment setup. The public field-inversion workflow starts from the retained observation tables and native-grid operators, not a fresh reconstruction of the historical raw-waveform processing pipeline.

The primary experiment directory is
\begin{quote}
\ttfamily\small experiments/image\_foundation\_prior\_128\_20260814/.
\end{quote}
The canonical sequence is:
\begin{Verbatim}[fontsize=\small,breaklines=true,breakanywhere=true]
python download_data.py --root data/raw
python experiment.py --config config_full.json --out_dir run_full_128
python experiment.py --config config_full_12.json --out_dir run_full_12
python experiment.py --config config_full_64.json --out_dir run_full_64
python evaluate_external_geophysics.py --checkpoint run_full_128/foundation_prior_inference.pt
python evaluate_openfwi.py --checkpoint run_full_128/foundation_prior_inference.pt
python evaluate_external_geophysics.py \
  --checkpoint run_full_12/foundation_prior_inference.pt \
  --out-dir external_test_run_12
python evaluate_external_geophysics.py \
  --checkpoint run_full_64/foundation_prior_inference.pt \
  --out-dir external_test_run_64
python evaluate_openfwi.py \
  --checkpoint run_full_12/foundation_prior_inference.pt \
  --out-dir openfwi_test_run_12
python evaluate_openfwi.py \
  --checkpoint run_full_64/foundation_prior_inference.pt \
  --out-dir openfwi_test_run_64
python compare_latent_ablation.py
python run_revision_controls.py
python run_random_decoder_multiseed.py
python run_multiphysics_posterior.py \
  --checkpoint run_full_128/foundation_prior_inference.pt \
  --out-dir multiphysics_posterior_run_128
python project_heldout_regional_truth.py
python audit_wang_projection.py
python run_shared_coordinate_sensitivity.py --truth-regime heldout_regional \
  --out-dir shared_coordinate_sensitivity_heldout_run \
  --device cpu --steps 1200 --posterior-samples 128
python rerun_hmc_full_preconditioned.py
python analyze_hmc_diagnostics.py
python analyze_hmc_draw_stability.py
python analyze_hmc_spatial_pit.py
python analyze_vi_hmc_dispersion.py
python make_revision_manuscript_figures.py
python ../real_socal_foundation128_20260817/run_field_inversion.py \
  --checkpoint run_full_128/foundation_prior_inference.pt \
  --out-dir ../real_socal_foundation128_20260817/run_4_verified
python ../real_socal_foundation128_20260817/make_manuscript_field_figure.py
python ../two_step_surface_wave_qiu_20260822/download_qiu_data.py
python ../two_step_surface_wave_qiu_20260822/run_two_step_inversion.py \
  --config ../two_step_surface_wave_qiu_20260822/config_blocked.json \
  --out-dir ../two_step_surface_wave_qiu_20260822/run_blocked
python ../two_step_surface_wave_qiu_20260822/validate_results.py \
  --run-dir ../two_step_surface_wave_qiu_20260822/run_blocked
python ../pnas_strengthening_20260825/run_field_parameterization_counterfactuals.py
python ../pnas_strengthening_20260825/run_grouped_field_holdout.py
python ../pnas_strengthening_20260825/run_shared_state_transfer.py \
  --device cpu --steps 1200 --posterior-samples 128
python ../pnas_strengthening_20260825/compare_surface_wave_operators.py \
  --config ../two_step_surface_wave_qiu_20260822/config_blocked.json \
  --profile ../two_step_surface_wave_qiu_20260822/run_blocked/profile_dispersion.npz \
  --out-dir ../pnas_final_20260826/surface_operator_comparison_blocked
python ../pnas_final_20260826/audit_surface_wave_chain.py
python ../pnas_final_20260826/run_shared_state_mismatch_calibration.py \
  --device cpu --steps 1200 --posterior-samples 128
python ../pnas_final_20260826/run_coupling_uncertainty_guard.py \
  --device cpu --steps 1200 --posterior-samples 128
python ../pnas_final_20260826/run_openfwi_inverse_benchmark.py \
  --device auto --steps 500 --posterior-samples 128 --truths-per-family 4
python ../real_socal_foundation128_20260817/run_field_inversion.py \
  --out-dir ../pnas_strengthening_20260825/field_vi_rank8_seed0 --vi-rank 8
python ../real_socal_foundation128_20260817/run_field_inversion.py \
  --out-dir ../pnas_strengthening_20260825/field_vi_rank24_seed2 --vi-rank 24 --seed-offset 2
python ../real_socal_foundation128_20260817/run_field_inversion.py \
  --out-dir ../pnas_strengthening_20260825/field_vi_rank48_seed0 --vi-rank 48
python ../pnas_strengthening_20260825/analyze_field_vi_sensitivity.py
python ../pnas_strengthening_20260825/make_final_pnas_figures.py
\end{Verbatim}

The retained artifacts include image identity manifests, source metadata, configs, checkpoint hashes, regional and OpenFWI projections, the matched complex-geometry inverse benchmark, counterfactual fields, typical-shell coordinates, variational states, forward matrices, noise values, HMC latent chains, transition diagnostics, decoded posterior fields, spatial PIT maps and interval curves, the VI--HMC dispersion audit, field-level chain comparisons, shared-state mismatch tests and coupling guard, measured-observation tables, fixed and grouped split audits, matched field parameterizations, blocked surface-operator comparisons, posterior-rank reruns, field posterior predictive draws, and figure payloads. The measured-data experiment is in \path{experiments/real_socal_foundation128_20260817/}; strengthening and final-audit experiments are in \path{experiments/pnas_strengthening_20260825/} and \path{experiments/pnas_final_20260826/}. Each formal result records the applicable checkpoint or source inputs, observation definitions, and numerical summary.

The public release uses repository-relative paths and includes the 128-dimensional training and inference code, all declared configurations, the compact frozen checkpoint, measured-data observation operators, HMC latent chains and summary fields, the complete physical two-step Rayleigh implementation, fixed partitions, and stored numerical reports. The recomputable 5.9-GB decoded-HMC cache and public raw image archives are excluded; the retained latent chains and frozen checkpoint regenerate that cache. Running \texttt{python experiments/verify\_release\_package.py} verifies required files, SHA-256 identities, dimensions, result gates, and the four-figure evidence chain before any expensive recomputation.

The validated scope comprises three scalar Earth properties, three differentiable 2.5-dimensional operators, and blind prediction of measured southern-California differential arrivals, gravity, and phase dispersion. Full-wave and three-dimensional likelihoods define the next physical test without changing the reusable-coordinate construction.

\clearpage
\section*{S16. Spatial covariance, coordinate sensitivity, and data weighting}

\subsection*{A geophysical interpretation of the generative prior}

The coordinates parameterize coordinated changes in a field, not independent physical values at every grid cell. Each 128-input scalar generator produces 65,536 values on its native grid. This construction restricts admissible structures to the generator's image; it does not assert that every possible Earth field has 128 degrees of freedom. The 12/64/128 comparison in Section S2 measures representational capacity, whereas the sensitivity calculation below measures what a specified acquisition can resolve within the deployed representation. These are different questions.

Let $\bm m=g(\boldsymbol{\zeta})$ denote the full three-property mapping, including the frozen decoder, shared/private assembly, fixed physical adapters, and bounds. The prior on $\bm m$ is the pushforward of $\boldsymbol{\zeta}\sim\N(\bm0,\bm I)$. Its covariance is
\begin{equation}
\bm C_m=\mathbb E[(g(\boldsymbol{\zeta})-\bar{\bm m})
                 (g(\boldsymbol{\zeta})-\bar{\bm m})^\mathsf T],
\qquad \bar{\bm m}=\mathbb E[g(\boldsymbol{\zeta})].
\end{equation}
Independent latent coordinates therefore do not imply independent grid values. In a neighborhood of a reference coordinate, a first-order approximation gives
\begin{equation}
\bm C_m^{\rm local}\simeq \bm J_g\bm C_\zeta\bm J_g^\mathsf T,
\qquad \bm J_g=\frac{\partial g}{\partial\boldsymbol{\zeta}}.
\label{si:eq:localcovariance}
\end{equation}
For the latent prior, $\bm C_\zeta=\bm I$; a posterior has its own covariance. Equation~\ref{si:eq:localcovariance} is a local approximation, not an identity for the nonlinear ensemble covariance. In particular, its rank bound must not be applied to the global covariance of a curved nonlinear image.

The generative prior incorporates the role played by spatial model regularization in conventional inversion. It need not be Gaussian or describable by one stationary correlation length. Its structure comes jointly from image-trained weights, the latent distribution, the shared/private construction, and the adapters; images alone do not establish a geophysical joint distribution. The current implementation adds the standard-normal latent prior once, rather than adding a second physical-space Gaussian penalty. Private coordinates allow departures from shared patterns. No $V_P/V_S$ interval or velocity--density regression is imposed in the joint generative inversion. Fixed conversions used by the separate local Rayleigh solver belong to that operator experiment, not to generator training.

\begin{figure}[H]
\centering
\includegraphics[width=\linewidth]{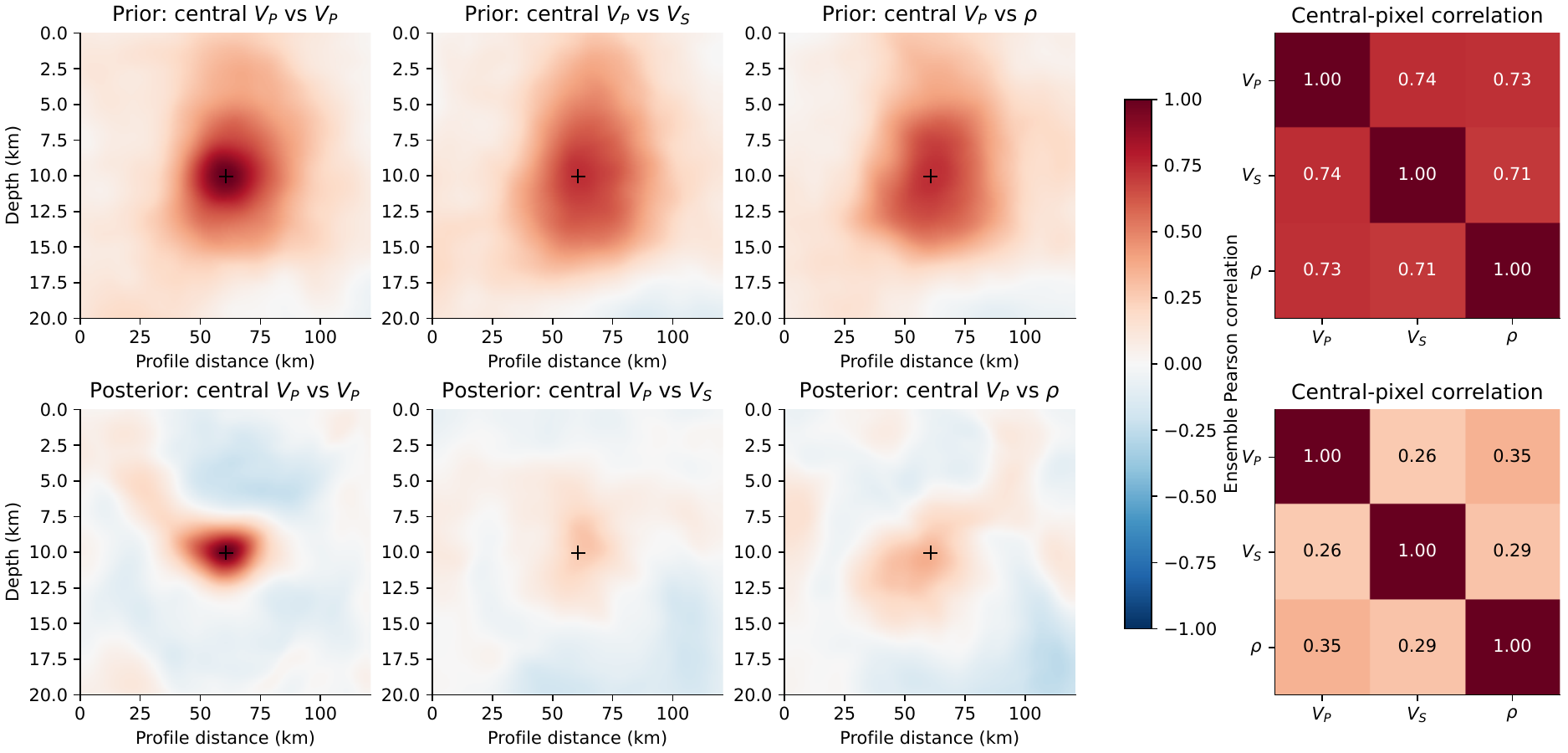}
\caption{Spatial and cross-property correlations induced by the generator. Top: 512 standard-normal joint-state draws. Bottom: 512 draws from the saved field VI distribution. The first three columns correlate each field with $V_P$ at the fixed central cell (cross). The last column compares the three properties at that cell. All panels use the same $[-1,1]$ scale. Correlations describe the model distribution under the declared adapters and coupling, not measured petrophysical relationships.}
\label{si:fig:generatorcovariance}
\end{figure}

We illustrate the induced correlations with 512 prior draws and 512 fresh draws from the archived field VI distribution, all using seed 20260911. The reference is the fixed central grid cell, index $(128,128)$ on the zero-indexed $256\times256$ profile, not a selected well or geological target. Figure~\ref{si:fig:generatorcovariance} shows correlation with its $V_P$ value throughout all three fields and the three-property correlation at the same cell. Off-diagonal central correlations span 0.710--0.736 in the prior and 0.258--0.349 in the fitted VI posterior. The posterior changes both the spatial pattern and cross-property dependence. These are conditional correlations across generated models, not empirical rock-property regressions or evidence of calibration. Using the first 256 rather than all 512 draws changes the correlation maps by RMS 0.043 in the prior and 0.047 in the posterior.

\subsection*{How observations constrain structural coordinates}

At the archived field VI mean $\boldsymbol{\zeta}_*$, we differentiate the full native-grid mapping and the three original field forward operators in all 256 directions. Centered differences use steps 0.01 and 0.005; their field-Jacobian relative $L_2$ difference is $7.13\times10^{-4}$. Six fixed random scalar projections of the data derivative agree with reverse-mode automatic differentiation to relative $L_2$ error at most $7.22\times10^{-4}$. No reduced grid or newly trained network is used. Derivatives include the fixed physical adapters.

For modality $k$, let $\bm D_k$ contain the effective observation SDs on its inference rows and let $\bm Q_k$ have orthonormal columns spanning its whitened nuisance design. Define
\begin{equation}
\bm A_k=(\bm I-\bm Q_k\bm Q_k^\mathsf T)\bm D_k^{-1}
\left.\frac{\partial F_k(g(\boldsymbol{\zeta}))}{\partial\boldsymbol{\zeta}}\right|_{\boldsymbol{\zeta}_*},
\qquad
\bm H_{\rm data}=\sum_k\bm A_k^\mathsf T\bm A_k.
\end{equation}
This calculation uses only the 192 differential-arrival, 143 group-velocity, and 192 gravity inference rows. Nuisance projection is identical to the original fit. An eigenvalue $\lambda>1$ means that local data curvature exceeds the unit latent-prior precision along that combination of coordinates. The continuous effective dimension is $d_{\rm eff}=\sum_j\lambda_j/(1+\lambda_j)$.

\begin{table}[H]
\centering
\caption{Local data information at the saved VI mean. Counts and effective dimensions refer to coordinate combinations, not individually identified latent variables, independent geological parameters, or global resolution.}
\label{si:tab:localinformation}
\begin{adjustbox}{max width=\linewidth}\begin{tabular}{lrrr}
\toprule
Observations & Inference rows & Directions with $\lambda>1$ & $d_{\rm eff}$\\
\midrule
Differential P arrivals & 192 & 4 & 4.47\\
Group velocities & 143 & 1 & 1.22\\
Gravity & 192 & 6 & 6.36\\
Joint & 527 & 9 & 9.93\\
\bottomrule
\end{tabular}\end{adjustbox}
\end{table}

\begin{figure}[H]
\centering
\includegraphics[width=0.94\linewidth]{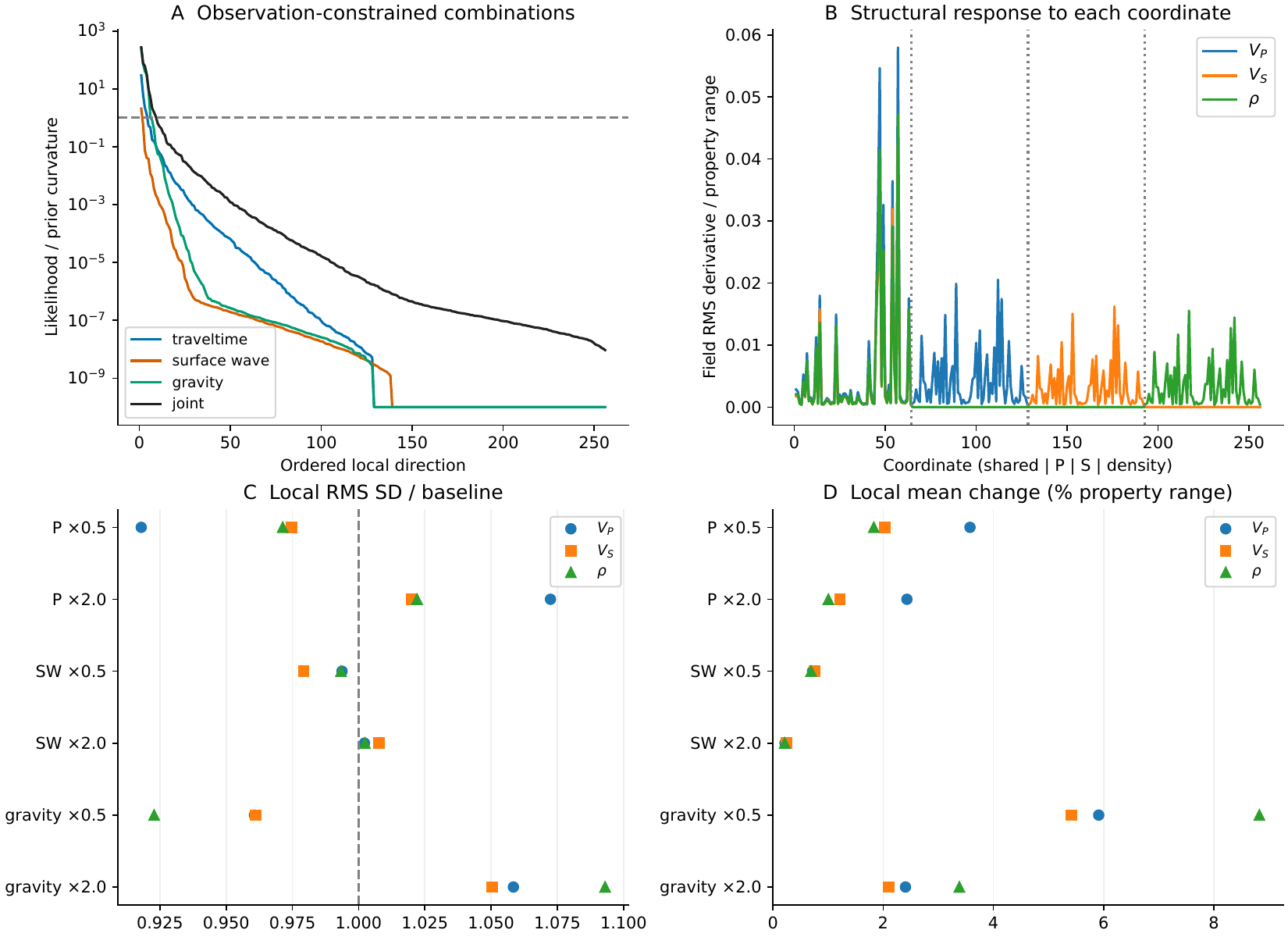}
\caption{Native-grid sensitivity and local error-scale perturbations. (A) Eigenvalues of each modality's and the joint whitened, nuisance-projected data information. The dashed line is unit latent-prior precision; values below $10^{-10}$ are displayed at that floor and have no rank interpretation. (B) RMS physical-field derivative divided by its fixed property range. Dotted lines delimit the shared and three private blocks. (C,D) RMS-SD ratios and mean changes in the affine-linearized problem when one modality's error SD is multiplied by 0.5 or 2; other scales remain fixed. These are local Gaussian diagnostics, not nonlinear VI/HMC reruns or new holdout validations.}
\label{si:fig:latentinformation}
\end{figure}

The joint likelihood constrains more local combinations than any individual modality, while most directions remain prior-dominated. Representation capacity and observational resolution therefore remain distinct: 128 inputs per property provide structural flexibility, not a claim that this survey independently resolves 128 parameters. The small group-velocity information pertains to the original simplified period--depth operator and its profiled period offsets, not to the separate layered-Rayleigh update. Figure~\ref{si:fig:latentinformation} also shows the RMS field response to each coordinate, including the exact absence of response in unrelated private blocks. These local spectra are not a rank determination for the nonlinear Earth manifold or a replacement for posterior sampling.

\subsection*{Where the relative data weights enter}

With fixed error scales, the implemented negative log likelihood, up to coordinate-independent constants, is
\begin{equation}
\Phi(\boldsymbol{\zeta})=\frac12\sum_k
\left\|(\bm I-\bm Q_k\bm Q_k^\mathsf T)\bm D_k^{-1}
       [F_k(g(\boldsymbol{\zeta}))-\bm d_k]\right\|_2^2 .
\label{si:eq:weightedlikelihood}
\end{equation}
The MAP objective adds $\|\boldsymbol{\zeta}\|^2/2$; VI adds the corresponding latent KL term once. There is no extra division by modality sample count. Relative influence depends on observation uncertainty, retained data count, physical sensitivity, and nuisance projection. A unit conversion leaves the residual unchanged when both observations/predictions and their SDs are converted consistently.

Base SDs combine the released measurement uncertainties and declared approximation floors. Before posterior fitting, development rows set an additional modality multiplier using $\max\{1,\operatorname{median}(|r/\sigma|)/0.67448975\}$, subject to the configured cap. Residuals here are evaluated at the fixed zero-coordinate prediction, with nuisance coefficients fitted only on inference rows. The frozen multipliers are 5.938 for differential arrivals, 1.471 for group velocities, and 6.262 for gravity. They are effective residual-scale adjustments; they do not uniquely separate instrument noise, forward error, and prior mismatch. Holdout rows select none of these values.

The current model uses diagonal observation errors. Nuisance projection removes specified offsets and trends; grouped holdouts test predictions on separated observation groups. Neither supplies a full observation-error covariance. Hierarchical Bayesian inversion can instead infer noise scales and correlations jointly with Earth structure \cite{BodinEtAl2012}. Such an extension would require the likelihood normalization and hyperpriors; it is not implemented here. The present posterior is conditional on the frozen scales.

To quantify local sensitivity to those choices, we multiply one modality's effective SD by 0.5 or 2 while retaining all observations, adapters, nuisance subspaces, and the standard-normal prior. Write $\bm r_k$ for the whitened, nuisance-projected residual at $\boldsymbol{\zeta}_*$. For scale factors $a_k$, the affine-linearized problem has
\begin{equation}
\begin{aligned}
\bm C_a&=\left(\bm I+\sum_k a_k^{-2}\bm A_k^\mathsf T\bm A_k\right)^{-1},\\
\boldsymbol{\delta}_a&=-\bm C_a\left(\boldsymbol{\zeta}_*+
                      \sum_k a_k^{-2}\bm A_k^\mathsf T\bm r_k\right).
\end{aligned}
\end{equation}
All changes are relative to the optimum of this same linearized problem at $a_k=1$, not relative to a purported refit of the nonlinear posterior. Field mean changes use $\bm J_g(\boldsymbol{\delta}_a-\boldsymbol{\delta}_1)$; RMS field SD uses Equation~\ref{si:eq:localcovariance} with $\bm C_a$. The baseline Newton displacement has norm 7.49, underscoring that the VI mean is not the optimum of this affine surrogate and that finite local changes must not be interpreted as validated nonlinear solutions.

Across all six scale scenarios and three properties, the local RMS-SD ratio is 0.918--1.093. Mean sensitivity is greatest when gravity SD is halved: RMS changes are 0.207 km s$^{-1}$ in $V_P$, 0.108 km s$^{-1}$ in $V_S$, and 0.106 g cm$^{-3}$ in density. These correspond to 5.9\%, 5.4\%, and 8.8\% of the fixed property ranges. The result quantifies local trade-offs; it does not establish global weight invariance. No decoder weight, archived posterior, or main-text predictive score is changed.

The protocol, derivative arrays, source hashes, ensemble summaries, figure payloads, and deterministic validation are in \path{experiments/prior_geometry_20260911/}. Run \texttt{./reproduce.sh geometry} from the paper root. The derivative calculation and ensemble analysis are reproducible additions to the existing experiment, not new observations or repeated-truth calibration.

\clearpage
\section*{S17. Empirical map of the tested conditions}

Table~\ref{si:tab:evidencemap} separates representability, inverse recovery, and information transfer. It synthesizes existing results; it is not an additional experiment or a fitted decision rule for choosing a prior. The crossed rankings depend on the tested structures, acquisition, and inference approximations. The map identifies comparisons that the current evidence can resolve, rather than extrapolating a universal applicability boundary.

\begin{table}[H]
\centering
\small
\caption{Tested conditions, comparative outcomes, and supported interpretations. RMSE units are given for physical predictions; structural errors are normalized unless otherwise stated. Sections S3--S4 and S9--S12 provide complete protocols and results.}
\label{si:tab:evidencemap}
\begin{adjustbox}{max width=\linewidth}\begin{tabular}{>{\raggedright\arraybackslash}p{3.2cm}>{\raggedright\arraybackslash}p{6.2cm}>{\raggedright\arraybackslash}p{6.3cm}}
\toprule
Tested condition & Outcome & Interpretation \\
\midrule
Representation: 256 OpenFWI targets & Projection RMSE: image 0.0524, DCT 0.2130, frozen random 0.3775. & Pretrained weights encode useful Earth-like spatial patterns. This is representability, not 256 inverse recoveries or field replications. \\[5pt]
Inverse recovery: four CurveVel-A targets & Image/RBF field RMSE 0.078/0.113; interface F1 0.404/0.247. & Image coordinates recover these curved-velocity targets better under the fixed line-integral acquisition. \\[5pt]
Inverse recovery: four CurveFault-A targets & Image/RBF field RMSE 0.089/0.056; interface F1 0.510/0.555. Aggregate errors across both families: 0.083/0.085. & RBF recovers these faulted targets better. The nearly tied aggregate masks opposite family rankings; four targets per family do not define all curved or faulted geology. \\[5pt]
One measured 120.9-km profile & Original image/RBF holdout errors: 0.726/0.703 s and 4.278/4.006 mGal. Grouped: 1.014/0.915 s and 4.313/3.994 mGal. & The frozen image decoder supports measured-data inference; RBF predicts better. The 43\%/44\% gains compare image posterior predictions with the pretrained image prior before observation updating. \\[5pt]
Measured P-and-gravity-only sharing & Shear-mediated Rayleigh error: image 0.149$\to$0.213 km s$^{-1}$; smooth control 0.147$\to$0.097 km s$^{-1}$. Independent-image prediction is unchanged. & Fixed image coupling transmits information in the wrong direction here. Decoder reuse and beneficial cross-property transfer are separate claims. \\[5pt]
Direct layered-Rayleigh update & Phase error 0.208$\to$0.151 km s$^{-1}$ over 85 values at five blind sites. & Direct shear-sensitive constraints improve prediction using the same decoder. Correlated periods within five sites do not establish independent-region generalization. \\
\bottomrule
\end{tabular}\end{adjustbox}
\end{table}

The prior-prediction baseline already contains pretrained image weights and fixed physical adapters. It is not the frozen-random (untrained) network control. Both prior and posterior predictions use the same inference-row nuisance-adjustment procedure and frozen error scales. Here ``prior'' refers to the latent-field distribution before its observational update, not a wholly data-free predictive model. Improvements over this baseline quantify field-coordinate updating within the representation; comparisons against RBF quantify relative predictive performance.

The covariance and derivative diagnostics in Section S16 show how this learned map constrains fields and interacts with each likelihood. They do not isolate a causal role for image spectra, edge statistics, or layered textures. No matched geological-image-trained generator is evaluated. Thus the study demonstrates that structural knowledge can originate outside geological training models, without claiming that this source is superior to geological pretraining. The field map is conditional on the simplified 2.5-dimensional joint operators and the separate modular layered-Rayleigh update, not a validation of fully three-dimensional or full-wave inference.

\end{document}